\documentclass[11pt]{amsart}

\usepackage{amsmath,amssymb}

\newcommand{\N}{\ensuremath{\mathbb{N}}}

\newcommand{\R}{\ensuremath{\mathbb{R}}}

\newcommand{\order}{\ensuremath{\mathcal{O}}}

\newcommand{\abs}[1]{\ensuremath{\left\lvert #1 \right\rvert}}
\newcommand{\norm}[1]{\ensuremath{\lvert\lvert #1\rvert\rvert}}

\newcommand{\bsym}[1]{\ensuremath{\boldsymbol{#1}}}
\renewcommand{\vec}[1]{\ensuremath{\mathbf{#1}}}
\newcommand{\set}[1]{\ensuremath{\mathcal{#1}}}

\newcommand{\inv}{\ensuremath{^{-1}}}
\newcommand{\Prob}{\ensuremath{\mathbb{P}}}
\newcommand{\Expect}{\ensuremath{\mathbb{E}}}

\newcommand{\diag}{\ensuremath{\mathrm{diag}}}

\newcommand{\esssup}{\ensuremath{\mathrm{ess}\sup}}

\newcommand{\mxm}{\ensuremath{\text{maximize}}}

\newcommand{\wrt}{\ensuremath{\mathrm{with \; respect \; to}}}

\newcommand{\ith}{\ensuremath{^\text{th}}}

\newtheorem{theorem}{Theorem}[section]

\newtheorem{lemma}[theorem]{Lemma}

\newtheorem{corollary}[theorem]{Corollary}
\newtheorem{definition}{Definition}[section]
\newtheorem{assumption}{Assumption}[section]

\title[On the Existence of Equilibrium Prices]{On the Existence of Bertrand-Nash Equilibrium Prices Under Logit Demand}

\author{W. Ross Morrow}
\author{Steven J. Skerlos}

\address{Iowa State University}
\email[W. Ross Morrow]{wrmorrow@iastate.edu}

\address{The University of Michigan, Department of Mechanical Engineering}
\email[Steven J. Skerlos]{skerlos@umich.edu}

\thanks{Significant portions of this research were undertaken while W. Ross Morrow was a Ph.D. student in Mechanical Engineering at the University of Michigan. The National Science Foundation, the University of Michigan Transportation Research Institute's Doctoral Studies Program, and Iowa State University provided support for this research. The authors wish to thank Fred Feinberg, Erin MacDonald, Jong-Shi Pang, Che-Lin Su, and Norman Shiau for helpful suggestions.}

\begin{document}
\maketitle

\begin{abstract}
This article proves the existence of equilibrium prices in Bertrand competition with multi-product firms using the Logit model of demand. The most general proof, an application of the Poincare-Hopf Theorem, does not rely on restrictive assumptions such as single-product firms, firm homogeneity or symmetry, homogeneous product costs, or even concavity of the utility function with respect to prices. This proof relies on new conditions for the indirect utility function, along with fixed-point equations derived from the first-order conditions and a direct analysis of the second-order conditions that proves the uniqueness of profit-maximizing prices. The degree to which our conditions are as weak as possible is discussed. Models with finite purchasing power and convex total costs are also addressed. Analysis of equilibrium prices for multi-product firms with constant unit costs suggests that Bertrand-Nash equilibrium cannot adequately describe multi-product pricing in differentiated product markets. 
\end{abstract}

% \keywords{Bertrand-Nash Equilibrium, Differentiated Product Markets, Logit, Poincare-Hopf Theorem}

\maketitle

% \tableofcontents

%%%%%%%%%%%%%%%%%%%%%%%%%%%%%%%%%%%%%%%%%%%%%%%%%%%%%
%%%%%%%%%%%%%%%%%%%%%%%%%%%%%%%%%%%%%%%%%%%%%%%%%%%%%
%%%%%%%%%%%%%%%%%%%%%%%%%%%%%%%%%%%%%%%%%%%%%%%%%%%%%

\section{Introduction}

	%%%%%%%%%%%%%%%%%%%%%%%%%%%%%%%%%%%%%%%%%%%%%%%%%%%%%%%%%%
	%%%%%%%%%%%%%%%%%%%%%%%%%%%%%%%%%%%%%%%%%%%%%%%%%%%%%%%%%%
	%%%%%%%%%%%%%%%%%%%%%%%%%%%%%%%%%%%%%%%%%%%%%%%%%%%%%%%%%%
	
	Bertrand competiton has been a prominent paradigm for the empirical study of differentiated product markets for over twenty years \cite{Bresnahan87, Berry95, Goldberg95, Verboven99, Hausman02, Goldberg98, Sudhir01, Petrin02, Berry04, Smith04, Aguirregabiria06}. Most of these empirical applications have been undertaken without theoretical assurances of the existence, uniqueness, and even ``plausibility'' of Bertrand-Nash equilibrium prices. This article proves the existence of equilibrium prices for differentiated product market models based on the Logit model under weak conditions on the (indirect) utility function and convex total costs (i.e., unit costs that increase with volume). Further analysis reveals some counter-intuitive properties of equilibrium prices that suggest more complex models than Logit are required to adequately model differentiated product markets in price competition. 
	
	Most existing theoretical analyses of Bertrand competition are based on assumptions too restrictive to suit empirical applications of Bertrand competition and that obscure potentially counterintuitive properties of equilibrium. For example, there are few theoretical studies that consider multi-product firms (see, e.g., \cite{Shaked90, Anderson92b, Anderson06}), but real firms in differentiated product markets almost always offer more than one product. Theoretical analyses of Bertrand-Nash equilibrium prices have also typically relied on homogeneity or ``symmetry'' between firms with respect to the costs and ``values''\footnote{Authors in the theoretical literature use the term ``quality'' to describe the non-price utility of a product \cite{Mussa78}. This use of the term is confounded with the way it would be interpreted by many engineers, marketers, operations researchers, or laypeople as a measure of reliability.} of the products offered. Real markets are heterogeneous with respect to the number of products offered, the values consumers derive from these products, and the costs with which these products are produced. In one analysis, Anderson \& dePalma \cite{Anderson06} state that
	\begin{quote}
		``empirical application[s] would have to relax the symmetry assumptions and allow firms to produce products of different qualities, allow for heterogeneity across firms, and differing costs to introducing products.'' \cite[pg. 98]{Anderson06} 
	\end{quote}
	Thus, existing theoretical analyses currently offer limited support to empirical applications of differentiated product market models or other models in which Logit models might be useful, such as those described by Gallego et. al \cite{Gallego06}.

	A theoretical understanding of Bertrand-Nash equilibrium prices begins with the conditions under which equilibrium prices exist. Perloff \cite{Perloff85} provided an early existence proof for Bertrand-Nash equilibrium under a general Random Utility Maximization (RUM) model. Firms in Perloff's model are `systematically homogeneous' in that product differentiation exists only through random brand preference, rather than differentiated product characteristics and unit costs. Anderson \& dePalma \cite{Anderson92a} undertake an analysis of equilibrium with single-product firms focusing on the linear-in-price utility Logit model. They characterize equilibrium prices with a closed-form expression when there is no outside good, and as solutions to a fixed-point equation when an outside good exists. Milgrom \& Roberts \cite{Milgrom90}m Caplin \& Nalebuff \cite{Caplin91}, and Gallego et. al \cite{Gallego06} have also provided equilibrium existence proofs for Bertrand competition between single-product firms that apply to the Logit RUM, assuming utility is linear in price. Such results have been used to ensure that empirical single-product firm models based on Bertrand competition are well-posed \cite{Pakes94, Anderson95, Gowrisankaran97, Anderson01a}. More recently, Sandor \cite{Sandor01} and Konovalov \& Sandor \cite{Konovalov10} have proven the existence and uniqueness of equilibrium prices with multi-product firms and the linear-in-price utility Logit model. Beyond models with single-product firms and linear utility functions, the literature lacks general conditions under which equilibrium exists. Without this understanding, it is not known if empirical examples cannot have equilibrium prices. This article provides one example where this has already occurred. 
	
	Unfortunately the mathematical methods employed in these works cannot be extended to establish the existence of equilibrium prices for models with multi-product firms and general non-linear utility functions. Perloff's \cite{Perloff85} and Anderson \& dePalma's \cite{Anderson92a} analyses are specific to symmetric equilibrium between homogeneous single-product firms. While Milgrom \& Roberts \cite{Milgrom90} apply a general property $-$ ``supermodularity'' $-$ to prove the existence (and uniqueness) of single-product firm equilibrium prices under Logit, Sandor \cite{Sandor01} has shown that multi-product firm profit functions under linear-in-price utility Logit fail to be supermodular {\em arbitrarily near} equilibrium prices, ruling out the application of this property for multi-product firms; Appendix \ref{SUBSEC:Supermodular} extends Sandor's proof to any Logit model within the class studied here. Similarly, the proofs from Caplin \& Nalebuff \cite{Caplin91} and Gallego et. al \cite{Gallego06} rely on quasi-concavity of the firms' profit functions. Hanson \& Martin \cite{Hanson96}, however, have observed that multi-product firm profits are not quasi-concave under the Logit model. Thus new mathematical tools are needed to prove the existence of equilibrium prices for Bertrand competition between multi-product firms.

	The remainder of this article proceeds as follows. Section \ref{SEC:Framework} presents a framework for Bertrand competition under an arbitrary RUM model assuming unit costs are constant (equivalently, total costs depend linearly on quantity sold). Special attention is paid to the interpretation of an RUM model as the generator of a stochastic choice process leading to random demand, highlighting some implicit assumptions that may not be commonly acknowledged. Section \ref{SEC:LogitModels} specializes this framework to the Logit RUM model, equivalent to the ``attraction demand model'' being used by some researchers in revenue management (e.g., \cite{Gallego06}). Specifically, Section \ref{SEC:LogitModels} presents a new set of utility specifications, defines the Logit choice probabilities, and identifies when firm profits under the Logit model are bounded. 
	
	Section \ref{SEC:LogitPriceEquilibrium} proves the existence of Bertrand-Nash equilibrium prices using fixed-point equations derived from the first-order or ``Simultaneous Stationarity'' condition when unit costs are constant. Most existing analyses of equilibrium prices also rely on the first-order condition. Moreover, fixed-point expressions have already been utilized to characterize equilibrium under linear-in-price utility Logit models \cite{Anderson92a, Berry95, Besanko98, Sandor01, Konovalov10} and even for more complex Mixed Logit models \cite{Berry95, Morrow08, Morrow10a, Morrow10b}. Here three fixed-point equations for Logit models are derived, two of which generalize to Mixed Logit models; see \cite{Morrow08, Morrow10a, Morrow10b}. As is common in analyses of equilibrium, the existence proof has two parts: proving that (i) there exist simultaneously stationary prices and (ii) simultaneously stationary prices are in fact equilibria. Existence proofs using both Brouwer's fixed-point theorem and the Poincare-Hopf Theorem \cite[Chapter 6]{Milnor65} are given. The Poincare-Hopf approach to this problem was first taken in \cite{Morrow08}, and has also been used by Konovalov \& Sandor \cite{Konovalov10}, restricting the utility function to be linear-in-price. Simsek et. al \cite{Simsek07} provide several references to applications of the Poincare-Hopf Theorem in general equilibrium models. Identifying simultaneously stationary prices with equilibria requires a direct analysis of the second-order conditions, combined with a second application of the Poincare-Hopf Theorem to prove the uniqueness of profit-maximizing prices, circumventing the lack of quasi-concavity in Logit profits for multi-product firms. These results are based on new, general conditions on the utility function that are weaker than most assumptions currently applied in theoretical economics, econometrics, and marketing. 
	
	Sections \ref{SEC:QuantityCosts} and \ref{SEC:FiniteIncomes} generalize the analysis in Section \ref{SEC:LogitPriceEquilibrium} to address non-constant unit costs and populations with finite purchasing power, respectively. The treatment of non-constant unit costs is a fairly straightforward extension of the analysis for the constant unit cost case, at least when total costs are convex (that is, unit costs are increasing in volume). Limits on purchasing power qualitatively change the behavior that may occur in equilibrium: While every product has a non-zero (if small) probability of being purchased in equilibrium with no such limit, some products can be profit-optimally ``priced out of the market'' when there is a finite limit on purchasing power. The fixed-point approach from the traditional, no-limit case is extended to characterize equilibrium prices when there is a limit, and existence is proved with essentially the same methods. 
	
	Finally, Section \ref{SEC:LogitStructure} identifies several structural properties of equilibrium prices under the Logit model. 
%	First, profit-optimal pricing under Logit with is a ``single-parameter problem'' even when firms offer many products. That is, {\em all} of a firm's profit-optimal prices can be derived from the knowledge of a {\em single} parameter such as optimal profits or the price of an arbitrary product in their portfolio. From this property one can conclude that the Logit model lacks the flexibility to truly model multi-product firm pricing behavior, a supply-side analogue to the reductive properties of Logit, like the Independence of Irrelevant Alternatives (IIA), well known for the demand side. 
	First, if the consumer population systematically values some product's characteristics more than the same firm's other offerings, that product must be given {\em lower} profit-optimal markup by the firm in equilibrium when unit costs are constant. This counterintuitive result cannot be observed in analyses with single-product firms, and relies only on the common assumptions that (i) utility is concave in price and separable in price and characteristics and (ii) unit costs, though constant, increase with the value of product characteristics \cite{Mussa78, Bresnahan87}. As a consequence, Bertand competition under Logit with conventional utility specifications and constant unit costs cannot have fixed percentage markups as an equilibrium outcome; i.e. ``cost plus pricing'' \cite{Nagle87} is not rationalized by Bertrand competition under conventional Logit models. Second, there exists a {\em portfolio effect} for Bertrand-Nash equilibrium under Logit with constant unit costs: equilibrium prices for an identical product offered at the same unit cost by two distinct firms depends on the profitability of the entire portfolio of products offered by these firms. In other words, heterogeneous portfolios can lead to {\em distinct} equilibrium prices for {\em otherwise identical} products. This property would not be observed from analyses that assume firms are homogeneous. 
	
	The most limiting assumption in this article is the absence of consumer heterogeneity in the choice model. The Logit RUM model does allows some random variance in the utilities individuals in the population derive from the differentiated products, and thus contains some degree of population heterogeneity. However, the degree to which this is expressed with the Logit model has long been known to generate patterns of substitution that are unrealistic form many empirical applications \cite{Berry95, Train03}. The techniques used in this paper to prove the existence of simultaneously stationary prices can be extended to a large class of Mixed Logit models with some ease; see \cite{Morrow10a, Morrow10b}. Moreover, the fixed-point equations used here are very useful in computations of equilibrium prices under such models \cite{Morrow10a}. However, the central element of the existence results established here is a generalization of quasi-concavity property that ensures profits under Logit have unique profit-maximizing prices. The conditions under which such a condition holds for Mixed Logit models are not obvious and will be nontrivial \cite{Morrow10a}. 
	
	A review of mathematical notation, several examples, and additional results are provided in the appendices.

%%%%%%%%%%%%%%%%%%%%%%%%%%%%%%%%%%%%%%%%%%%%%%%%%%%%%%%%%
% FRAMEWORK  %%%%%%%%%%%%%%%%%%%%%%%%%%%%%%%%%%%%%%%%%%%%%%%%
%%%%%%%%%%%%%%%%%%%%%%%%%%%%%%%%%%%%%%%%%%%%%%%%%%%%%%%%%

\section{Bertrand Competition Under an Arbitrary Random Utility Model}
\label{SEC:Framework}
	
	This section presents a mathematical framework for Bertrand competition under an arbitrary Random Utility Maximization (RUM) model. This generalizes the discussion in \cite{Baye08} to multi-product firms and RUM demand. Conceptually, a fixed number of firms decide on prices for a fixed set of products prior to some time period in which these prices must remain fixed. During this purchasing period, a fixed number of individuals independently choose to purchase one of the products offered by these firms, or to forgo purchase of any of these products, following a given RUM model. Verboven \cite{Verboven99} describes this as a two-stage stochastic game, where in the first stage the firms choose prices and in the second stage individuals choose products to maximize their own utility after sampling, or ``drawing,'' from the distribution of random utilities.\footnote{In the second stage, all consumers have dominant strategies.} 

	%%%%%%%%%%%%%%%%%%%%%%%%%%%%%%%%%%%%%%%%%%%%%%
	%%%%%%%%%%%%%%%%%%%%%%%%%%%%%%%%%%%%%%%%%%%%%%
	%%%%%%%%%%%%%%%%%%%%%%%%%%%%%%%%%%%%%%%%%%%%%%
	
	\subsection{Random Utility Models and Demand for Products}
	\label{SEC:RUMsAndDemand}

	RUM models provide a means to describe selection from a {\em choice set}, a collection of $J \in \N$ products that individuals may choose to purchase along with a no-purchase option (or ``outside good'') indexed by 0. Each product $j \in \N(J)$ is characterized by its price $p_j \in [0,\infty)$ and vector of {\em characteristics} $\vec{y}_j \in \set{Y}$, where $\set{Y} \subset \R^K$ for some $K \in \N$. 
	
	The random variable $U_{i,j}(\vec{y}_j,p_j)$ gives the utility individual $i$ receives by purchasing product $j \in \N(J)$, while the random variable $U_{i,0}$ gives the utility received by {\em not} purchasing any of the products (i.e. ``purchasing the outside good''). Conditional on the values of $\{ U_{i,0} \} \cup \{U_{i,j}(\vec{y}_j,p_j)\}_{j \in \N(J)}$, individual $i$ chooses the option $j \in \{0\} \cup \N(J)$ with the highest utility. The choice variable $C_i(\vec{Y},\vec{p})$ encapsulates this selection, taking values in $\{0\} \cup \N(J)$ following the distribution
	\begin{equation*}
		\Prob ( C_i(\vec{Y},\vec{p}) = j ) 
			= \left\{ \begin{aligned}
				&\Prob \left ( U_{i,j}(\vec{y}_j,p_j) = \max \left\{ \; U_{i,0} \; , \; \max_{k \in \N(J)} U_{i,k}(\vec{y}_k,p_k) \; \right\} \right )
					&&\quad \text{if } j \in \N(J) \\
				&\Prob \left ( U_{i,0} = \max \left\{ \; U_{i,0} \; , \; \max_{k \in \N(J)} U_{i,k}(\vec{y}_k,p_k) \; \right\} \right )
					&&\quad \text{if } j = 0
			\end{aligned} \right.
	\end{equation*}
	The distribution of these random utilities assures that ``ties'' occur with probability zero. Let 
	\begin{equation*}
		\vec{U}_i(\vec{Y},\vec{p}) = (U_{i,0},U_{i,1}(\vec{Y},\vec{p}),\dotsc,U_{i,J}(\vec{Y},\vec{p})),
	\end{equation*}
	and make the following assumption:
	
	\begin{assumption}
		\label{IndependentUtilities}
		For any $(\vec{Y},\vec{p}) \in \set{Y}^J \times \R_+^J$ and $i,i^\prime \in \N(I)$, $\vec{U}_i(\vec{Y},\vec{p})$ and $\vec{U}_{i^\prime}(\vec{Y},\vec{p})$ are independent and identically distributed. 
	\end{assumption}
	Under this common assumption, the individual index on utilities and the choice variable can be dropped. Note also that this does {\em not} imply that $U_j(\vec{y}_j,p_j)$ and $U_k(\vec{y}_k,p_k)$ are independent. 
	
	In practice, models often take the form
	\begin{equation*}
		% \label{StraightSpecification}
		U_0 = \vartheta + \set{E}_0
		\quad\quad\text{and}\quad\quad
		U_j(\vec{y}_j,p_j) = u(\vec{y}_j,p_j) + \set{E}_j
		\text{ for all } j \in \N(J)
	\end{equation*}
	for some (conditional, indirect) {\em utility function} $u : \set{Y} \times [0,\infty) \to \R$, $\vartheta \in [-\infty,\infty)$ and ``error'' vector $\bsym{\set{E}} = \{ \set{E}_j \}_{j=0}^J$. When $\bsym{\set{E}}$ is given an i.i.d. extreme value distribution, we have the Logit RUM \cite{Anderson92a,Train03}.\footnote{This independence assumption on $\bsym{\set{E}}$ is distinct from our independence assumption on the random utilities in that now it is independence {\em across products in the choice set}, rather than {\em across individuals in the population}.} Letting $\bsym{\set{E}}$ have a Generalized Extreme Value distribution we have a GEV RUM like the Nested Logit model \cite{Train03,Bierlaire03}, or taking $\bsym{\set{E}}$ multivariate normal gives the Probit RUM \cite{Train03}. Either of these latter two forms can have a $\bsym{\set{E}}$ with correlated components.\footnote{More generally, we can let $\set{T}$ be a space of individual characteristics or ``demographics'' and define
	\begin{equation*}
		% \label{MixedSpecification}
		U_0 = \vartheta(\bsym{\Theta}) + \set{E}_0
		\quad\quad\text{and}\quad\quad
		U_j(\vec{y}_j,p_j) = u(\bsym{\Theta},\vec{y}_j,p_j) + \set{E}_j
		\text{ for all } j \in \N(J)
	\end{equation*}
	where $u : \set{T} \times \set{Y} \times [0,\infty) \to \R$, $\vartheta : \set{T} \to [-\infty,\infty)$ where $\bsym{\Theta}$ is a $\set{T}$-valued random variable with the distribution $\mu$, ostensibly representing the distribution of demographic variables over the population. With the same error distribution, we obtain a ``mixed'' RUM. Particularly, taking $\bsym{\set{E}}$ i.i.d. extreme value gives the ``random coefficients'' or Mixed Logit RUM class \cite[Chapter 6]{Train03}.}
	
	% \subsubsection{Demands}

	{\em Demands}, the total quantity of each product purchased during the purchasing period, must be defined to define firms' profits. Extrapolating demands from the stochastic choice model above requires the following assumption. 
	\begin{assumption}
		\label{IdenticalChoiceSet}
		Every individual $i\in\N(I)$ observes the same choice set, $\{0\} \cup \N(J)$, during the purchase period. 
	\end{assumption}
%	\begin{assumption}
%		\label{IndependentChoices}
%		A population of $I \in \N$ individuals independently choose either to purchase one of $J$ products described by $(\vec{Y},\vec{p})$ according to some choice probabilities $\vec{P}(\vec{Y},\vec{p})$, or not to purchase any of these products with probability $1 - \vec{P}(\vec{Y},\vec{p})^\top\vec{1}$. That is, $C_i(\vec{Y},\vec{p}) \sim C(\vec{Y},\vec{p})$ where $\Prob( C(\vec{Y},\vec{p}) = j ) = P_j(\vec{Y},\vec{p})$ for all $j \in \N(J)$ and $\Prob( C(\vec{Y},\vec{p}) = 0 ) = 1 - \vec{P}(\vec{Y},\vec{p})^\top\vec{1}$. 
%	\end{assumption}
	Under this assumption, the demand $Q_j(\vec{Y},\vec{p})$ for each product $j \in \N(J)$ can be expressed simply as $Q_j(\vec{Y},\vec{p}) = \sum_{i=1}^I 1_{ \{ C_i(\vec{Y},\vec{p}) = j \} }$, where here $\{ C_i(\vec{Y},\vec{p}) \}_{i \in \N(I)}$ are $I$ i.i.d. ``copies'' of $C(\vec{Y},\vec{p})$. The primary benefit of Assumption \ref{IdenticalChoiceSet} is that $\{ Q_0(\vec{Y},\vec{p}) \} \cup \{ Q_j(\vec{Y},\vec{p})\}_{j \in \N(J)}$ is a multinomial family of variables with parameter $I$ and probabilities $\{ P_j(\vec{Y},\vec{p}) \}_{j=0}^J$, and thus expected demands for each product are given simply by $\Expect[Q_j(\vec{Y},\vec{p})] = I P_j(\vec{Y},\vec{p})$ \cite{Feller68}. 
	
	A more serious implication of Assumption \ref{IdenticalChoiceSet} is there must be at least $I$ units of every product available for the individuals to choose during the purchasing period.\footnote{We could alternatively interpret this condition in terms of consumers ``ordering'' products during the purchasing period and assuming delivery schedules do not impact demand.} Specifically, no product can ``sell out,'' or if it does, ``backordering'' does not impact utilities. If any firm commits (or is forced by capacity constraints) to only produce $I^\prime < I$ units of some product they offer during the purchasing period and individuals do not ``backorder'', then with some positive probability Assumption \ref{IdenticalChoiceSet} will be violated. 
	
%	For suppose there were only $I^\prime < I$ units of some product $j$ that could be made available to the population during the purchasing period, and that the RUM has $P_j(\vec{Y},\vec{p}) > 0$. Given that individuals $\set{I}^\prime = \{ i_1,...,i_{I^\prime} \} \subset \N(I)$ purchase product $j$ during the purchasing period, the probability that any individual $i^\prime \in \N(I) \setminus \set{I}^\prime (\neq \{ \emptyset\})$ purchases product $j$ is zero. However, both $\Prob \big( C_{^\prime}(\vec{Y},\vec{p}) = j \big) > 0$ and $\Prob\big( C_i(\vec{Y},\vec{p}) = j \text{ for all } i \in \set{I}^\prime \big) > 0$ hold, and thus
%	\begin{align*}
%		&\Prob \big( C_{^\prime}(\vec{Y},\vec{p}) = j \big) \Prob\big( C_i(\vec{Y},\vec{p}) = j \text{ for all } i \in \set{I}^\prime \big) \\
%			&\quad\quad\quad\quad
%				\neq 0 = \Prob \big( \{ C_{^\prime}(\vec{Y},\vec{p}) = j \} \; \cap \; \{C_i(\vec{Y},\vec{p}) = j \text{ for all } i \in \set{I}^\prime \} \big). 
%	\end{align*}

	%%%%%%%%%%%%%%%%%%%%%%%%%%%%%%%%%%%%%%%%%%%%%%%
	%%%%%%%%%%%%%%%%%%%%%%%%%%%%%%%%%%%%%%%%%%%%%%%
	%%%%%%%%%%%%%%%%%%%%%%%%%%%%%%%%%%%%%%%%%%%%%%%
	
	\subsection{Firms, Product Portfolios, Costs, and Profits}
	\label{SEC:FirmsPortfoliosCostsAndProfits}

	Let $F \in \N$ denote the number of firms. For each $f \in \N(F)$, there exists a set $\set{J}_f \subset \N(J)$ of indices that corresponds to the $J_f = \abs{\set{J}_f}$ products offered by firm $f$. The collection of all these sets, $\{\set{J}_f\}_{f=1}^F$, forms a partition of $\N(J)$. Subsequently, in writing ``$f(j)$'' for some $j \in \N(J)$, we mean the unique $f \in \N(F)$ such that $j \in \set{J}_f$. The vector $\vec{p}_f \in \R^{J_f}$ refers to the vector of prices of the products offered by firm $f$. Negative subscripts denote competitor's variables as in, for instance, $\vec{p}_{-f} \in \R^{J_{-f}}$, where $J_{-f} = \sum_{g \neq f} J_g$, is the vector of prices for products offered by all of firm $f$'s competitors. Firm-specific choice probability functions are denoted by $\vec{P}_f(\vec{p})$. 
	
	Additional assumptions concerning costs and production are required to complete the definition of firms' profits. 
	\begin{assumption}
		\label{ASS:ConstUnitCostsAssum}
		There exists a unit cost function $c_f^U : \set{Y} \to \R_+$ and a fixed cost function $c_f^F : \mathfrak{F}(\set{Y}) \to \R_+$ for all $f \in \N(F)$ that depend only on the collection of product characteristics chosen by the firm. 
	\end{assumption}
	Particularly, unit and fixed costs are independent of the quantity sold, ruling out dependence of {\em unit} and fixed costs on production volumes. This assumption is relaxed in Section \ref{SEC:QuantityCosts} below. Bertrand competition also entails the following ``comittment'' assumption on the quantities produced \cite{Baye08}. 
	\begin{assumption}[Bertrand Production Assumption]
		\label{BertrandProductionAssumption}
		Each firm commits to producing exactly $Q_j(\vec{Y},\vec{p})$ units of each product $j \in \set{J}_f$ during the purchasing period. 
	\end{assumption}
	Again, this implies that the firm has no production capacity constraints that limit a firm's ability to meet {\em any} demands that arise during the purchase period. The random variable $\sum_{j \in \set{J}_f} c_f^U(\vec{y}_j)Q_j(\vec{Y},\vec{p}) + c_f^F(\vec{Y}_f)$ gives the total cost firm $f$ incurs in producing $Q_j(\vec{Y},\vec{p})$ units of product $j$, for all $j \in \set{J}_f$, during the purchasing period. We let $\vec{c}_f^U(\vec{Y}_f)$ be the vector of these unit costs for the products offered by firm $f$. 
	
	Under Assumption \ref{BertrandProductionAssumption}, the random variable $\Pi_f(\vec{Y},\vec{p}) = \vec{Q}_f(\vec{Y},\vec{p})^\top ( \vec{p}_f - \vec{c}_f^U(\vec{Y}_f) ) - c_f^F(\vec{Y}_f)$ gives firm $f$'s profits for the production period as a function of product characteristics and prices. Following most of the theoretical and empirical literature in both marketing and economics, we assume that firms take expected profits, 
	\begin{equation}
		\label{ExpectedProfits}
		% \Expect[\Pi_f(\vec{Y},\vec{p})] = 
			\pi_f(\vec{Y},\vec{p}) = I\hat{\pi}_f(\vec{Y},\vec{p}) - c_f^F(\vec{Y}_f)
		\quad\text{where}\quad
		\hat{\pi}_f(\vec{Y},\vec{p}) = \vec{P}_f(\vec{Y},\vec{p})^\top ( \vec{p}_f - \vec{c}_f^U(\vec{Y}_f) ),
	\end{equation}
	as the metric by which they optimize their pricing decisions in this stochastic optimization problem. 
	
%	One way the assumption that unit costs do not depend on production volume can be relaxed without overly complicating the resulting equilibrium problem is to suppose that firms decide on pricing based on the total costs of {\em expected} demands, as opposed to the actual expected total costs. That is, if $c_f^U : \set{Y} \times (0,\infty) \to (0,\infty)$, we assume firms aim to maximize 
%	\begin{equation*}
%		\Expect[\vec{Q}(\vec{Y},\vec{p})]^\top \big( \vec{p}_f - \vec{c}_f^U(\vec{Y}_f,\Expect[\vec{Q}(\vec{Y},\vec{p})]) \big)
%	\end{equation*}
%	This changes the first-order conditions to
%	\begin{equation*}
%		(D_f\vec{P}_f)(\vec{p})^\top \big( \vec{p}_f - \vec{c}_f^U(\vec{p}) - N (D_f^q\vec{c}_f^U)(\vec{p})^\top \vec{P}_f(\vec{p}) \big) + \vec{P}_f(\vec{p})
%	\end{equation*}
%	For logit, our fixed-point characterization becomes
%	\begin{equation*}
%		\vec{p}_f 
%			= \vec{c}_f^U + \vec{d}_f(\vec{p}) 
%				+ ( \hat{\pi}_f(\vec{p}) - \vec{P}_f(\vec{p})^\top \vec{d}_f(\vec{p}) ) \vec{1}
%				+ \abs{ ( D_f\vec{w}_f)(\vec{p}_f) }\inv \vec{1}
%	\end{equation*}
	
	Eqn. (\ref{ExpectedProfits}) demonstrates that neither the total firm fixed costs $c_f^F$ nor the population size $I$ play a role in determining the prices that maximize expected profits. Therefore we only consider the ``population-normalized gross expected profits'' $\hat{\pi}_f(\vec{p})$, referred to below as simply ``profits''. We also consider $\vec{Y}$ fixed, and cease to include this characteristic matrix as an argument. Finally, we write $\vec{c}_f = \vec{c}_f^U$ as these are the only relevant costs for the price equilibrium problem. Henceforth we write simply
	\begin{equation*}
		\hat{\pi}_f(\vec{p}) = \vec{P}_f(\vec{p})^\top ( \vec{p}_f - \vec{c}_f ).
	\end{equation*}
	
%		\begin{equation} 
%			\label{SecondDerivatives}
%			(D_f\nabla_f \hat{\pi}_f)(\vec{p}) 
%				= \vec{A}_f(\vec{p}) + (D_f \vec{P}_f)(\vec{p})^\top + (D_f\vec{P}_f)(\vec{p})
%		\end{equation}

	The following adaptation of well-known necessary conditions for the local maximization of an unconstrained, continuously differentiable function (e.g., \cite{Munkres91}) informs our derivation of the Simultaneous Stationarity Condition. 
	\begin{lemma}
		\label{ProfitStationarity}
		Suppose $\vec{P}_f(\cdot,\vec{p}_{-f})$ is continuously differentiable on some open $\set{A} \subset (\vec{0},\bsym{\infty}) \subset \R^{J_f}$. If $\vec{p}_f \in \set{A}$ is a local maximizer of $\hat{\pi}_f(\cdot,\vec{p}_{-f})$, then 
		\begin{equation} 
			\label{ProfitGradientFormula}
			(\nabla_f \hat{\pi}_f)(\vec{p}) 
				= (D_f \vec{P}_f)(\vec{p})^\top (\vec{p}_f - \vec{c}_f) + \vec{P}_f(\vec{p})
				= \vec{0}. 
		\end{equation}
	\end{lemma}
	
	%%%%%%%%%%%%%%%%%%%%%%%%%%%%%%%%%%%%%%%%%%%%%%%%%%%%%%
	%%%%%%%%%%%%%%%%%%%%%%%%%%%%%%%%%%%%%%%%%%%%%%%%%%%%%%
	%%%%%%%%%%%%%%%%%%%%%%%%%%%%%%%%%%%%%%%%%%%%%%%%%%%%%%
	
	\subsection{Local Equilibrium and the Simultaneous Stationarity Conditions}
	\label{SEC:EquilibriumAndTheSSCs}
	
	As in much of the existing literature, our analysis relies on local conditions for optimality of prices and thus must rely on the following {\em local} definition of equilibrium. 
	
	\begin{definition} 
		A price vector $\vec{p} \in [\vec{0},\bsym{\infty}]$ is called a {\em local equilibrium} if $\vec{p}_f$ is a local maximizer of $\hat{\pi}_f(\cdot,\vec{p}_{-f})$ for all $f \in \N(F)$. A price vector $\vec{p} \in [\vec{0},\bsym{\infty}]$ is called an {\em equilibrium} if $\vec{p}_f$ is a maximizer of $\hat{\pi}_f(\cdot,\vec{p}_{-f})$ for all $f \in \N(F)$. 
%		The set of all local equilibria, as a function of all product characteristics and unit costs, will be denoted by $\mathfrak{E}^L(\vec{Y},\vec{c}^U)$. The set of all equilibria will be denoted by $\mathfrak{E}(\vec{Y},\vec{c}^U)$. 
	\end{definition}
	
	Finally, the following {\em Simultaneous Stationarity Condition} is a generic necessary condition for local equilibrium if the RUM choice probabilities are continuously differentiable in prices. 
	\begin{definition}
		Let $(\tilde{\nabla}\hat{\pi})(\vec{p})$ denote the ``combined gradient'' with components $((\tilde{\nabla}\hat{\pi})(\vec{p}))_j = (D_j\hat{\pi}_{f(j)})(\vec{p})$. Let $(\tilde{D}\vec{P})(\vec{p})$ be the sparse matrix corresponding to the intra-firm price derivatives of choice probabilities; that is, 
		\begin{equation*}
			\big( (\tilde{D}\vec{P})(\vec{p}) \big)_{j,k} 
				= \left\{ \begin{aligned}
					&(D_kP_j)(\vec{p})	&&\quad\text{if } f(j) = f(k) \\
					&\quad\quad 0		&&\quad\text{if } f(j) \neq f(k)
				\end{aligned} \right. .
		\end{equation*}
	\end{definition}
	
	\begin{lemma}[Simultaneous Stationarity Condition]
		\label{SSCProp}
		Suppose $\vec{P}$ is continuously differentiable on some open $\set{A} \subset (\vec{0},\bsym{\infty})$. If $\vec{p} \in \set{A}$ is a local equilibrium, then 
		\begin{equation}
			\label{SSCs}
			(\tilde{\nabla}\hat{\pi})(\vec{p})
				= (\tilde{D}\vec{P})(\vec{p})^\top (\vec{p} - \vec{c}) + \vec{P}(\vec{p}) 
				= \vec{0}. 
		\end{equation}
	\end{lemma}
	Prices satisfying Eqn. (\ref{SSCs}) are called ``simultaneously stationary.'' In principle, simultaneously stationary prices  need not be equilibria. Additional analysis is required to link stationarity with local optimality of profits with respect to changes to the prices of a firm's own products. 
	
	The necessity of the Simultaneous Stationarity Condition does not depend on the RUM type, but only on the continuous differentiability of the choice probabilities (with respect to price) and the cost assumption. Furthermore, much of this development is the same for an arbitrary demand function, rather than a RUM; see, e.g., \cite{Bresnahan87,Hausman02}. Thus, Eqn. (\ref{SSCs}) has appeared in many different studies using alternative RUM specifications such as Logit models \cite{Besanko98}, Generalized Extreme Value models \cite{Goldberg95, Goldberg98, Besanko98, VillasBoas05}, and Mixed Logit models \cite{Berry95, Berry04, Nevo00a, Nevo01, Sudhir01, Petrin02, Smith04, Aguirregabiria06, Beresteanu08, Morrow10a}. In most of these studies, Eqn. (\ref{SSCs}) has not been investigated far beyond Lemma \ref{SSCProp}. 
	
	Eqn. (\ref{SSCs}) has been consistently used through the corresponding {\em BLP markup equation} $\vec{p} = \vec{c} + \bsym{\eta}(\vec{p})$ where
	\begin{equation}
		\label{MarkupEquation}
		\bsym{\eta}(\vec{p})  = - (\tilde{D}\vec{P})(\vec{p})^{-\top} \vec{P}(\vec{p})
	\end{equation}
	assuming $(\tilde{D}\vec{P})(\vec{p})^\top$ is nonsingular.\footnote{For competing single-product firms, this reduces to the famous ``negative reciprocal of elasticity'' form for the Lerner index (i.e. percent markups); see \cite{Perloff07}.} This equation is typically used to estimate costs assuming prices are in equilibrium. However, the markup equation $\vec{p} = \vec{c} + \bsym{\eta}(\vec{p})$ is a fixed-point equation satisfied by all simultaneously stationary prices. In Section \ref{SEC:LogitPriceEquilibrium}, we give a specific form for $\bsym{\eta}$ under the Logit model and derive a new fixed-point equation for simultaneously stationary prices by factoring out the gradient of the inclusive value from $(\tilde{\nabla}\hat{\pi})(\vec{p})$; this fixed-point equation is a specialization of the $\bsym{\zeta}$-markup equation used by Morrow \& Skerlos for large-scale computations of equilibrium prices \cite{Morrow10a, Morrow10b}. 
	
\section{Logit Models}
\label{SEC:LogitModels}
	
	This section reviews the Logit model, providing the groundwork for the analysis in later sections. Section \ref{SUBSEC:LogitUtilitySpecifications} defines a new class of nonlinear utility functions for which equilibrium prices can be shown to exist. Section \ref{SUBSEC:LogitModel} derives the corresponding Logit choice probabilities and their derivatives. Finally, Section \ref{SUBSEC:BoundingLogitProfits} provides conditions under which profit-maximizing prices are finite, a pre-requisite for the existence of (finite) equilibrium prices. 
	
	%%%%%%%%%%%%%%%%%%%%%%%%%%%%%%%%%%%%%%%%%%%%%%%%%%%%%%
	%%%%%%%%%%%%%%%%%%%%%%%%%%%%%%%%%%%%%%%%%%%%%%%%%%%%%%
	%%%%%%%%%%%%%%%%%%%%%%%%%%%%%%%%%%%%%%%%%%%%%%%%%%%%%%
	
	\subsection{Systematic Utility Specifications}
	\label{SUBSEC:LogitUtilitySpecifications}
	
	The random utility any individual receives by purchasing any particular product is parameterized by its characteristic vector and price through some function $u : \set{Y} \times [0,\infty) \to [-\infty,\infty)$. We consider specifications of the following form.
	\begin{assumption}
		\label{LogitUtilityAssumption}
		There are functions $w : \set{Y} \times [0,\infty) \to (-\infty,\infty)$ and $v : \set{Y} \to (-\infty,\infty)$ such that utility can be written $u(\vec{y},p) = w(\vec{y},p) + v(\vec{y})$. Concerning the behavior of $w$, we assume that, for all $\vec{y} \in \set{Y}$, $w(\vec{y},\cdot) : [0,\infty) \to (-\infty,\infty)$ is (a) strictly decreasing, and (b) continuously differentiable on $(0,\infty)$. We also assume that (c) $\lim_{p \uparrow \infty} w(\vec{y},p) = -\infty$, and subsequently set $w(\vec{y},\infty) = -\infty$. 
	\end{assumption}
	
	Writing $u(\vec{y},p) = w(\vec{y},p) + v(\vec{y})$ is completely general so long as utility is defined for all $p \in [0,\infty)$. This form is convenient to define the ``value'' of a product as that component of utility that does not vary with price, and to define ``separable'' utilities, the most common class of utility functions used in practice. 
	
	\begin{definition}
		We say $v(\vec{y})$ is the {\bfseries value} of any product with characteristic vector $\vec{y}$, and that {\bfseries utility is separable in price and characteristics} (or simply {\bfseries separable}) if $w(\vec{y},p) = w(p)$ for all $\vec{y} \in \set{Y}$. We call $\abs{(Dw)(\vec{y},p)}\inv$ the {\bfseries (local) willingness to pay (for product value)}.
	\end{definition}
	
	The class formed by Assm. \ref{LogitUtilityAssumption} encompasses the majority of utility functions used in the theoretical and empirical literature. Assm. (a) is required of a suitable indirect utility function, and (b) is required for an analysis of equilibrium based on the first-order conditions. The assumption (c) is a natural condition that ensures that the choice probabilities vanish as prices increase without bound. In fact, by including utility functions that are not concave-in-price, this class is larger than that typically studied. A number of examples are given in Appendix \ref{APP:LogitExamples}. 
	
%	The motivation for this specification is the Caplin \& Nalebuff \cite{Caplin91} have given a prescription on which we base our specification. In our notation, they assume that $w(\vec{y},p) = \alpha(\vec{y})r(\varsigma - p)$ for $p \in [0, \varsigma)$, where $\varsigma \in (0,\infty)$ is intended to represent income, $r$ is some strictly increasing and concave function, and $\alpha(\vec{y}) > 0$ for all $\vec{y} \in \set{Y}$; $w$ is not defined for $[\varsigma,\infty]$. This prescription implies that, where defined, $w(\vec{y},p)$ is strictly decreasing and concave in price. Caplin \& Nalebuff's \cite{Caplin91} framework is more general in that it does not need to be defined on all of $[0,\infty)$ and, strictly speaking, allows for heterogeneous values. See \cite{Morrow08} for extensions to models of this type. 
		
	Concave-in-price utilities are certainly an important special case often considered in economics. However, concavity turns out to be a stronger assumption than is required to ensure the existence of finite equilibrium prices under Logit. Instead, the following weaker property of the utility price derivatives is sufficient. 
	\begin{definition}
		\label{EDSQ}
		$w$ {\bfseries eventually decreases sufficiently quickly at} $\vec{y} \in \set{Y}$ if there exists some $r(\vec{y}) > 1$ and some $\bar{p}(\vec{y}) \in [0,\infty)$ such that $(Dw)(\vec{y},p) \leq -r(\vec{y})/p = - r(\vec{y}) D[ \log p ]$ for all $p > \bar{p}(\vec{y})$. $w$ itself {\bfseries eventually decreases sufficiently quickly} if $w$ eventually decreases sufficiently quickly at all $\vec{y} \in \set{Y}$. 
	\end{definition}
	The most commonly used finite utility functions, particularly strictly decreasing and concave in price utility functions, satisfy $\lim_{p \to \infty} (Dw)(\vec{y},p) < 0$, and hence eventually decrease sufficiently quickly with any $r$. Note also that if $w$ does {\em not} eventually decrease sufficiently quickly at $\vec{y}$, then necessarily $\abs{(Dw)(\vec{y},p)} \to 0$ as $p \to \infty$. The example $w(\vec{y},p) = - \alpha(\vec{y}) \log p$ ($\alpha(\vec{y}) > 0$) shows that this does not contradict Assm. \ref{LogitUtilityAssumption}, (c). 
	
	A distinct requirement on the second derivatives of utility is synonymous with the sufficiency of stationarity under Logit. 
	\begin{definition}
		\label{SQSD}
		Suppose $w(\vec{y},\cdot)$ is twice continuously differentiable for all $\vec{y} \in \set{Y}$. We say that $w$ has {\bfseries sub-quadratic second derivatives} at $(\vec{y},p) \in \set{Y} \times [0,\infty)$ if $\omega(\vec{y},p) = (D^2w)(\vec{y},p)/(Dw)(\vec{y},p)^2 < 1$. We say that $w$ itself has {\bfseries sub-quadratic second derivatives} if $w$ has sub-quadratic second derivatives at all $(\vec{y},p) \in \set{Y} \times [0,\infty)$. 
	\end{definition}
	Note that if $w(\vec{y},\cdot)$ is concave, then $w$ trivially has sub-quadratic second derivatives. However, $w(\vec{y},\cdot)$ can be convex and still have sub-quadratic second derivatives. For example, $w(\vec{y},p) = - \alpha(\vec{y})\log p$ for $\alpha(\vec{y}) > 1$ has $\omega(\vec{y},p) = 1 / \alpha < 1$. 
%	Clearly
%	\begin{equation*}
%		\frac{(D^2w)(\vec{y},p)}{\abs{(Dw)(\vec{y},p)}^2}
%			= - D \left[ \frac{1}{(Dw)(\vec{y},p)} \right]
%			= D \left[ \frac{1}{\abs{(Dw)(\vec{y},p)}} \right]
%	\end{equation*}
%	and thus
%	\begin{equation*}
%		\left( \frac{1}{\abs{(Dw)(\vec{y},p)}} \right) - \left( \frac{1}{\abs{(Dw)(\vec{y},q)}} \right)
%			= \int_q^p D \left[ \frac{1}{\abs{(Dw)(\vec{y},p)}} \right]_s ds
%			< \int_q^p ds = p - q
%	\end{equation*}
%	for any $p > q$. If, in addition, 
	
	With any collection of fixed product characteristic vectors $\{\vec{y}_j\}_{j=1}^J$, we set $w_j(p) = w(\vec{y}_j,p)$ and $v_j = v(\vec{y}_j)$ and thus generate a collection of product-specific utility functions, $u_j(p) = w_j(p) + v_j$, that depend on price alone. Vector functions $\vec{w} : [0,\infty]^J \to [-\infty,\infty)^J$ and $\vec{u} : [0,\infty]^J \to [-\infty,\infty)^J$ are constructed from these product-specific components by taking $(\vec{w}(\vec{p}))_j = w_j(p_j)$ and $(\vec{u}(\vec{p}))_j = u_j(p_j)$. In particular, $\vec{u}(\vec{p}) = \vec{w}(\vec{p}) + \vec{v}$. Firm-specific product values $\vec{v}_f$ and utilities $\vec{u}_f(\vec{p}_f) = \vec{w}_f(\vec{p}_f) + \vec{v}_f$ are also defined in the natural way. 
	
	%%%%%%%%%%%%%%%%%%%%%%%%%%%%%%%%%%%%%%%%%%%%%%%%%%%%%%
	%%%%%%%%%%%%%%%%%%%%%%%%%%%%%%%%%%%%%%%%%%%%%%%%%%%%%%
	%%%%%%%%%%%%%%%%%%%%%%%%%%%%%%%%%%%%%%%%%%%%%%%%%%%%%%
 
	\subsection{Logit Choice Probabilities}
	\label{SUBSEC:LogitModel} 

	 The Logit model \cite[Chapter 3]{Train03} takes the utility any individual receives when purchasing product $j$ to be the random variable $U_j(\vec{y}_j,p_j) = u(\vec{y}_j,p_j) + \set{E}_j$ and the utility of the outside good to be the random variable $U_0 = \vartheta + \set{E}_0$, where $\bsym{\set{E}} = \{ \set{E}_j \}_{j=0}^J$ is a family of i.i.d. standard extreme value variables and $\vartheta \in [-\infty,\infty)$ is a number representing the utility of the outside good.
	
	The i.i.d. standard extreme value specification for $\bsym{\set{E}}$ generates the following choice probabilities (see, e.g., \cite{Train03}):
	\begin{equation}
		\label{LogitChoiceProbabilities}
		P_j^L(\vec{p}) 
			= \frac{ e^{u_j(p_j)} }{ e^{\vartheta} + \sum_{k =1}^J e^{u_{k}(p_{k})} }
	\end{equation}
	The equivalent formula
	\begin{equation*}
		P_j^L(\vec{p}) = \frac{ e^{(u_j(p_j)-\vartheta)} }{ 1 + \sum_{k=1}^J e^{(u_{k}(p_{k})-\vartheta)} }
	\end{equation*}
	corresponding to setting $\vartheta = 0$ is often seen in the literature, but offers no substantial advantage to the analysis in this article. When $\vartheta = -\infty$, 
	\begin{equation*}
		P_j^L(\vec{p}) = \frac{ e^{u_j(p_j)} }{ \sum_{k=1}^J e^{u_{k}(p_{k})} }. 
	\end{equation*}
	
	The following basic properties of the Logit choice probabilities are used throughout. 
	\begin{lemma}
		\label{BasicLogitProperties}
		The following hold under Assumption \ref{LogitUtilityAssumption}, for any $j$ and $f$: (i)  $0 < P_j^L(\vec{p}) < 1$ and $\vec{P}_f^L(\vec{p})^\top\vec{1} < 1$ for all $\vec{p} \in [0,\infty)^J$. (iii) If $\vartheta > - \infty$ and $\vec{q} \in [0,\infty]^J$, $\lim_{ \vec{p} \to \vec{q} } P_j^L(\vec{p})$ exists. Moreover, $\lim_{ \vec{p} \to \vec{q} } P_j^L(\vec{p}) = 0$ if $q_j = \infty$, and $\vec{P}_f^L(\vec{p})^\top\vec{1} < 1$ for all $\vec{p} \in [0,\infty]^J$. (iv) If $\vartheta = -\infty$, then for any $\vec{x} \in [0,1]^J$, $\sum_{j=1}^J x_j = 1$, there exists some sequence $\{\vec{p}^{(n)}\}_{n \in \N} \subset [0,\infty)^J$ with $\vec{p}^{(n)} \to \bsym{\infty}$ such that $\lim_{n \to \infty} \vec{P}^L(\vec{p}^{(n)}) = \vec{x}$. 
	\end{lemma}
	\proof
		(i), (ii), and (iii) follow easily from Eqn. \ref{LogitChoiceProbabilities}. To prove (iv), first note that $\vec{P} : [0,\infty)^J \to \triangle(J)$ is onto when $\vartheta = -\infty$, where $\triangle(J) = \{ \vec{x} \in [0,1]^J : \sum_{j=1}^J x_j = 1\}$. Let $\vec{x} \in \triangle(J)$. It suffices to solve $u_j(p_j) = \log x_j$ for $p_j$, for all $j$, for then 
		 \begin{align*}
		 	P_j^L(\vec{p})
				= \frac{e^{u_j(p_j)}}{\sum_{k=1}^J e^{u_k(p_k)}}
				= \frac{e^{\log x_j}}{\sum_{k=1}^J e^{\log x_k}}
				= \frac{x_j}{\sum_{k=1}^J x_k}
				= x_j. 
		 \end{align*}
		 So long as $\log x_j \geq u_j(0)$, such a $p_j$ exists and is unique. Assuming, without loss of generality, that $u_j(0) \geq 0$ for all $j$ ensures that this condition holds for all $x_j \in [0,1]$. The existence of a sequence tending to infinity with $\lim_{n \to \infty} \vec{P}^L(\vec{p}^{(n)}) = \vec{x}$ then follows from the invariance result in Lemma \ref{InvariantSets} below. 
	\endproof
	
%	Claim (i) proves that $\hat{\pi}_f(\vec{p}) > 0$ for all $\vec{p}_f > \vec{c}_f$. Thus, profit-maximizing and equilibrium prices always generate positive {\em gross} expected profits. Some authors have taken this to imply that competition under the Logit model will generate ``endless entry''; e.g., see \cite{Perloff07} or \cite{Berry08}. However, if there are $F$ firms an $(F+1)$st firm could be reasonably expected to enter the market only if 
%	\begin{equation*}
%		\hat{\pi}_{F+1}(\vec{p}_*) > \frac{c_{F+1}^F}{I}
%	\end{equation*}
%	where $\vec{p}_*$ denote equilibrium prices in the resulting market. Similarly, some incumbent firm $f$ would exit if
%	\begin{equation*}
%		\hat{\pi}_{f}(\vec{p}_*) < \frac{c_f^F}{I}. 
%	\end{equation*}
%	Thus the fixed costs and the population size can control entry and exit from the market. 
	
	Claim (iv) amounts to the fact that the Logit choice probabilities without an outside good cannot be both continuous and single valued on $[0,\infty]^J$, and suggests that the presence of an outside good ``purchased'' with positive probability is very important to optimization and equilibrium problems under Logit. As noted in the proof, this claim is a consequence of the following generalization of the ``invariance of uniform price shifts'' property of the linear in price utility Logit model to the class of utility functions specified by Assumption \ref{LogitUtilityAssumption}:
	\begin{lemma}
		\label{InvariantSets}
		Suppose $w$ satisfies Assumption \ref{LogitUtilityAssumption}. For any $p \in (0,\infty)$ and each $j \in \N(J)$, define $\chi_{j,p} : [1,\infty) \to [p,\infty)$ by $\chi_{j,p}(\lambda) = w_j\inv( w_j(p) - \log \lambda )$, and define $\bsym{\chi}_{\vec{p}} : [1,\infty) \to [\vec{p},\bsym{\infty})$ componentwise by $(\bsym{\chi}_{\vec{p}}(\lambda))_j = \chi_{j,p_j}(\lambda)$. (i) $\bsym{\chi}_{\vec{p}}(\lambda)$ is well-defined, strictly increasing, and $\lim_{\lambda \to \infty} \bsym{\chi}_{\vec{p}}(\lambda) = \bsym{\infty}$. (ii) If $\vartheta = -\infty$, $\vec{P}^L$ is invariant on $\bsym{\chi}_{\vec{p}}([1,\infty))$; i.e., $\vec{P}^L(\bsym{\chi}_{\vec{p}}(\lambda)) \equiv \vec{P}^L(\vec{p})$. (iii) If $\vartheta > -\infty$, $\vec{P}^L(\bsym{\chi}_{\vec{p}}(\lambda))$ is strictly decreasing in $\lambda$, and $\vec{P}^L(\bsym{\chi}_{\vec{p}}(\lambda)) \to \vec{0}$ as $\lambda \to \infty$. 
	\end{lemma}
	\proof
		(i): By definition, $w_j(\chi_{j,p}(\lambda)) = w_j(p) - \log \lambda$. Because $w_j$ is strictly decreasing and $w_j(\cdot) : [p,\infty) \to (-\infty,w_j(p)]$ is onto, $\chi_{j,p}(\lambda)$ is uniquely defined for all $\lambda \geq 1$ and strictly increasing. Because $w_j(p) - \log \lambda \downarrow -\infty$ as $\lambda \uparrow \infty$, $\lim_{\lambda \uparrow \infty} \chi_{j,p}(\lambda) = \infty$. 
		
		(ii): Note that 
		\begin{equation*}
			e^{u_j(\chi_{j,p}(\lambda))}
				= e^{w_j(\chi_{j,p}(\lambda)) + v_j}
				= e^{w_j(p) - \log \lambda + v_j}
				= \lambda\inv \big( e^{w_j(p) + v_j} \big). 
		\end{equation*}
		Thus if $\vartheta = -\infty$, 
		\begin{equation*}
			P_j^L(\bsym{\chi}_{\vec{p}}(\lambda)) 
				= \frac{ e^{u_j(\chi_{j,p_k}(\lambda))} }{ \sum_{k=1}^J e^{u_{k}(\chi_{k,p_k}(\lambda))} }
				= \frac{ \lambda\inv e^{w_j(p_j) + v_j} }{ \lambda\inv \sum_{k=1}^J  e^{w_k(p_k) + v_k} }
				= P_j^L(\vec{p}). 
		\end{equation*}
		(iii): Similarly, if $\vartheta > -\infty$, 
		\begin{equation*}
			P_j^L(\bsym{\chi}_{\vec{p}}(\lambda)) 
				= \frac{ e^{u_j(\chi_{j,p_k}(\lambda))} }{ e^{\vartheta} + \sum_{k=1}^J e^{u_{k}(\chi_{k,p_k}(\lambda))} }
				= \frac{ e^{w_j(p_j) + v_j} }{ \lambda e^{\vartheta} + \sum_{k=1}^J  e^{w_k(p_k) + v_k} }
				< P_j^L(\vec{p})
		\end{equation*}
		for all $\lambda > 1$ and $P_j^L(\bsym{\chi}_{\vec{p}}(\lambda)) \to 0$ as $\lambda \to \infty$. 
	\endproof
	
	The invariance of the Logit choice probabilities over sequences of prices that tend to infinity should be viewed as an unacceptable property for realistic market models.\footnote{Mizuno \cite{Mizuno03} makes explicit use of this unrealistic property in proving the existence and uniqueness of equilibrium prices under Logit with single-product firms and linear in price utilities.} Individuals are sure to make purchasing decisions based on the absolute value of product prices, rather than just the relative value. It is easy also fairly easy to see that Lemma \ref{InvariantSets}, (iv) extends beyond Logit to any Generalized Extreme Value model without an outside good. 
	
%	\begin{lemma} 
%		\label{LCPExtension}
%		Suppose $w$ satisfies (a-c). 
%		\begin{itemize}
%			\item[(i)] If $\vartheta > -\infty$, $\vec{P}^L : [\vec{0},\bsym{\infty}) \to \triangle(J)$ can be continuously extended to a single-valued map $\vec{P}^L : [\vec{0},\bsym{\infty}] \to \triangle(J)$ with the property that $(\vec{P}^L(\vec{p}))_j = P_j^L(\vec{p}) = 0$ if $p_j = \infty$. 
%			\item[(ii)] If $\vartheta = -\infty$, $\vec{P}^L : [\vec{0},\bsym{\infty}) \to \mathbb{S}(J)$ cannot be continuously extended to a single-valued function on $[\vec{0},\bsym{\infty}]$. Any extension of $\vec{P}^L : [\vec{0},\bsym{\infty}) \to \mathbb{S}(J)$ to an outer semi-continuous set-valued map $\vec{P}^L : [\vec{0},\bsym{\infty}] \rightsquigarrow \mathbb{S}(J)$ must satisfy $\vec{P}^L(\bsym{\infty}) = \mathbb{S}(J)$. 
%		\end{itemize}
%	\end{lemma}
	
	The following form for the price derivatives of the Logit choice probabilities is also required. 
	\begin{lemma}
		\label{LogitChoiceProbabilityDerivatives}
		If $w$ satisfies Assm. \ref{LogitUtilityAssumption} (b), then $\vec{P}^L$ is continuously differentiable for all $\vec{p} \in (0,\infty)^J$ with
		\begin{equation}
			\label{ComponentFormula}
			\begin{aligned}
			(D_kP_j^L)(\vec{p}) 
				= P_j^L(\vec{p})(\delta_{j,k} - P_k^L(\vec{p}))(Dw_k)(p_k)
				= (\delta_{j,k} - P_j^L(\vec{p}))\lambda_k(\vec{p})
			\end{aligned}
		\end{equation}
		where $\lambda_k(\vec{p}) = (Dw_k)(p_k)P_k^L(\vec{p})$. In other words, 
		\begin{equation}
			\label{IntraFirmLogitChoiceProbabilityDerivatives}
			(D_f\vec{P}^L_f)(\vec{p})
				= \left( \vec{I} - \vec{P}^L_f(\vec{p})\vec{1}^\top \right)\bsym{\Lambda}_f(\vec{p})
			\quad\text{and}\quad
			(D\vec{P}^L)(\vec{p})
				= \left( \vec{I} - \vec{P}^L(\vec{p})\vec{1}^\top \right)\bsym{\Lambda}(\vec{p})
		\end{equation}
		where $\bsym{\Lambda}_f(\vec{p}) = \mathrm{diag}(\bsym{\lambda}_f(\vec{p}))$ and $\bsym{\Lambda}(\vec{p}) = \mathrm{diag}(\bsym{\lambda}(\vec{p}))$. When $w$ is twice differentiable on $(0,\infty)$, $\vec{P}^L$ is as well and the second derivatives of the Logit choice probabilities are given by
		\begin{equation}
			\label{SecondComponentFormula}
			\begin{aligned}
			(D_lD_kP_j^L)(\vec{p})
				&= \delta_{k,l} \big( (D^2w_k)(p_k) + (Dw_k)(p_k)^2 \big) P_k^L(\vec{p}) \big( \delta_{j,k} - P_j^L(\vec{p})  \big) \\
				&\quad\quad\quad\quad
					+ \lambda_k(\vec{p}) \big( 2P_j^L(\vec{p}) - \delta_{j,k} - \delta_{j,l} \big) \lambda_l(\vec{p}). 
			\end{aligned}
		\end{equation}
	\end{lemma}
	\proof
		These follow directly from Eqn. (\ref{LogitChoiceProbabilities}). 
	\endproof

	%%%%%%%%%%%%%%%%%%%%%%%%%%%%%%%%%%%%%%%%%%%%
	%%%%%%%%%%%%%%%%%%%%%%%%%%%%%%%%%%%%%%%%%%%%
	%%%%%%%%%%%%%%%%%%%%%%%%%%%%%%%%%%%%%%%%%%%%
	
	\subsection{Bounded and Vanishing Logit Profits}
	\label{SUBSEC:BoundingLogitProfits}
	
	 An understanding of when profits are bounded over the set of all non-negative prices is a pre-requisite to a general analysis of profit-optimal prices and corresponding price equilibrium. One might expect that because Assumption \ref{LogitUtilityAssumption} (c) implies that the choice probabilities vanish as prices increase without bound that profits should also, but this is not true: $w(\vec{y},p) = - \alpha \log p$, a specification derived by Allenby \& Rossi to represent ``asymmetric brand switching under price changes'' \cite{Allenby91}, can generate unbounded profits even though the choice probabilities vanish. The following property of utility functions guarantees not only the finiteness of Logit profits, but that these profits vanish as prices increase without bound.\footnote{The constant $\kappa(\vec{y})$ is convenient, but not necessary; it is easy to show that $w$ is eventually log bounded with $(r(\vec{y}),\bar{p}(\vec{y}),\kappa(\vec{y}))$ where $\kappa(\vec{y}) \neq 0$ if and only if it is so with some $(r^\prime(\vec{y}),\bar{p}^\prime(\vec{y}),0)$.}
	\begin{definition}
		\label{ELB}
		$w$ is {\bfseries eventually log bounded} at $\vec{y} \in \set{Y}$ if there exists some $r(\vec{y}) > 1$, $\kappa(\vec{y})$, and some $\bar{p}(\vec{y}) \in [0,\infty)$ such that $w(\vec{y},p) \leq - r(\vec{y}) \log p + \kappa(\vec{y})$ for all $p > \bar{p}(\vec{y})$. $w$ itself is {\em eventually log bounded} if $w$ is eventually log bounded at all $\vec{y} \in \set{Y}$. 
	\end{definition}
	Note that if $w$ is eventually log bounded then Assumption \ref{LogitUtilityAssumption} (c) necessarily holds. Furthermore, if $w$ eventually decreases sufficiently quickly then the fundamental theorem of calculus implies that $w$ is also eventually log bounded. Appendix \ref{APP:LogitExamples} contains a somewhat pathological example demonstrating that the converse need not hold. 
	
	\begin{lemma}
		\label{BoundedLogitProfits}
		Suppose $w$ satisfies Assumption \ref{LogitUtilityAssumption}. (i) Let $\vartheta > -\infty$, $\vec{q} \in [0,\infty]^J$, and suppose that there exists $r : \set{Y} \to [1,\infty)$, $\bar{p} : \set{Y} \to [0,\infty)$, and $\kappa : \set{Y} \to \R$ such that $w(\vec{y},p) \leq - r(\vec{y}) \log p + \kappa(\vec{y})$ for all $p > \bar{p}(\vec{y})$. Then $\lim_{\vec{p} \to \vec{q}} \hat{\pi}_f(\vec{p}) < \infty$. 
		(ii) If in fact $w$ is eventually log bounded, i.e. $r(\vec{y}) : \set{Y} \to (1,\infty)$, then $\lim_{\vec{p} \to \vec{q}} \hat{\pi}_f(\vec{p}) = 0$ if $\vec{q}_f = \bsym{\infty}$. 
	\end{lemma}
	\proof
		The following inequality always holds: 
		\begin{equation*}
			P_j^L(\vec{p})p_j
				\leq \frac{ e^{u_j(p_j)} }{ e^\vartheta + e^{u_j(p_j)} } p_j
				= \frac{ e^{w_j(p_j) + v_j - \vartheta} }{ 1 + e^{w_j(p_j) + v_j - \vartheta} } p_j
				= p_j e^{w_j(p_j) + v_j - \vartheta}. 
		\end{equation*}
		Under the hypothesis of (i),
		\begin{equation*}
			P_j^L(\vec{p})p_j
				\leq p_j^{1-r_j} e^{\kappa_j + v_j - \vartheta}
				\leq e^{\kappa_j + v_j - \vartheta}
		\end{equation*}
		for all $p_j$ sufficiently large. Claim (i) is a consequence of this bound. Moreover, $r_j < 1$ for all $j$, then $P_j^L(\vec{p})p_j \downarrow 0$ as $p_j \uparrow \infty$. Claim (ii) is a consequence. 
	\endproof
	Appendix \ref{APP:LogitExamples} contains an example demonstrating that the converse to the second claim is false. That is, bounded and vanishing Logit profits need not imply that $w$ is eventually log bounded. If eventual log boundedness is {\em strongly} violated in the sense of the hypothesis in the following lemma, then profits must {\em increase} without bound as prices do. 
	\begin{lemma}
		\label{UnboundedLogitProfits}
		Let $\vartheta > -\infty$ and Assumption \ref{LogitUtilityAssumption} hold. Suppose that for some $\vec{y}_* \in \set{Y}$ there exists $r(\vec{y}_*) \in (0,1)$, $\kappa(\vec{y}_*)$, and $\bar{p} \in [0,\infty)$ such that for all $p > \bar{p}$, $w(\vec{y}_*,p) \geq - r \log p + \kappa(\vec{y}_*)$. Suppose further that $\vec{y}_j = \vec{y}_*$ for some $j \in \set{J}_f$. Then $\lim_{\vec{p} \to \vec{q}} \hat{\pi}_f(\vec{p}) = \infty$ for any $\vec{q} \in [0,\infty]^J$ with $q_j = \infty$. 
	\end{lemma}
	\proof
		Under the hypothesis, $p_j e^{u_j(p_j)} \geq (p_j)^{1-r}e^{\kappa + v_j}$ for all sufficiently large $p_j$. Thus $p_j e^{u_j(p_j)} \to \infty$ as $p_j \uparrow \infty$ because $r < 1$. Clearly then $P_j^L(p_j,\vec{p}_{-j}) p_j \to \infty$ as $p_j \uparrow \infty$. The claim follows. 
	\endproof
	
	The results above establish when optimal profits are positive and finite, and when profit-optimal prices are not {\em all} infinite. Showing that profit maximizing prices are {\em all} finite is proved in Section \ref{SEC:LogitPriceEquilibrium} with the slightly strengthened hypothesis that $w$ eventually decreases sufficiently quickly. 
	\begin{lemma}
		\label{FiniteBRPrices}
		Suppose $w$ satisfies Assumption \ref{LogitUtilityAssumption} and is eventually log bounded.
		(i) If $\vartheta > -\infty$ and $\vec{p}_f \in [0,\infty]^{J_f}$ locally maximizes $\hat{\pi}_f(\cdot,\vec{p}_{-f})$ for any $\vec{p}_{-f} \in [0,\infty]^{J_{-f}}$, then $\vec{p}_f \neq \bsym{\infty}$. 
		(ii) If $\vartheta = -\infty$ and $\vec{p}_f \in [0,\infty]^{J_f}$ locally maximizes $\hat{\pi}_f(\cdot,\vec{p}_{-f})$ for any $\vec{p}_{-f} \in [0,\infty]^{J_{-f}} \setminus \{ \bsym{\infty} \}$, then $\vec{p}_f \neq \bsym{\infty}$. However, $\hat{\pi}_f(\cdot,\bsym{\infty})$ is maximized only by $\vec{p}_f = \bsym{\infty}$, and thus $\bsym{\infty}$ is always an equilibrium. 
	\end{lemma}
	\proof
		(i): Profit maximizing prices $\vec{p}_f$ are not {\em all} infinite because $\hat{\pi}_f(\bsym{\infty},\vec{p}_{-f}) = 0$, and any prices $\vec{p}_f > \vec{c}_f$ give $\hat{\pi}_f(\vec{p}) > 0$. (ii): The same holds when $\vartheta = -\infty$ and $\vec{p}_{-f} \neq \bsym{\infty}$, because some product's utility is finite. However, if $\vec{p}_{-f} = \bsym{\infty}$, Lemma \ref{InvariantSets} proves that
		\begin{align*}
			\hat{\pi}_f(\bsym{\chi}_{f,\vec{p}_f}(\lambda),\bsym{\infty})
				= \vec{P}_f^L(\bsym{\chi}_{f,\vec{p}}(\lambda),\bsym{\infty})^\top
					( \bsym{\chi}_{f,\vec{p}_f}(\lambda) - \vec{c}_f )
				= \vec{P}_f^L(\vec{p}_f,\bsym{\infty})^\top
					( \bsym{\chi}_{f,\vec{p}_f}(\lambda) - \vec{c}_f )
				\to \infty
		\end{align*}
		as $\lambda \uparrow \infty$, for any finite $\vec{p}_f$. 
	\endproof

%%%%%%%%%%%%%%%%%%%%%%%%%%%%%%%%%%%%%%%%%%%%%%%%%%%%%%%%%%%%%%%%%%
%%%%%%%%%%%%%%%%%%%%%%%%%%%%%%%%%%%%%%%%%%%%%%%%%%%%%%%%%%%%%%%%%%
%%%%%%%%%%%%%%%%%%%%%%%%%%%%%%%%%%%%%%%%%%%%%%%%%%%%%%%%%%%%%%%%%%

\section{Equilibrium Prices Under Logit Models}
\label{SEC:LogitPriceEquilibrium}

	This section proves the following theorem regarding equilibrium prices for Bertrand competition under the Logit model as described in Sections \ref{SEC:Framework}-\ref{SEC:LogitModels}: 
	\begin{theorem}
		\label{FPTheorem}
		Suppose that $\vartheta > -\infty$, Assumption \ref{LogitUtilityAssumption} holds, $w$ eventually decreases sufficiently quickly (Defn. \ref{EDSQ}) and $w$ has sub-quadratic second derivatives (Defn. \ref{SQSD}). There is at least one equilibrium $\vec{p}$, and any equilibrium satisfies $\vec{c} < \vec{p} < \bsym{\infty}$. 
	\end{theorem}
	For further clarity, the key results for both profit maximization and equilibrium problems are outlined with their assumptions in Tables \ref{TAB:ImportantProfitResults} and \ref{TAB:ImportantEquilibriumResults}. 
	
	Three fixed-point characterizations are applied to prove Theorem \ref{FPTheorem}. One fixed-point characterization is a generalization of existing results, while two are apparently novel. The first of these novel characterizations states that markups are equal to profits plus the (local) willingness to pay for product value. In essence, this equation is derived by factoring out the gradient of the ``inclusive value,'' or expected maximum utility, from $(\tilde{\nabla}\hat{\pi})(\vec{p})$. This new fixed-point equation also proves that multi-product firm optimal pricing problems under Logit are always ``one-parameter'' problems. Specifically, all profit-maximizing prices are determined uniquely from a knowledge of profits a single price. This observation yields our second novel fixed-point characterization, a ``reduced-form'' characterization in terms of equilibrium profits alone. 
	
	As is common in theoretical economics, the proof has two parts: First, the existence of simultaneously stationary prices is proved, followed by a proof that simultaneously stationary prices are equilibria. The most general proof of the existence of finite simultaneously stationary prices is accomplished using the Poincare-Hopf theorem \cite{Milnor65,Simsek07}. Brouwer's theorem can also be applied under stronger assumptions. Proving that simultaneously stationary prices are in fact equilibria is somewhat more involved. In the past appeals to quasi-concavity have been used to prove that profits have unique maximizers (see, e.g., \cite{Caplin91}). While the multi-product firm Logit profit functions are not quasi-concave \cite{Hanson96}, under utilities with sub-quadratic second derivatives first-order stationarity of profits in fact implies local concavity, the second-order sufficiency condition. A distinct application of the Poincare-Hopf theorem then implies that Logit profits have unique stationary points which must be unique global profit maximizers for fixed competitor's prices, effectively circumventing the difficulties with profits that are not quasi-concave. Note that while this argument establishes that simultaneously stationary prices are equilibria, it does {\em not} necessarily imply that equilibria are unique. The analysis in this section concerning models with constant unit costs and no finite limit on purchasing power serves as a prototype for the analysis of the more general cases presented in Sections \ref{SEC:QuantityCosts} and \ref{SEC:FiniteIncomes}. 
	
	\begin{table}
		\begin{center}
		\begin{small}
		\caption[Assumptions required for important profit maximizations results.]{Assumptions required for important profit maximizations results.}
		\label{TAB:ImportantProfitResults}
		\begin{tabular}{ll}
			\\ \hline
			Stationarity is necessary 
				& Assumption \ref{LogitUtilityAssumption} \\
			Stationarity is sufficient 
				& Assumption \ref{LogitUtilityAssumption}, Defn. \ref{SQSD} \\
			Optimal profits are finite
				& Defn. \ref{ELB} \\ 
			Profit-maximizing prices are finite 
				& $\vartheta > -\infty$, Assumption \ref{LogitUtilityAssumption}, Defn. \ref{EDSQ} \\
			Profit-maximizing prices are unique
				& $\vartheta > -\infty$, Assumption \ref{LogitUtilityAssumption}, Defn. \ref{EDSQ}, Defn. \ref{SQSD} \\ \hline
		\end{tabular}
		\caption[Assumptions required for important equilibrium results.]{Assumptions required for important equilibrium results. }
		\label{TAB:ImportantEquilibriumResults}
		\begin{tabular}{ll}
			\\ \hline
			Simultaneous stationarity is necessary
				& Assumption \ref{LogitUtilityAssumption} \\
			Simultaneously stationary prices are local equilibria
				& Assumption \ref{LogitUtilityAssumption}, Defn. \ref{SQSD} \\
			Simultaneously stationary prices exist
				& $\vartheta > -\infty$, Assumption \ref{LogitUtilityAssumption}, Defn. \ref{EDSQ} \\
			Local equilibria are equilibria
				& Assumption \ref{LogitUtilityAssumption}, Defn. \ref{SQSD} \\ \hline
		\end{tabular}
		\end{small}
		\end{center}
	\end{table}

	\subsection{Fixed-Point Characterizations of Price Equilibrium}
	\label{SUBSEC:FixedPointCharacterizations} 
	
	 This section characterizes simultaneous stationarity in terms of fixed-point equations. 
	 
	%%%%%%%%%%%%%%%%%%%%%%%%%%%%%%%%%%%%%%%%%%%%
	 
	 \subsubsection{The BLP Markup Equation}
	 \label{SUBSEC:MarkupEqn}
	 
	One fixed-point characterization is derived by noting that
	\begin{equation*}
		(D_f\vec{P}_f)(\vec{p})^\top
			= \bsym{\Lambda}_f(\vec{p}) ( \vec{I} - \vec{1}\vec{P}_f(\vec{p})^\top )
	\end{equation*}
	and hence $(\nabla_f\hat{\pi}_f)(\vec{p}) = \vec{0}$ for $\vec{p} \in (0,\infty)^J$ if, and only if, 
	\begin{equation}
		\label{TmpEqn}
		 ( \vec{I} - \vec{1}\vec{P}_f(\vec{p})^\top )(\vec{p}_f - \vec{c}_f )
%			= - [ (D_f\vec{w}_f)(\vec{p}_f)\diag(\vec{P}_f(\vec{p})) ] \inv \vec{P}_f(\vec{p})
			= - (D_f\vec{w}_f)(\vec{p}_f)\inv \vec{1}. 
	\end{equation}
	This is a direct generalization of the fixed-point equations derived under ``constant coefficient'' linear in price utility (i.e., $w(\vec{y},p) = - \alpha p$ for some $\alpha > 0$) for single-product firms by Anderson \& de Palma \cite{Anderson92a} and for multi-product firms by Besanko et al \cite{Besanko98}. 
	
	The following statements formalize this result. 
	\begin{lemma}
		\label{FPENecessaryEta}
		Suppose $w$ satisfies Assumption \ref{LogitUtilityAssumption}. (i) $( \vec{I} - \vec{1}\vec{P}^L_f(\vec{p})^\top)\inv$ exists whenever $\vartheta > -\infty$ or, if $\vartheta = -\infty$, when $\vec{p}_{-f} \neq \bsym{\infty}$, but not otherwise. 
			(ii) If $\vec{p}_f \in (0,\infty)^{J_f}$ locally maximizes $\hat{\pi}_f(\cdot,\vec{p}_{-f})$, then $\vec{p}_f = \vec{c}_f + \bsym{\eta}_f(\vec{p})$ where
			\begin{equation}
				\label{GenLogitFPE}
				\bsym{\eta}_f(\vec{p}) = - ( \vec{I} - \vec{1}\vec{P}^L_f(\vec{p})^\top)\inv (D\vec{w}_f)(\vec{p}_f)\inv \vec{1}. 
			\end{equation}
			(iii) If $\vec{p} \in (0,\infty)^J$ is a local equilibrium, then $\vec{p} = \vec{c} + \bsym{\eta}(\vec{p})$ where
			\begin{equation}
				\bsym{\eta}_f(\vec{p})
					= - (D\vec{w}_f)(\vec{p}_f)\inv \vec{1}
						- \left( \frac{\vec{P}^L_f(\vec{p})^\top(D\vec{w}_f)(\vec{p}_f)\inv \vec{1}}
								{1 - \vec{P}^L_f(\vec{p})^\top \vec{1} } \right) 
							\vec{1}
			\end{equation}
			(iv) For any $\vec{p}\in(0,\infty)^J$, $\bsym{\eta}_f(\vec{p}) > \vec{0}$ and $\bsym{\eta}_f(\vec{p}) > \vec{0}$. As a consequence, equilibrium prices have positive markups. 
	\end{lemma}
	The fixed-point equation in (iii) is a specialization of the ``markup'' equation Eqn. (\ref{MarkupEqn}) popularized for Mixed Logit models by Berry, Levinsohn, \& Pakes \cite{Berry95}; see also \cite{Morrow08, Morrow10a}. 
	\proof
		(i): The Sherman-Morrison-Woodbury formula for the inverse of a rank-one perturbation of the identity \cite[Chapter 2, pg. 50]{Ortega70} implies that
		\begin{equation}
			\label{EQN:RankOneInverse}
			\left( \vec{I} - \vec{1}\vec{P}_f^L(\vec{p})^\top \right)\inv 
				= \vec{I} + \left( \frac{1}{1-\vec{P}_f^L(\vec{p})^\top\vec{1}} \right) \vec{1}\vec{P}_f^L(\vec{p})^\top; 
		\end{equation}
		so long as $\vec{P}_f^L(\vec{p})^\top\vec{1} < 1$. This last condition will hold if either $\vartheta > -\infty$ or, if $\vartheta = -\infty$, $\vec{p}_{-f} \neq \bsym{\infty}$. 
		
		(ii): We can write
		\begin{align*}
			(\nabla_f\hat{\pi}_f)(\vec{p})
				&= \bsym{\Lambda}_f(\vec{p})\Big( ( \vec{I}-\vec{1P}_f^L(\vec{p})^\top )(\vec{p}_f-\vec{c}_f) + (D_f\vec{w}_f)(\vec{p}_f)\inv\vec{1} \Big). 
		\end{align*}
		Stationarity then requires $( \vec{I}-\vec{1P}_f^L(\vec{p})^\top )(\vec{p}_f-\vec{c}_f) +(D_f\vec{w}_f)(\vec{p}_f)\inv\vec{1} = \vec{0}$. (iii) is a consequence of (ii). 
		
		(iv): Eqn. (\ref{EQN:RankOneInverse}) proves that $\bsym{\eta}_f(\vec{p}) > \vec{0}$ so long as $(Dw_j)(p_j) < 0$. 
	\endproof
	 
	%%%%%%%%%%%%%%%%%%%%%%%%%%%%%%%%%%%%%%%%%%%%
	 
	 \subsubsection{A New Equation}
	 \label{SUBSEC:ZetaEqn}
	
	Another fixed-point characterization follows by multiplying $\vec{p}_f - \vec{c}_f$ through $\vec{I} - \vec{1}\vec{P}_f(\vec{p})^\top$, instead of inverting $\vec{I} - \vec{1}\vec{P}_f(\vec{p})^\top$ as a whole, yielding
	\begin{equation*}
		\vec{p}_f - \vec{c}_f
%			= ( \vec{P}_f(\vec{p})^\top(\vec{p}_f - \vec{c}_f) ) \vec{1} - (D_f\vec{w}_f)(\vec{p}_f)\inv \vec{1}
			= \hat{\pi}_f(\vec{p}) \vec{1} - (D_f\vec{w}_f)(\vec{p}_f)\inv \vec{1}
	\end{equation*}
	$\vec{1}\vec{P}_f(\vec{p})^\top$ could be considered the ``contractive'' part of $(\vec{I} - \vec{1}\vec{P}_f(\vec{p})^\top)$ because $\norm{\vec{1}\vec{P}_f(\vec{p})^\top}_\infty = \norm{\vec{P}_f(\vec{p})}_1 < 1$. 
	
	This derivation proves the following result. 
	\begin{lemma}
		\label{FPENecessary}
		Suppose $w$ satisfies Assumption \ref{LogitUtilityAssumption} (a) and (b). Define $\bsym{\zeta} : (0,\infty)^J \to \R^J$ by $\bsym{\zeta}(\vec{p}) = \tilde{\bsym{\pi}}(\vec{p}) - (D\vec{w})(\vec{p})\inv\vec{1}$ where $\tilde{\bsym{\pi}}(\vec{p}) \in \R^J$ is the vector with components $( \tilde{\bsym{\pi}}(\vec{p}) )_j = \hat{\pi}_{f(j)}(\vec{p})$. $\bsym{\zeta}$ has components $\zeta_j(\vec{p}) = \hat{\pi}_f(\vec{p}) - (Dw_j)(p_j)\inv$ where $j \in \set{J}_f$, and ``intra-firm'' components $\bsym{\zeta}_f(\vec{p}) = \hat{\pi}_f(\vec{p})\vec{1} - (D\vec{w}_f)(\vec{p}_f)\inv\vec{1}$. 
		
		(i) For any $\vec{p} \in (0,\infty)^J$, $(\nabla_f\hat{\pi}_f)(\vec{p}) = \bsym{\Lambda}_f(\vec{p}) \bsym{\varphi}_f(\vec{p})$ and $(\tilde{\nabla}\hat{\pi})(\vec{p}) = \bsym{\Lambda}(\vec{p}) \bsym{\varphi}(\vec{p})$ where
		\begin{equation}
			\label{LogitFixedPointGradient}
			\bsym{\varphi}_f(\vec{p}) = \vec{p}_f - \vec{c}_f - \bsym{\zeta}_f(\vec{p})
			\quad\quad\text{and}\quad\quad
			\bsym{\varphi}(\vec{p}) = \vec{p} - \vec{c} - \bsym{\zeta}(\vec{p}). 
		\end{equation}
		(ii) If $\vec{p}_f \in (0,\infty)^{J_f}$ locally maximizes $\hat{\pi}_f(\cdot,\vec{p}_{-f})$, then $\bsym{\varphi}_f(\vec{p}) = \vec{0}$; i.e., $\vec{p}_f = \vec{c}_f + \bsym{\zeta}_f(\vec{p}_f,\vec{p}_{-f})$. 
		(iii) If $\vec{p} \in (0,\infty)^J$ is a local equilibrium, then $\bsym{\varphi}(\vec{p})$; i.e. $\vec{p}  = \vec{c} + \bsym{\zeta}(\vec{p})$. 
	\end{lemma}
	
	In Appendix \ref{APP:LogitExamples}, Eqn. (\ref{LogitFixedPointGradient}) is used to show that profits under Logit with $w(\vec{y},p) = - \alpha \log p$, a model first posed by Allenby \& Rossi \cite{Allenby91}, have no finite profit-maximizing prices when $\alpha \leq 1$. Sandor \cite{Sandor01} has also made this observation. Notably, Allenby \& Rossi {\em do} undertake price optimization exercises. Such exercises thus rely on estimating a coefficient $\alpha$ that is statistically significantly strictly greater than one, a question not addressed in \cite{Allenby91}. 
	
	Positivity and finiteness of equilibrium prices can be considered important regularity properties, and follow from the $\bsym{\zeta}$ characterization. 
	\begin{lemma}
		Suppose $w$ satisfies Assumption \ref{LogitUtilityAssumption}, $w$ eventually decreases sufficiently quickly, and either $\vartheta > -\infty$ or $\vec{p}_{-f} \neq \bsym{\infty}$ if $\vartheta = -\infty$. Then no $\vec{p}_f$ with some $p_j < c_j$ or $p_j = \infty$ maximizes $\hat{\pi}_f(\cdot,\vec{p}_{-f})$. 
	\end{lemma}
	\proof
		Lemma \ref{FPENecessaryEta}, (iv), proves that $p_j > c_j$ if $p_j \neq 0$. We complete the claim by proving that no price $p_j = 0$ in equilibrium. For $p_j > 0$, the profit derivatives are
		\begin{align*}
			(D_j\hat{\pi}_f)(\vec{p})
				= \lambda_j(\vec{p}) ( p_j - c_j - \hat{\pi}_f(\vec{p}) ) + P_j^L(\vec{p})
				= \lambda_j(\vec{p})( p_j - c_j - \zeta_j(\vec{p}) )
		\end{align*}
		If $\lim_{p\downarrow 0}(Dw_j)(p) = 0$, then the first equation here proves that $\lim_{p_j \downarrow 0} (D_j\hat{\pi}_f)(\vec{p}) = P_j^L(\vec{p}) > 0$, and thus $p_j = 0$ cannot be profit-maximizing. If $\lim_{p\downarrow 0}(Dw_j)(p) > 0$, $(D_j\hat{\pi}_f)(\vec{p}) \leq 0$ if, and only if, $p_j - c_j - \zeta_j(\vec{p}) \geq 0$. As $p_j \downarrow 0$, $p_j - c_j - \zeta_j(\vec{p}) \geq 0$ if, and only if,
		\begin{align*}
			- \hat{\pi}_f(\vec{p}) \geq ( 1 - P_j^L(\vec{p}) )c_j + \frac{1}{\abs{(Dw_j)(p_j)}} . 
		\end{align*}
		Because profit-optimal prices are positive, the left hand side is negative while the right hand side is positive. By contradiction, $p_j = 0$ cannot be profit-maximizing. The finiteness of equilibrium prices follows from Lemma \ref{LEM:ZetaBound}. 
	\endproof
	 
	%%%%%%%%%%%%%%%%%%%%%%%%%%%%%%%%%%%%%%%%%%%%
	 
	 \subsubsection{A Single-Parameter Equation}
	 \label{SUBSEC:MarkupEqn}
	
	The $\bsym{\zeta}$ characterization also illustrates that price equilibrium problems with the Logit model and constant unit costs are ``single-parameter problems.'' Define $\psi_j(p) = p - c_j + (Dw_j)(p)\inv$, and write $p_j = c_j + \zeta_j(\vec{p})$ as $\psi_j(p_j) = \hat{\pi}_f(\vec{p})$. Note that for fixed $f$, the right hand side of this equation is the same for all $j \in \set{J}_f$. Thus if the right hand side is known and $\psi_j$ is invertible, all prices are uniquely defined. Conversely if a {\em single} price is known the right hand side can be computed, thus generating all prices. 
	
	The following characteristics of the maps $\psi_j : [c_j,\infty) \to [0,\infty)$ formalize this logic. 
	\begin{lemma}
		\label{LEM:PsiLem}
		Suppose Assumption \ref{LogitUtilityAssumption}  holds. (i) $\psi_j(c_j) < 0$. (ii) If $w$ also eventually decreases sufficiently quickly, then $\psi_j(p) \to \infty$ as $p \to \infty$. (iii) Finally, $\psi_j$ is differentiable and strictly increasing if, and only if, $w$ is twice differentiable and has sub-quadratic second derivatives. 
	\end{lemma}
	\proof
		(i): $\psi_j(c_j) = (Dw_j)(c_j)\inv < 0$. (ii): By assumption, there exists some $r_j > 1$ and $\bar{p}_j > 0$ such that $(Dw_j)(p_j)\inv \geq - r_j / p_j$ for all $p_j \geq \bar{p}_j$. Then
		\begin{equation*}
			\psi_j(p_j) = p_j - c_j + \frac{1}{(Dw_j)(p_j)}
				\geq \left( 1 - \frac{1}{r_j} \right)p_j - c_j
				\to 0
				\quad\text{as}\quad p_j \uparrow \infty. 
		\end{equation*}
		(iii): $\psi_j$ is continuously differentiable if $w_j$ is twice continuously differentiable and strictly decreasing. Specifically, $(D\psi_j)(p) = 1 - \omega_j(p)$, and $\psi_j$ is increasing if, and only if, $w$ has sub-quadratic second derivatives. 
	\endproof
	
	\begin{corollary}
		Let $w$ satisfy Assumption \ref{LogitUtilityAssumption}, eventually decrease sufficiently quickly, and be twice continuously differentiable with sub-quadratic second derivatives. Then for all $j$ the equation $\psi_j(p) = \pi$ has a unique solution $\Psi_j(\pi) > c_j$ for any $\pi > 0$. 
	\end{corollary}
	
	Equilibrium prices under Logit models can thus be characterized in terms of a fixed-point equation for equilibrium {\em profits} alone. Let $\bsym{\Psi} : \set{P}^F \to \prod_{j=1}^J [c_j,\infty)$ be defined component-wise by $( \bsym{\Psi}(\bsym{\pi}) )_j = \Psi_j(\pi_{f(j)})$, where $\bsym{\pi} \in \set{P}^F$ and $f(j) \in \{1,\dotsc,F\}$ denotes the (unique) index of the firm offering product $j$. Next, let $\hat{\bsym{\pi}} : \prod_{j=1}^J [c_j,\infty) \to \set{P}^F$ have profits $\hat{\pi}_f$ as component functions; i.e. $\hat{\pi}_f : \set{P}^J \to \R$. Equilibrium profits satisfy the fixed-point equation 
	\begin{equation*}
		\bsym{\pi}
			= \hat{\bsym{\pi}} ( \bsym{\Psi} (\bsym{\pi}) ) 
			= ( \hat{\bsym{\pi}} \circ \bsym{\Psi} )(\bsym{\pi}) 
			= \bsym{\phi}(\bsym{\pi}) . 
	\end{equation*}
	Given any such fixed-point $\bsym{\pi}$, all equilibrium prices can be recovered by evaluating $\bsym{\Psi}$. 

	%%%%%%%%%%%%%%%%%%%%%%%%%%%%%%%%%%%%%%%%%%%%
	%%%%%%%%%%%%%%%%%%%%%%%%%%%%%%%%%%%%%%%%%%%%
	%%%%%%%%%%%%%%%%%%%%%%%%%%%%%%%%%%%%%%%%%%%%
	
	\subsection{Existence of Simultaneously Stationary Prices}
	\label{SUBSEC:LogitExistence}
	
	This section provides three proofs of the existence of simultaneously stationary prices using each of the fixed-point characterizations given above. Brouwer's fixed-point theorem is the typical tool, and is used for proofs based on the $\bsym{\eta}$ and $\bsym{\phi}$ characterizations. However the most general result applies the $\bsym{\zeta}$ characterization and the Poincare-Hopf theorem. Throughout this section it is assumed that $\vartheta > -\infty$. 
	
	%%%%%%%%%%%%%%%%%%%%%%%%%%%%%%%%%%%%%%%%%%%%%%%%%%%%%%%
	\subsubsection{A proof based on the $\bsym{\eta}$ map.}
	
	Brouwer's theorem can be applied to the $\bsym{\eta}$ characterization. First recall that $\bsym{\eta}(\vec{p}) \geq \vec{0}$. Next, assume that $\tau = \sup_{\vec{p} \in (\vec{0},\bsym{\infty})} \norm{ \bsym{\eta}(\vec{p}) }_\infty < \infty$;  conditions for this are given below. It then follows that $\vec{c} + \bsym{\eta}(\vec{p})$ maps $[\vec{c},\tau \vec{1}] \subset \R^J$, a compact convex set, into itself. Since $\vec{c} + \bsym{\eta}(\cdot)$ is also continuous, Brouwer's fixed-point theorem implies the existence of a fixed-point $\vec{p} = \vec{c} + \bsym{\eta}(\vec{p})$. 
	
	\begin{lemma}
		\label{EtaBounded}
		Suppose $\vartheta > -\infty$, Assumption \ref{LogitUtilityAssumption} holds, and $w$ is concave in price. Then $\tau < \infty$. 
	\end{lemma} 
	\proof
		Because $\norm{ ( \vec{I} - \vec{1P}_f^L(\vec{p})^\top )\inv }_\infty \leq ( 1-\norm{\vec{P}_f^L(\vec{p})}_1 )\inv$, the bound
		\begin{align*}
			\norm{ \bsym{\eta}_f(\vec{p}) }_\infty 
%				&= \norm{ ( \vec{I} - \vec{1P}_f(\vec{p})^\top )\inv ( D\vec{w}_f )(\vec{p}_f)\inv\vec{1} }_\infty \\
%				& \leq \norm{ ( \vec{I} - \vec{1P}_f(\vec{p})^\top )\inv _\infty \norm{ ( D\vec{w}_f )(\vec{p}_f)\inv\vec{1} }_\infty \\
%				&\leq \frac{ \norm{ ( D\vec{w}_f )(\vec{p}_f)\inv\vec{1} }_\infty }{ 1 - \norm{\vec{P}_f^L(\vec{p})}_1 } \\
				&= \frac{ \max_{j \in \set{J}_f} \abs{(Dw_j)(p_j)}\inv }{ 1 - \sum_{j\in\set{J}_f} P_j^L(\vec{p}) }
		\end{align*}
		controls the growth in $\bsym{\eta}$. If $w$ satisfies Assumption \ref{LogitUtilityAssumption} and is concave in price, then $\abs{(Dw_j)(p_j)}\inv \leq \abs{(Dw_j)(c_j)}\inv$ for all $p_j \geq c_j$. This implies that $\max_{j \in \set{J}_f} \abs{(Dw_j)(p_j)}\inv$ is bounded over $\vec{p}_f \in [\vec{c}_f,\bsym{\infty}) \subset \R^{J_f}$. If $\vartheta > -\infty$, $\sup_{\vec{p} \in (\vec{0},\bsym{\infty})} ( \sum_{j\in\set{J}_f} P_j^L(\vec{p}) ) < 1$ for all $f$. Under these assumptions, $\tau < \infty$. 
	\endproof
	
	\begin{corollary}
		If $\vartheta > -\infty$ and $w$ satisfies Assumption \ref{LogitUtilityAssumption}  and is concave in price, then there exists a fixed-point $\vec{p} = \vec{c} + \bsym{\eta}(\vec{p})$. 
	\end{corollary}
	
	%%%%%%%%%%%%%%%%%%%%%%%%%%%%%%%%%%%%%%%%%%%%%%%%%%%%%%%
	
	\subsubsection{A proof based on $\bsym{\phi}$.}
	
	Section \ref{SUBSEC:FixedPointCharacterizations} defined simultaneously stationary {\em profits} as a fixed-point of the map $\bsym{\phi} = \hat{\bsym{\pi}} \circ \bsym{\Psi}$. This characterization and Brouwer's theorem can be used to prove the existence of simultaneously stationary prices. 
	
	\begin{lemma}
		\label{PiFPExists}
		Suppose $w$ satisfies Assumption \ref{LogitUtilityAssumption}, eventually decreases sufficiently quickly, has sub-quadratic second derivatives and $\vartheta > -\infty$. Then there exists at least one fixed-point $\bsym{\pi} = \bsym{\phi}(\bsym{\pi})$. 
	\end{lemma}
	\proof $\Psi_j$ can be {\em continuously} extended to $[0,\infty]$ as $\Psi_j(\infty) = \infty$ under the condition that $w$ eventually decreases sufficiently quickly. Because $w$ is then also eventually log bounded, letting $\vartheta > -\infty$ ensures that for any $f$, $\hat{\pi}_f(\vec{p}) < \infty$ for all $\vec{p} \in [0,\infty]^J$. Thus $\bsym{\phi}$ is continuous and maps the compact, convex set $[0,\infty]^J$ strictly into itself. By Brouwer's fixed-point theorem, there exists a fixed-point $\bsym{\pi}$ on $[0,\infty]^F$. Furthermore, this fixed-point has no infinite components, by Lemma \ref{FiniteBRPrices}. \endproof
	
	While the restriction $\vartheta > -\infty$ could be removed through an application of Kakutani's extension of Brouwer's theorem \cite{Kakutani41,Caplin91}, the fact that $\bsym{\infty}$ is always an equilibrium makes this approach uninformative. 
	
	\subsubsection{A proof based on the $\bsym{\zeta}$ map.}
	
	The Poincare-Hopf theorem requires that sum of the indices of the vector field $\bsym{\varphi}(\vec{p}) = \vec{p} - \vec{c} - \bsym{\zeta}(\vec{p})$ over all zeros of $\bsym{\varphi}$ equals one, so long as $\bsym{\varphi}$ points outward on the boundary of some compact hyper-rectangle $[\vec{c},\bar{\vec{p}}] \subset \set{P}^J$ for some $\bar{\vec{p}} < \bsym{\infty}$. Particularly, this sum of indices cannot be empty and hence there must be {\em at least one} zero of $\bsym{\varphi}$, and thus at least one simultaneously stationary point. 
	
	The hypotheses required in the Poincare-Hopf Theorem follow from the next lemma. 
	\begin{lemma}
		\label{LEM:ZetaBound}
		Suppose $\vartheta > -\infty$ and $w$ satisfies Assumption \ref{LogitUtilityAssumption}. (i) If $\vec{p} \geq \vec{c}$ and $p_j = c_j$, $\varphi_j(\vec{p}) < 0$. (ii) If $w$ eventually decreases sufficiently quickly, there exists some $\bar{\vec{p}} \in (\vec{c},\bsym{\infty})$ such that $\varphi_j(\vec{p}) > 0$ whenever $p_j \geq \bar{p}_j$, regardless of $\vec{p}_{-j}$.
	\end{lemma}
	\proof
		(i): When $\vec{p} \geq \vec{c}$ and $p_j = c_j$, 
		\begin{align*}
			p_j - c_j - \zeta_j(\vec{p})
				= - \hat{\pi}_f(\vec{p}_f,\vec{p}_{-f}) - \frac{1}{\abs{(Dw_j)(c_j)}}
				< 0. 
		\end{align*}
		(ii): Observe that the following bound is valid for large enough $p_j$ because $w$ eventually decreases sufficiently quickly: 
		\begin{align*}
			p_j - c_j - \zeta_j(\vec{p})
				\geq \left( 1 - \frac{1}{r_j} \right) p_j - ( c_j + \hat{\pi}_f(\vec{p}_f,\vec{p}_{-f}) ).  
		\end{align*}
		Because $\hat{\pi}_f(\cdot,\vec{p}_{-f})$ is bounded ($w$ is eventually log bounded), and $r_j > 1$, $(1 - 1/r_j)p_j \to \infty$ as $p_j \to \infty$, $p_j$ can always be chosen large enough to make $p_j - c_j - \zeta_j(\vec{p}) > 0$. When $\vartheta > -\infty$, $\hat{\pi}_f(\cdot)$ itself is bounded and hence $\bar{\vec{p}}_f$ can be chosen independently of $\vec{p}_{-f}$. 
	\endproof
	
	\begin{theorem}
		\label{FPExists}
		Suppose $\vartheta > -\infty$, $w$ satisfies Assumption \ref{LogitUtilityAssumption} and eventually decreases sufficiently quickly. There exists at least one $\vec{p} \in (\vec{c},\bsym{\infty})$ such that $\vec{p} = \vec{c} + \bsym{\zeta}(\vec{p})$. 
	\end{theorem}
	\proof
		Let $\bar{\vec{p}}$ be as in Lemma \ref{LEM:ZetaBound}. $\bsym{\varphi}(\vec{p})$ is a vector field on $[\vec{c},\bar{\vec{p}}]$ that points outward on the boundary of $[\vec{c},\bar{\vec{p}}]$. Let the set of zeros of $\bsym{\varphi}$ be denoted by $\mathfrak{Z} = \{ \vec{p} \in (\vec{c},\bar{\vec{p}}) : \bsym{\varphi}(\vec{p}) = \vec{0} \}$ and let $\mathrm{index}_{\vec{p}}(\bsym{\varphi})$ denote the index of $\bsym{\varphi}$ at $\vec{p} \in \mathfrak{Z}$. The Poincare-Hopf Theorem states that $\sum_{\vec{p} \in \mathfrak{Z}} \mathrm{index}_{\vec{p}}(\bsym{\varphi}) = 1$, where the value of the sum on the left is taken to be $0$ if $\mathfrak{Z} = \{ \emptyset \}$. Hence there is at least one zero of $\bsym{\varphi}$. 
	\endproof
	Note that it is not required that $w$ be concave in price {\em or} have sub-quadratic second derivatives in order for simultaneously stationary prices to exist. 
	
	Lemma \ref{LEM:ZetaBound} also shows that $\vec{c} + \bsym{\zeta}(\cdot)$ maps $[\vec{c},\bar{\vec{p}} + \bsym{\epsilon}]$ into itself, for any $\bsym{\epsilon} \geq \vec{0}$. Thus Brouwer's Theorem could be applied just as easily to achieve this existence result. This is not the case in Section \ref{SEC:QuantityCosts} below, however.

	\subsection{Sufficiency of Stationarity}
	\label{SUBSEC:LogitSufficiency}
	
	A general approach to multi-product firm equilibrium problems can rely on quasi-concavity to establish the uniqueness of profit-maximizing prices \cite{Hanson96}. However, something {\em like} quasi-concavity is required to be able to connect fixed-points $\vec{p}_f = \vec{c}_f + \bsym{\zeta}_f(\vec{p})$ to profit maximizers, and thus fixed-points $\vec{p} = \vec{c} + \bsym{\zeta}(\vec{p})$ to {\em local} equilibria. Furthermore, uniqueness of profit maximizing prices is required to ensure the existence of equilibria proper. 
	
	Logit profits have the surprising property that stationarity, the first-order necessary condition, implies local concavity, the second-order sufficient condition when $w$ has sub-quadratic second derivatives. The Poincare-Hopf theorem again serves to commute this local result on the second derivatives of profits to a global property, the uniqueness of profit-maximizing prices. 
	
	\begin{lemma}
		\label{FPESufficient}
		Suppose $\vartheta > -\infty$ and $w$ satisfies Assumption \ref{LogitUtilityAssumption} and has sub-quadratic second derivatives. 
		(i) Satisfaction of the first-order condition $(\nabla_f\hat{\pi}_f)(\vec{p}_f,\vec{p}_{-f}) = \vec{0}$ is sufficient for $\vec{p}_f \in (\vec{0},\bsym{\infty}) \subset \R^{J_f}$ to be a local maximizer of $\hat{\pi}_f(\cdot,\vec{p}_{-f})$.
		(ii) Satisfaction of the simultaneous stationarity condition $(\tilde{\nabla}\hat{\pi})(\vec{p}) = \vec{0}$ is sufficient for $\vec{p} \in (\vec{0},\bsym{\infty}) \subset \R^J$ to be a local equilibrium. 
		(iii) If, in addition, $w$ also eventually decreases sufficiently quickly then there is a unique stationary point that is a finite maximizer of $\hat{\pi}_f(\cdot,\vec{p}_{-f})$. 
		(iv) When $w$ eventually decreases sufficiently quickly, the simultaneous stationarity condition $(\tilde{\nabla}\hat{\pi})(\vec{p}) = \vec{0}$ is sufficient for $\vec{p} \in (\vec{0},\bsym{\infty}) \subset \R^J$ to be an equilibrium. 
	\end{lemma}

	\proof Claim (ii) is an obvious corollary to (i), and claim (iv) is an obvious corollary to (iii). 
	
		Claim (i) is a consequence of the following componentwise formula for the intra-firm profit price-Hessians $(D_f\nabla_f\hat{\pi}_f)(\vec{p})$ when $\vec{p}_f$ makes $\hat{\pi}_f(\cdot,\vec{p}_{-f})$ stationary: for $k,l \in \set{J}_f$, 
		\begin{align*}
			(D_lD_k\hat{\pi}_f^L)(\vec{p})
				&= (D_k\lambda_k)(\vec{p}) \varphi_k(\vec{p}) + \lambda_k(\vec{p}) (D_l\varphi_k)(\vec{p}) \\
				&= \lambda_k(\vec{p}) \Bigg( \delta_{k,l} - (D_l\hat{\pi}_f)(\vec{p}) - \omega_k(p_k)\delta_{k,l} \Bigg) \\
				&= \lambda_k(\vec{p}) (1-\omega_k(p_k))\delta_{k,l}
		\end{align*}
%		\begin{equation}
%			\label{LogitProfitsSecondDers}
%			\begin{aligned}
%			(D_lD_k\hat{\pi}_{f(k)})(\vec{p})
%				&= \delta_{k,l} (Dw_k)(p_k) P_k^L(\vec{p}) \\
%				&\quad\quad\quad\quad
%					\times\left( \left( \frac{(D^2w_k)(p_k) + (Dw_k)(p_k)^2}{(Dw_k)(p_k)} \right) ( p_k - c_k - \hat{\pi}_{f(k)}(\vec{p}) ) + 2 \right) \\
%				&\quad\quad\quad\quad
%					- (D_k\hat{\pi}_{f(k)})(\vec{p}) P_l^L(\vec{p}) (Dw_l)(p_l) \\
%				&\quad\quad\quad\quad
%					- (Dw_k)(p_k) P_k^L(\vec{p}) (D_l\hat{\pi}_{f(l)})(\vec{p}) 
%			\end{aligned}
%		\end{equation}
%		Particularly, when $\vec{p}_f$ makes $\hat{\pi}_f(\cdot,\vec{p}_{-f})$ stationary, 
		In matrix form, 
		\begin{equation*}
			(D_f\nabla_f\hat{\pi}_f)(\vec{p})
				= \bsym{\Lambda}_f(\vec{p})(\vec{I} - \bsym{\Omega}_f(\vec{p}_f))
		\end{equation*}
		where $\bsym{\Omega}_f(\vec{p}_f)$ is a diagonal matrix with entries $\omega_j(p_j) = (D^2w_j)(p_j) / (Dw_j)(p_j)^2$. Thus, the Hessians are diagonal matrices with negative diagonal entries when $w$ has sub-quadratic second derivatives and $\hat{\pi}_f(\cdot,\vec{p}_{-f})$ is locally concave at any stationary prices. 
		
		Claim (iii) is a consequence of the Poincare-Hopf Theorem: Because $\hat{\pi}_f(\cdot,\vec{p}_{-f})$ is maximized at $\vec{p}_f = \vec{c}_f + \bsym{\zeta}_f(\vec{p})$, 
		\begin{align*}
			(-1)^{J_f}
				&= \mathrm{index}_{\vec{p}_f}( (\nabla_f\hat{\pi}_f)(\cdot,\vec{p}_{-f}) )
				= \mathrm{sign}\det(D_f\nabla_f\hat{\pi}_f)(\vec{p}) \\
				&= \mathrm{sign}\det \bsym{\Lambda}_f(\vec{p})
					\cdot \mathrm{sign}\det (D_f\bsym{\varphi}_f)(\vec{p})
				= (-1)^{J_f} \cdot \mathrm{sign}\det (D_f\bsym{\varphi}_f)(\vec{p}); 
		\end{align*}
		see Chapter 6 in \cite{Milnor65} for some of the basic results invoked here. Because of these equalities, 
		\begin{equation*}
			\mathrm{index}_{\vec{p}_f}( \bsym{\varphi}_f(\cdot,\vec{p}_{-f}) )
				= \mathrm{sign}\det (D_f\bsym{\varphi}_f)(\vec{p}) 
				= 1. 
		\end{equation*}
		But the Poincare-Hopf theorem require the sum of indices of all zeros of $\bsym{\varphi}_f(\cdot,\vec{p}_{-f})$ over $[\vec{c}_f,\bar{\vec{p}}_f]$ to be 1. The zero must, therefore, be unique. 
	\endproof

\section{Quantity-Dependent Unit Costs}
\label{SEC:QuantityCosts}

Much of the theoretical literature allows unit costs to depend on sales volumes. This section extends the techniques used in the previous section to this case. 

Specifically, the analysis in this section proves the following theorem:
\begin{theorem}
	Let $\vartheta > -\infty$, Assumption \ref{LogitUtilityAssumption} holds with a $w$ that eventually decreases sufficiently quickly with sub-quadratic second derivatives, and total costs satisfy Assumption \ref{ASS:NonConstUnitCostsAssum}. Then there exists a vector of equilibrium prices $\vec{p}$ satisfying $\vec{c}(\vec{P}(\vec{p})) < \vec{p} < \bsym{\infty}$, and no equilibrium prices that do not satisfy these bounds. 
\end{theorem}

\subsection{Assumptions} 

Assumption \ref{ASS:ConstUnitCostsAssum} restricted attention to constant unit costs for the following reason: Suppose that unit costs {\em did} depend on the quantity sold, and let $c_f^U(\vec{y}_j,Q_j(\vec{Y},\vec{p}))$ give the unit costs to firm $f$ for offering product $\vec{y}_j$ in the market of products with characteristics $\vec{Y}$, prices $\vec{p}$, and resultant demand $Q_j(\vec{Y},\vec{p})$. Then firm $f$'s random profits are
\begin{equation*}
	\Pi_f(\vec{Y},\vec{p}) 
		= \vec{Q}_f(\vec{Y},\vec{p})^\top ( \vec{p}_f - \vec{c}_f^U(\vec{Y}_f,\vec{Q}_f(\vec{Y},\vec{p})) ) 
			- c_f^F(\vec{Y}_f). 
\end{equation*}
Notice that profits are no longer a {\em linear} function of the demands, and thus expected profits need not be a linear function of the expected demands. Computing these expected costs could be quite difficult, as this would involve sums over the space of realizable demands.

One way to relax the assumption that unit costs do not depend on production volume without overly complicating the resulting equilibrium conditions is to suppose that firms decide on pricing based on the total costs corresponding to {\em expected} demands, as opposed to the {\em actual} expected total costs. The following assumption then generalizes Assumption \ref{ASS:ConstUnitCostsAssum} to unit costs that may depend on the quantity sold: 
\begin{assumption}
	\label{ASS:NonConstUnitCostsAssum}
	Firm $f$ has a normalized cost function $C_f : \set{Y} \times [0,1] \to [0,\infty)$ so that the total cost of offering a product with characteristics $\vec{y} \in \set{Y}$ for any demand $q$ is $I C_f( \vec{y} , q / I )$, where again $I$ is the market size. Assume also that for any $\vec{y} \in \set{Y}$, $C_f(\vec{y},\cdot) : [0,1] \to \R$ is twice continuously differentiable, strictly increasing and convex on $[0,1)$, $C_f(\vec{y},0) = 0$, and $\sup_{P \in [0,1]} c_f(\vec{y},P) < \infty$ where $c_f(\vec{y},P) = (D^PC_f)(\vec{y},P)$. 
\end{assumption}
 The simplest example is, of course, $C_f(\vec{y}, P) = c_f^U(\vec{y}) P$ as studied in Sections \ref{SEC:Framework}-\ref{SEC:LogitPriceEquilibrium}. Given $\vec{Y}_f \in \set{Y}^{J_f}$, define $C_j : [0,1] \to [0,\infty)$ by $C_j(P) = C_f(\vec{y}_j,P)$ and $c_j(P) = (DC_f)(\vec{y}_j,P)$ where the derivative is with respect to $P$. 

Under Assumption \ref{ASS:NonConstUnitCostsAssum}, firms choose prices by solving 
\begin{equation*}
	\mxm \quad \Expect\Pi_f(\vec{p}) 
					= I \sum_{j \in \set{J}_f}
						\Big( P_j^L(\vec{p})p_j - C_j( \Expect [ Q_j(\vec{p}) ] / I ) \Big)
	\quad \wrt \quad p_j \in \set{J}_f
\end{equation*}
equivalent to
\begin{equation}
	\label{EQN:QuantityCostProblem}
	\mxm \quad \hat{\pi}_f(\vec{p})
					= \sum_{j \in \set{J}_f} \big( P_j^L(\vec{p})p_j - C_j(P_j^L(\vec{p})) \big)
	\quad \wrt \quad p_j \in \set{J}_f
\end{equation}

The following observation is an extension of Lemma \ref{FiniteBRPrices}. 
\begin{lemma}
	Suppose $\vartheta > -\infty$, Assumption \ref{LogitUtilityAssumption} holds with a $w$ that is eventually log bounded, and Assumption \ref{ASS:NonConstUnitCostsAssum} holds. For any $\vec{p}_{-f}$, the optimal profits for Prob. (\ref{EQN:QuantityCostProblem}) are positive and finite. 
\end{lemma}
\proof
	Note that, for any $j$, 
	\begin{equation*}
		\lim_{p_j \uparrow \infty} \left( \frac{ C_j(P_j^L(\vec{p})) }{ P_j^L(\vec{p}) } \right)
			= \lim_{p_j \uparrow \infty} \left( \frac{ c_j(P_j^L(\vec{p})) (D_jP_j^L)(\vec{p}) }{ (D_jP_j^L)(\vec{p}) } \right)
			= c_j(0)
			< \infty
	\end{equation*}
	by L'Hopital's rule. Thus, as $p_j \uparrow \infty$, $p_j - C_j(P_j^L(\vec{p})) / P_j^L(\vec{p}) \to \infty$. Thus for all $j$, $p_j$ can be chosen large enough so that $p_j - C_j(P_j^L(\vec{p})) / P_j^L(\vec{p}) > 0$. Writing
	\begin{equation*}
		\hat{\pi}_f(\vec{p})
				= \sum_{j \in \set{J}_f} \big( P_j^L(\vec{p})p_j - C_j(P_j^L(\vec{p})) \big)
				= \sum_{j \in \set{J}_f} P_j^L(\vec{p}) \left( p_j - \frac{ C_j(P_j^L(\vec{p})) }{ P_j^L(\vec{p}) } \right)
	\end{equation*}
	proves that there exists $\vec{p}_f$, for any $\vec{p}_{-f}$ such that $\hat{\pi}_f(\vec{p}_f,\vec{p}_{-f}) > 0$. If $w$ is eventually log bounded, then $P_j(\vec{p})( p_j - C_j(P_j^L(\vec{p})) / P_j^L(\vec{p}) ) \to 0$ as $p_j \uparrow \infty$. Because $\hat{\pi}_f(\vec{p}_f,\vec{p}_{-f}) \to 0$ as $\vec{p}_f \to \bsym{\infty}$, and $\hat{\pi}_f(\vec{p}_f,\vec{p}_{-f})$ is finite if $p_j < \infty$ for {\em any} $j \in \set{J}_f$, optimal profits are finite. 
\endproof

Note that this proof does not say that all {\em prices} are finite; this is proved below using similar techniques to those used in Section \ref{SEC:LogitPriceEquilibrium}. 

\subsection{Stationarity} 

The stationarity conditions for Prob. (\ref{EQN:QuantityCostProblem}) are
\begin{align*}
	(D_k\hat{\pi}_f)(\vec{p})
		= \sum_{j \in \set{J}_f} (D_kP_j^L)(\vec{p})( p_j - c_j(P_j^L(\vec{p}))) + P_k^L(\vec{p}) 
		= 0
\end{align*}
for all $k \in \set{J}_f$ and all $f$, as can be checked. 

As before, stationarity can be written in either of two fixed-point forms: 
\begin{lemma}
	Suppose $\vartheta > -\infty$ and Assumptions \ref{LogitUtilityAssumption} and \ref{ASS:NonConstUnitCostsAssum} hold. At any simultaneously stationary prices $\vec{p} \in (0,\infty)^J$, $\vec{p} = \vec{c}(\vec{P}(\vec{p})) + \bsym{\eta}(\vec{p})$ where $\bsym{\eta}$ is as defined above and $\vec{p} = \vec{c}(\vec{P}(\vec{p})) + \bsym{\zeta}(\vec{p})$ where $\bsym{\zeta} : [0,\infty)^J \to \R^J$ is defined componentwise by
	\begin{align*}
		\zeta_k(\vec{p})
			= \sum_{j \in \set{J}_{f(k)}} P_j^L(\vec{p}) \big( p_j - c_j(P_j^L(\vec{p})) \big)
				+ \frac{1}{\abs{(Dw_k)(p_k)}}
	\end{align*}
\end{lemma}

As in Section \ref{SEC:LogitPriceEquilibrium}, these characterizations immediately establishes the positivity of markups in equilibrium:
\begin{lemma}
	Suppose $\vartheta > -\infty$, Assumption \ref{LogitUtilityAssumption} holds with $w$ eventually decreasing sufficiently quickly, and Assumption \ref{ASS:NonConstUnitCostsAssum} holds. If $\vec{p} \in [0,\infty]^J$ is a vector of equilibrium prices, then $\vec{c}(\vec{P}(\vec{p})) < \vec{p} < \bsym{\infty}$. 
\end{lemma}
\proof
	Lemma \ref{LEM:Inequalities} below proves that there exists some $\bar{p}_j$ such that $\hat{\pi}_f(\vec{p}) < 0$ whenever $p_j > \bar{p}_j$. Thus no price can be infinite in equilibrium. Moreover, no price can be zero: Suppose $\lim_{p_k \downarrow 0} (Dw_k)(p_k) = 0$. Then for $p_k > 0$, 
	\begin{equation*}
		(D_k\hat{\pi}_f)(\vec{p})
			= (Dw_k)(p_k)P_k^L(\vec{p})
				\left( p_k - c_k - \sum_{j \in \set{J}_{f(k)}} P_j^L(\vec{p}) \big( p_j - c_j(P_j^L(\vec{p})) \right)
				+ P_k^L(\vec{p})
			\to P_k^L(\vec{p}) > 0
	\end{equation*}
	as $p_k \downarrow 0$. Now suppose $\lim_{p_k \downarrow 0} (Dw_k)(p_k) < 0$. For $\vec{p}$ with $0 < p_k < \infty$, $(D_k\hat{\pi}_f)(\vec{p}) \leq 0$ if, and only if, 
	\begin{align*}
		p_k - c_k(P_k^L(\vec{p}))) - \sum_{j \in \set{J}_{f(k)}} P_j^L(\vec{p}) \big( p_j - c_j(P_j^L(\vec{p})) \big)
				- \frac{1}{\abs{(Dw_k)(p_k)}} \geq 0. 
	\end{align*}
	Taking the limit as $p_k \downarrow 0$ yields
	\begin{align*}
		- \sum_{j \in \set{J}_{f(k)} \setminus k} P_j^L(\vec{p}) \big( p_j - c_j(P_j^L(\vec{p})) \big) 
			\geq (1-P_k^L(\vec{p}))c_k(P_k^L(\vec{p}))) + \frac{1}{\abs{(Dw_k)(0)}}. 
	\end{align*}
	The right hand side is positive, but if profits are optimal, the left hand side is negative. Thus $(D_k\hat{\pi}_f)(\vec{p}) > 0$ for all $p_k$ sufficiently close to zero and all positive profits. Hence, $p_k = 0$ cannot be profit-optimal. 
	
	Knowing then that $\vec{p} \in (0,\infty)^J$, the equation $\vec{p} = \vec{c}(\vec{P}(\vec{p})) + \bsym{\eta}(\vec{p})$ applies. Because $\bsym{\eta}(\vec{p})$ is positive valued for all $\vec{p} \in (0,\infty)^J$, $\vec{p} > \vec{c}(\vec{P}(\vec{p}))$. 
\endproof

The remainder of the proof of equilibrium existence is analogous to the process for constant unit costs: First simultaneously stationary prices are shown to exist, followed by a proof that such prices are in fact always equilibrium prices. 

\subsection{Existence of Stationary Prices} 

The approach to establishing the existence of simultaneously stationary points in the following lemmas is as follows: First, a homeomorphism, $\bsym{\rho}$, between $\{ \vec{p} : \vec{p} \geq \vec{c}(\vec{P}(\vec{p})) \}$ and $[\vec{0},\bsym{\infty})$ is constructed. Second, the vector field $\bsym{\varphi} : [\vec{0},\bsym{\infty})$ defined componentwise by
\begin{align*}
	\varphi_k(\vec{p}) = p_k - c_k(P_k^L(\vec{p})) 
								- \sum_{j \in \set{J}_f} P_j^L(\vec{p})( p_j - c_j(P_j^L(\vec{p})))
								- \frac{1}{\abs{(Dw_k)(p_k)} }
\end{align*}
is transported from $\{ \vec{p} : \vec{p} \geq (D\vec{C})(\vec{P}(\vec{p})) \}$ to $[\vec{0},\bsym{\infty})$ by defining $\bsym{\psi} : [\vec{0},\bsym{\infty}) \to \R^J$ by $\bsym{\psi}(\bsym{\epsilon}) = \bsym{\varphi}(\bsym{\rho}(\bsym{\epsilon}))$ for any $\bsym{\epsilon} \in [\vec{0},\bsym{\infty})$. Third, a compact rectangle $[\vec{0},\bar{\bsym{\epsilon}}]$ is constructed, on which $\bsym{\psi}$ is continuous and points outward on the boundary. The Poincare-Hopf theorem then proves the existence of a zero $\bsym{\epsilon}_0 \in (\vec{0},\bar{\bsym{\epsilon}})$ for $\bsym{\psi}$, which maps to a zero $\vec{p}_0 = \bsym{\rho}(\bsym{\epsilon}_0)$ of $\bsym{\varphi}$ such that $\vec{p}_0 > (D\vec{C})(\vec{P}(\vec{p}_0))$. Such a point is necessarily simultaneously stationary because $(D_k\hat{\pi}_f)(\vec{p}) = \lambda_k(\vec{p})\varphi_k(\vec{p})$. 

\begin{lemma}
	\label{LEM:Homotopy}
	Suppose $\vartheta > -\infty$, Assumption \ref{LogitUtilityAssumption} holds with $w$ eventually decreasing sufficiently quickly, and Assumption \ref{ASS:NonConstUnitCostsAssum} holds. The fixed-point problem $\vec{p} = \vec{F}_{\bsym{\epsilon}}(\vec{p})$ where $\vec{F}_{\bsym{\epsilon}}(\vec{p}) = \vec{c}(\vec{P}(\vec{p})) + \bsym{\epsilon}$ has a unique solution $\bsym{\rho}(\bsym{\epsilon}) > \bsym{\epsilon}$ for every $\bsym{\epsilon} \in [\vec{0},\bsym{\infty}) \subset \R^J$. Moreover, the corresponding solution map, $\bsym{\rho} : [\vec{0},\bsym{\infty}) \to \{ \vec{p} : \vec{p} \geq \vec{c}(\vec{P}(\vec{p})) \}$ is a homeomorphism between $[\vec{0},\bsym{\infty})$ and $\{ \vec{p} : \vec{p} \geq \vec{c}(\vec{P}(\vec{p})) \}$. 
\end{lemma}
\proof Suppose that $\vec{c}(\vec{P}(\vec{p}))$ depends on $\vec{p}$, otherwise the claims are trivial. 

Assume the fixed-point $\bsym{\rho}(\bsym{\epsilon})$ is unique. $\bsym{\rho}(\bsym{\epsilon}) > \bsym{\epsilon}$ because $c_j(q) > 0$ for all $q < 1$ and $P_j^L(\vec{p}) < 1$ for all $\vec{p}$ when there is an outside good. The implicit function theorem guarantees the continuity of $\bsym{\rho}(\bsym{\epsilon})$, and the inverse map $\vec{p} \mapsto \bsym{\epsilon} = \vec{p} - \vec{c}(\vec{P}(\vec{p}))$ is continuous because $\vec{C}$ and $\vec{P}$ are continuously differentiable. $\bsym{\rho}(\bsym{\epsilon})$ is thus a homeomorphism. 

We now show that $\bsym{\rho}(\bsym{\epsilon})$ does indeed exist, as claimed. Because $\sup_{P \in [0,1]} c_j(P) = \kappa_j < \infty$, there are no fixed points with $p_j \geq \bar{p}_j = \kappa_j + \epsilon_j + \delta$ for any fixed $\delta > 0$. Moreover, there are no fixed points with $p_j = 0$ when $\inf_{\vec{p}} c_j(P_j^L(\vec{p})) > 0$; this will hold when there is an outside good, even if $c_j(1) = 0$, so long as $c_j(P) > 0$ for all $P \in [0,1)$. Let $\set{D} = [\vec{0},\bar{\vec{p}}]$, and consider 
\begin{equation*}
	(D\vec{F}_{\bsym{\epsilon}})(\vec{p})
		= (D^2\vec{C})(\vec{P}(\vec{p}))(D\vec{P})(\vec{p})
		= (D^2\vec{C})(\vec{P}(\vec{p}))(\vec{I} - \vec{P}(\vec{p})\vec{1}^\top)\bsym{\Lambda}(\vec{p})
\end{equation*}
$(D\vec{F}_{\bsym{\epsilon}})(\vec{p})$ has one as an eigenvalue only if there exists $\vec{x} \neq \vec{0}$ such that
\begin{align*}
	(D^2\vec{C})(\vec{P}(\vec{p}))(\vec{I} - \vec{P}(\vec{p})\vec{1}^\top)\bsym{\Lambda}(\vec{p})\vec{x}
		&= \vec{x}. 
%	\bsym{\Lambda}(\vec{p})\vec{x} - \Big( \vec{1}^\top\bsym{\Lambda}(\vec{p})\vec{x} \Big) \vec{P}(\vec{p}) 
%		&= (D^2\vec{C})(\vec{P}(\vec{p}))\inv\vec{x} \\
%	\Big( \bsym{\Lambda}(\vec{p}) - (D^2\vec{C})(\vec{P}(\vec{p}))\inv \Big) \vec{x} 
%		&= \Big( \vec{1}^\top\bsym{\Lambda}(\vec{p})\vec{x} \Big) \vec{P}(\vec{p})
\end{align*}
When $\abs{ \lambda_j(\vec{p}) } (D^2C_j)(P_j^L(\vec{p})) \neq -1$ for all $j$, this holds only if
\begin{align*}
	\sum_{j=1}^J \beta_j(\vec{p}) P_j^L(\vec{p}) = 1
	\quad\text{where}\quad
	\beta_j(\vec{p}) 
		= \frac{ \abs{\lambda_j(\vec{p})}(D^2C_j)(P_j^L(\vec{p})) }{ \abs{\lambda_j(\vec{p})}(D^2C_j)(P_j^L(\vec{p})) + 1 } 
\end{align*}
as can be checked. Clearly $\abs{ \lambda_j(\vec{p}) } (D^2C_j)(P_j^L(\vec{p})) \neq -1$ and $0 \leq \beta_j(\vec{p}) < 1$ for all $j$ when $C_j$ is convex. Thus, when $\vartheta \neq -\infty$, 
\begin{align*}
	\sum_{j=1}^J \beta_j(\vec{p}) P_j^L(\vec{p})
		< \sum_{j=1}^J \beta_j(\vec{p}) \left( \frac{ e^{u_j(p_j)} }{ \sum_{k=1}^J e^{u_k(p_k)} } \right)
		\leq  \max_{j = 1,\dotsc,J} \beta_j(\vec{p}) < 1. 
\end{align*}
Thus $(D\vec{F}_{\bsym{\epsilon}})(\vec{p})$ cannot have 1 as an eigenvalue. By Kellogg's Uniqueness Theorem \cite{Kellogg76}, $\bsym{\rho}(\bsym{\epsilon})$ is well-defined. \endproof

The assumption of convex costs $C_j$ would be difficult to relax in this proof: $\beta_j(\vec{p})$ cannot always be non-negative for concave costs such as $C_j(P_j^L(\vec{p})) = - \kappa_j P_j^L(\vec{p})^2$ because $\abs{ \lambda_j(\vec{p}) } \to 0$ as $p_j \uparrow \infty$, and thus $\abs{\lambda_j(\vec{p})} \kappa_j \leq 1$ for large enough $p_j$. Bounding the sums of $\beta$'s in a similar way can be done if we ensure that $\abs{ \lambda_j(\vec{p}) } \abs{ (D^2C_j)(P_j^L(\vec{p})) } > 1$, but this suggests assumptions that simultaneously restrict the behavior allowed in the utilities and costs. 

\begin{lemma}
	\label{LEM:Inequalities}
	Suppose $\vartheta > -\infty$ and Assumptions \ref{LogitUtilityAssumption} and \ref{ASS:NonConstUnitCostsAssum} hold. (i) If $\vec{p} \geq \vec{c}(\vec{P}(\vec{p}))$ and $p_k = c_k(P_k^L(\vec{p}))$, then $\varphi_k(\vec{p}) < 0$. (ii) When $w$ also eventually decreases sufficiently quickly, there exists $\bar{p}_k > 0$ such that $\varphi_k(\vec{p}) > 0$ for all $p_k \geq \bar{p}_k$ regardless of $\vec{p}_{-k}$. 
\end{lemma}
\proof
	(i) If $\vec{p} \geq \vec{c}(\vec{P}(\vec{p}))$ and $p_k = c_k(P_k^L(\vec{p}))$, 
	\begin{align*}
		\varphi_k(\vec{p}) 
			&= - \sum_{j \in \set{J}_f \setminus \{ k \} } P_j^L(\vec{p})( p_j - c_j(P_j^L(\vec{p})))
									- \frac{1}{\abs{(Dw_k)(p_k)} } \\
			&< - \sum_{j \in \set{J}_f \setminus \{ k \} } P_j^L(\vec{p})( p_j - c_j(P_j^L(\vec{p})))
			\leq 0. 
	\end{align*}
	(ii) Write $\varphi_k(\vec{p}) > 0$ as
	\begin{align}
		\label{EQN:TMP1}
		p_k - \frac{1}{\abs{(Dw_k)(p_k)} }
			> c_k(P_k^L(\vec{p})) 
				+ \sum_{j \in \set{J}_f} P_j^L(\vec{p})( p_j - c_j(P_j^L(\vec{p})))
	\end{align}
	Because there exists $r_k > 1$ and $\bar{p}_k$ such that
	\begin{align*}
		p_k - \frac{1}{\abs{(Dw_k)(p_k)} }
			\geq \left( 1 - \frac{1}{r_k} \right) p_k
	\end{align*}
	for all $p_k > \bar{p}_k$, the left-hand-side in (\ref{EQN:TMP1}) can be made as large as desired. Similarly, because $\sup_{q\in[0,1]} c_j(q) < \infty$ and $w_k$ is necessarily eventually log-bounded, the right-hand-side
	\begin{align*}
		c_k(P_k^L(\vec{p})) 
				+ \sum_{j \in \set{J}_f} P_j^L(\vec{p})( p_j - c_j(P_j^L(\vec{p})))
			\leq c_k(P_k^L(\vec{p})) + \sum_{j \in \set{J}_f} P_j^L(\vec{p})p_j
	\end{align*}
	is bounded over all $\vec{p}$. Thus there must exist $\bar{p}_k$ so large to make (\ref{EQN:TMP1}) hold for all $p_k \geq \bar{p}_k$. 
\endproof

\begin{corollary}
	Suppose $\vartheta > -\infty$, Assumption \ref{LogitUtilityAssumption} holds with $w$ eventually decreasing sufficiently quickly, and Assumption \ref{ASS:NonConstUnitCostsAssum} holds. $\bsym{\psi} = \bsym{\varphi} \circ \bsym{\rho} : [\vec{0},\bar{\vec{p}}] \to \R^J$ (i.e. $\bsym{\psi}(\bsym{\epsilon}) = \bsym{\varphi}(\bsym{\rho}(\bsym{\epsilon}))$) points outward on the boundary of $[\vec{0},\bar{\vec{p}}]$. 
\end{corollary}
\proof
	This follows immediately from Lemma \ref{LEM:Inequalities}, recalling that $\rho_k(\bsym{\epsilon}) > \bar{p}_k$ when $\epsilon_k = \bar{p}_k$. 
\endproof

\begin{theorem}
	There exists a zero, $\vec{p}$, of $\bsym{\varphi}$ such that $\vec{c}(\vec{P}(\vec{p})) < \vec{p} < \bsym{\infty}$. 
\end{theorem}
\proof
	Apply the Poincare-Hopf theorem to $\bsym{\psi}$, as described above, to establish the existence of $\bsym{\epsilon} \in (\vec{0},\bar{\vec{p}})$ such that $\bsym{\psi}(\bsym{\epsilon}) = \vec{0}$. Define $\vec{p} = \bsym{\rho}(\bsym{\epsilon}) > \vec{c}(\vec{P}(\vec{p}))$, observing that $\bsym{\varphi}(\vec{p}) = \bsym{\psi}(\bsym{\epsilon}) = \vec{0}$. 
\endproof

Note that the Poincare-Hopf Theorem is very useful here, relative to Brouwer's Theorem. Specifically, there is no obvious fixed-point equation for $\bsym{\epsilon}$, and it is not obvious when $\{ \vec{p} : \vec{p} \geq \vec{c}(\vec{P}(\vec{p})) \}$ (and thus $\{ \vec{p} : \vec{p} \geq \vec{c}(\vec{P}(\vec{p})) \} \cap [\vec{0},\bar{\vec{p}}]$) would be convex. 

\subsection{Existence of Equilibrium} 

The second component of the proof that equilibrium exists requires demonstrating the ``sufficiency of stationarity'' and the uniqueness of profit-maximizing prices. As in Section \ref{SEC:LogitPriceEquilibrium}, this is accomplished by proving that $(D_f\nabla_f\hat{\pi}_f)(\vec{p})$ is negative definite at any stationary prices and then applying the Poincare-Hopf Theorem. 

\begin{lemma}
	Suppose $\vartheta > -\infty$, Assumption \ref{LogitUtilityAssumption} holds with $w$ twice continuously differentiable, and Assumption \ref{ASS:NonConstUnitCostsAssum} holds. At any stationary point, $(D_f\nabla_f\hat{\pi}_f)(\vec{p}) = \bsym{\Lambda}_f(\vec{p})(\vec{I} - \bsym{\Omega}_f(\vec{p})) - \bsym{\Lambda}_f(\vec{p})\vec{H}_f(\vec{p})\bsym{\Lambda}_f(\vec{p})$ where
	\begin{equation*}
		\vec{H}_f(\vec{p})
			= \vec{K}_f(\vec{p})
				- \vec{K}_f(\vec{p})\vec{P}_f^L(\vec{p})\vec{1}^\top
				- \vec{1}\vec{P}_f^L(\vec{p})^\top\vec{K}_f(\vec{p})
				+ \left( \vec{P}_f^L(\vec{p})^\top\vec{K}_f(\vec{p})\vec{P}_f^L(\vec{p}) \right) \vec{1}\vec{1}^\top
	\end{equation*}
	and $\vec{K}_f(\vec{p}) = (D^2\vec{C}_f)(\vec{P}_f^L(\vec{p}))$. 
\end{lemma}
\proof
	Because $(D_k\hat{\pi}_f)(\vec{p}) = \lambda_k(\vec{p})\varphi_k(\vec{p})$ and $\varphi_k(\vec{p}) = 0$ at any stationary point, $(D_lD_k\hat{\pi}_f)(\vec{p}) = \lambda_k(\vec{p})(D_l\varphi_k)(\vec{p})$ for any $k,l \in \set{J}_f$. Moreover, 
	\begin{align*}
		(D_l\varphi_k)(\vec{p})
			&= (1-\omega_k(p_k))\delta_{k,l}
						- (D^2C_k)(P_k^L(\vec{p})) (D_lP_k^L)(\vec{p}) \\
			&\quad\quad\quad\quad\quad\quad\quad
						- \sum_{j\in\set{J}_f} (D_lP_j^L)(\vec{p}))
								( p_j - c_j(P_j^L(\vec{p}) ) 
						- P_l^L(\vec{p}) \\
			&\quad\quad\quad\quad\quad\quad\quad
						+ \sum_{j\in\set{J}_f} P_j^L(\vec{p})(D^2C_j)(P_j^L(\vec{p}))(D_lP_j^L)(\vec{p}) \\
			%%%%%%%%%%%%%%%%%%%%%%%%%%%%%%%%%%%%%%%%
			&= (1-\omega_k(p_k))\delta_{k,l}
						- (D^2C_k)(P_k^L(\vec{p})) (D_lP_k^L)(\vec{p}) \\
			&\quad\quad\quad\quad\quad\quad\quad
						+ \sum_{j\in\set{J}_f} P_j^L(\vec{p})(D^2C_j)(P_j^L(\vec{p}))(D_lP_j^L)(\vec{p})
	\end{align*}
	because $(D_l\hat{\pi}_f)(\vec{p}) = \sum_{j\in\set{J}_f} (D_lP_j^L)(\vec{p})) ( p_j - c_j(P_j^L(\vec{p}) ) - P_l^L(\vec{p}) = 0$. The result follows by substituting the definition of $(D_lP_k^L)(\vec{p})$ into this last equation and re-arranging terms. 
\endproof

A sufficient condition for $\hat{\pi}_f(\vec{p})$ to be locally concave at any stationary point follows:
\begin{lemma}
	Suppose $\vartheta > -\infty$, Assumptions \ref{LogitUtilityAssumption} and \ref{ASS:NonConstUnitCostsAssum} hold, and $w$ has sub-quadratic second derivatives. (i) $(D_f\nabla_f\hat{\pi}_f)(\cdot,\vec{p}_{-f})$ is negative definite at any stationary prices $\vec{p}_f$. (ii) As a consequence, profit-maximizing prices $\vec{p}_f$ are unique for any competitor's prices $\vec{p}_{-f}$ and (iii) any simultaneously stationary point is an equilibrium. 
\end{lemma}
\proof
	First note that if $w$ has sub-quadratic second derivatives and $\vec{H}_f(\vec{p})$ is positive semi-definite (at any stationary prices $\vec{p}_f$) then $(D_f\nabla_f\hat{\pi}_f)(\vec{p})$ is negative definite (at any stationary prices $\vec{p}_f$). For if $w$ has sub-quadratic second derivatives, then $\bsym{\Lambda}_f(\vec{p})(\vec{I} - \bsym{\Omega}_f(\vec{p}))$ is negative definite, and if $\vec{H}_f(\vec{p})$ is positive semi-definite, then $- \bsym{\Lambda}_f(\vec{p})\vec{H}_f(\vec{p})\bsym{\Lambda}_f(\vec{p})$ is negative semi-definite.

	We now show that when $C_j$ is convex, $\vec{H}_f(\vec{p})$ is positive semi-definite. Because $\vec{K}_f(\vec{p})$ is positive definite, define an inner product $\langle \vec{x} , \vec{y} \rangle_f = \vec{x}^\top\vec{K}_f(\vec{p})\vec{y}$ on $\R^{J_f}$ with $\norm{\vec{x}}_f = \sqrt{ \langle \vec{x} , \vec{x} \rangle_f }$ the corresponding norm. The Cauchy-Schwartz inequality states that 
	\begin{equation*}
		\big( \vec{P}_f(\vec{p})^\top\vec{K}_f(\vec{p})\vec{x} \big)^2
			= \abs{ \langle \vec{P}_f(\vec{p}) , \vec{x} \rangle_f }^2
			\leq \norm{\vec{P}_f(\vec{p})}_f^2\norm{\vec{x}}_f^2
			= \big( \vec{P}_f(\vec{p})^\top\vec{K}_f(\vec{p})\vec{P}_f(\vec{p}) \big)
					\big( \vec{x}^\top\vec{K}_f(\vec{p})\vec{x} \big)
	\end{equation*}
	for any vector $\vec{x} \in \R^{J_f}$. Note that
	\begin{equation*}
		\vec{x}^\top\vec{H}_f(\vec{p})\vec{x}
			= \big( \vec{P}_f(\vec{p})^\top\vec{K}_f(\vec{p})\vec{P}_f(\vec{p}) \big)
					\big( \vec{1}^\top\vec{x} \big)^2
				- 2 \big( \vec{P}_f(\vec{p})^\top\vec{K}_f(\vec{p})\vec{x} \big) 
					\big( \vec{1}^\top\vec{x} \big)
				+ \vec{x}^\top\vec{K}_f(\vec{p})\vec{x}
	\end{equation*}
	Because any convex quadratic $q(\xi) = a \xi^2 - 2 b \xi + c$ (i.e., where $a > 0$) is minimized at $\xi_* = b/a$ with value $q(\xi_*) = c - b^2 / a$, $\vec{x}^\top\vec{H}_f(\vec{p})\vec{x} \geq 0$ for all $\vec{x}$ if 
	\begin{equation*}
		\big( \vec{P}_f(\vec{p})^\top\vec{K}_f(\vec{p})\vec{x} \big)^2
			\leq \big( \vec{P}_f(\vec{p})^\top\vec{K}_f(\vec{p})\vec{P}_f(\vec{p}) \big)
				\big( \vec{x}^\top\vec{K}_f(\vec{p})\vec{x} \big)
	\end{equation*}
	for all $\vec{x}$, which follows from the Cauchy-Schwartz inequality. Strictly speaking, this inequality is only required for $\vec{x}$ satisfying $\vec{1}^\top\vec{x} = \vec{P}_f(\vec{p})^\top\vec{K}_f(\vec{p})\vec{x} / \vec{x}^\top\vec{K}_f(\vec{p})\vec{x}$ to prove the positive semi-definiteness of $\vec{H}_f(\vec{p})$; however the Cauchy-Schwartz inequality requires this to hold for all $\vec{x}$. 
	
	The discussion above proves that any vector of stationary prices $\vec{p}_f$ is in fact a {\em local} maximizer of firm $f$'s profits; it remains to prove that this is fact a unique, global maximizer of firm $f$'s profits. The homotopic construction in Lemma \ref{LEM:Homotopy} applies equally well to a single firm, given fixed competitor prices $\vec{p}_{-f}$. By the analogue of Lemma \ref{LEM:Inequalities}, the vector field $\bsym{\psi}_f = \bsym{\varphi}_f \circ \bsym{\rho}_f : [\vec{0},\bar{\vec{p}}_f] \to \R^{J_f}$ points outward on the boundary of $[\vec{0},\bar{\vec{p}}_f]$ and has at least one zero $\bsym{\epsilon}_f$. Assume that the index of any such zero is one. Then there can be only one such zero, because the Poincare-Hopf Theorem requires the sum of the indices of all zeros to be one. 
	
	The proof that the index of any zero $\bsym{\epsilon}_f$ of $\bsym{\psi}_f$ is one begins with the standard index formula:
	\begin{equation*}
		\mathrm{index}_{\bsym{\epsilon}_f}(\bsym{\psi}_f)
			= \mathrm{sign} \det (D_f\bsym{\psi}_f)(\bsym{\epsilon}_f)
			= \mathrm{sign} \det (D_f\bsym{\varphi}_f)(\bsym{\rho}_f(\bsym{\epsilon}_f))
				\cdot  \mathrm{sign} \det (D_f\bsym{\rho}_f)(\bsym{\epsilon}_f). 
	\end{equation*}
	Both determinants on the right-hand-side are positive, as we now prove. 
	
	First, consider $\mathrm{sign} \det (D_f\bsym{\varphi}_f)(\bsym{\rho}_f(\bsym{\epsilon}_f))$. The zero $\vec{p}_f = \bsym{\rho}_f(\bsym{\epsilon}_f)$ of $\bsym{\varphi}_f(\cdot,\vec{p}_{-f})$ is a local maximizer of $\hat{\pi}_f(\cdot,\vec{p}_{-f})$, and thus the index of the gradient vector field $(\nabla_f\hat{\pi}_f)(\cdot,\vec{p}_{-f})$ at $\vec{p}_f$ is $(-1)^{J_f}$. But then
	\begin{equation*}
		(-1)^{J_f} = \mathrm{sign}\det(D_f\nabla_f\hat{\pi}_f)
				= \mathrm{sign}\det \Big( \bsym{\Lambda}_f (D_f\bsym{\varphi}_f) \Big)
				= (-1)^{J_f} \mathrm{sign}\det(D_f\bsym{\varphi}_f), 
	\end{equation*}
	and $\mathrm{sign}\det(D_f\bsym{\varphi}_f)(\bsym{\rho}_f(\bsym{\epsilon}_f),\vec{p}_{-f}) = 1$ as claimed. 
	
	Next, consider $\mathrm{sign} \det (D_f\bsym{\rho}_f)(\bsym{\epsilon}_f)$. The Jacobian of $(D_f\bsym{\rho}_f)(\bsym{\epsilon}_f)$ is given by
	\begin{equation*}
		(D\bsym{\rho}_f)(\bsym{\epsilon}_f)
			= ( \vec{I} - (D^2\vec{C}_f)(D_f\vec{P}_f^L) )\inv
	\end{equation*}
	(where we neglect the arguments for simplicity); this inverse is well defined because $(D^2\vec{C}_f)(D_f\vec{P}_f^L)$ does not have one as an eigenvalue, as proved in Lemma \ref{LEM:Homotopy}. Thus zero is not an eigenvalue of $(D_f\bsym{\rho}_f)(\bsym{\epsilon}_f)$, and any eigenvalue $\mu$ ($\neq 0$) satisfies
	\begin{equation*}
		\Big( \vec{I} - (D^2\vec{C}_f)(D_f\vec{P}_f^L) \Big)\vec{x}
			= \left( \frac{1}{\mu} \right) \vec{x}. 
	\end{equation*}
	Rearranging this equation yields
	\begin{equation*}
%		\Big( \vec{I} - (D^2\vec{C}_f)(\vec{I}-\vec{P}_f^L\vec{1}^\top)\bsym{\Lambda}_f \Big)\vec{x}
%			= \left( \frac{1}{\mu} \right) \vec{x}
		\left[ \left( \frac{1}{\mu} - 1 \right) \vec{I} + (D^2\vec{C}_f)\bsym{\Lambda}_f \right] \vec{x}
			= \big( \vec{1}^\top\bsym{\Lambda}_f\vec{x} \big) (D^2\vec{C}_f)\vec{P}_f^L. 
	\end{equation*}
	It is straightforward to see that this can hold only if
	\begin{equation*}
		1 = \theta\left(1 - \frac{1}{\mu}\right)
		\quad\text{where}\quad
		\theta(\alpha) = \sum_{j \in \set{J}_f}
				\left( \frac{\abs{\lambda_j}(D^2C_j)}{\alpha + \abs{\lambda_j}(D^2C_j)} \right) P_j^L. 
	\end{equation*}
	Without loss of generality, suppose $\set{J}_f = \{ 1,\dotsc,J_f\}$,
	\begin{equation*}
		\abs{\lambda_1}(D^2C_1) \leq \dotsb \leq \abs{\lambda_{J_f}}(D^2C_{J_f}), 
	\end{equation*}
	let $N$ be the number of distinct values of $\abs{\lambda_j}(D^2C_j)$, $\kappa_1 < \dotsb < \kappa_N$ these values, and $M_n$ be the number of $j \in \set{J}_f$ for which $\abs{\lambda_j}(D^2C_j) = \kappa_n$. 
	
	Note that $\theta$ is not defined for $\alpha \in \set{S}_f = \{ - \abs{\lambda_j}(D^2C_j) : j \in \set{J}_f \} = \{ - \kappa_n \}_{n=1}^N$, and 
	\begin{equation*}
		D\theta(\alpha) 
			= - \sum_{j \in \set{J}_f}
				\left( \frac{\abs{\lambda_j}(D^2C_j)}{(\alpha + \abs{\lambda_j}(D^2C_j))^2} \right) P_j^L
			< 0, 
	\end{equation*}
	when defined. Moreover, $\theta(\alpha) < 0 < 1$ for all $\alpha \in (-\infty,\kappa_N)$; thus no solutions to $\theta(\alpha) = 1$ are less than $-\kappa_N$. Consider any interval $(-\kappa_{n+1},-\kappa_{n})$, $n \in \{1,\dotsc,N-1\}$. In such an interval $\theta$ is strictly decreasing, $\theta(\alpha) \uparrow \infty$ as $\alpha \downarrow - \kappa_{n+1}$, and $\theta(\alpha) \downarrow -\infty$ as $\alpha \uparrow - \kappa_n$. There is thus a unique $\alpha_n \in (-\kappa_{n+1},-\kappa_{n})$ such that $\theta(\alpha_n) = 1$. Finally, $\theta(\alpha) > 0$ for all $\alpha > - \kappa_1$, $D\theta(\alpha) < 0$ for all $\alpha > - \kappa_1$, and $\theta(0) = \sum_{j\in\set{J}_f} P_j^L < 1$ imply that there is a unique $\alpha_N \in (-\kappa_1,0)$ such that $\theta(\alpha_N) = 1$. 
	
	The $N$ solutions to this equation map to the $N$ distinct eigenvalues of $(D_f\bsym{\rho}_f)(\bsym{\epsilon}_f)$ (with multiplicities $M_n$) via $\alpha_n = 1 - 1/\mu_n$; that is, $\mu = 1/(1-\alpha_n)$. Because each $\alpha_n < 0$, each distinct eigenvalue of $(D_f\bsym{\rho}_f)(\bsym{\epsilon}_f)$ is positive. Thus $\mathrm{sign}\det(D_f\bsym{\rho}_f)(\bsym{\epsilon}_f) = 1$ as claimed. 
\endproof

Consider this proof from the perspective of concave costs $C_j$. 
\begin{equation*}
	\vec{x}^\top\vec{H}_f(\vec{p})\vec{x}
		= - \big( \vec{P}_f(\vec{p})^\top\abs{\vec{K}_f(\vec{p})}\vec{P}_f(\vec{p}) \big)
				\big( \vec{1}^\top\vec{x} \big)^2
			+ 2 \big( \vec{P}_f(\vec{p})^\top\abs{\vec{K}_f(\vec{p})}\vec{x} \big) 
				\big( \vec{1}^\top\vec{x} \big)
			- \vec{x}^\top\abs{ \vec{K}_f(\vec{p}) }\vec{x}
\end{equation*}
is now a {\em concave} quadratic in $\vec{1}^\top\vec{x}$, $q(\xi) = -a \xi^2 + 2 b \xi - c$ (where $a,c > 0$), {\em maximized} at $\xi_* = b/a$ with value $q(\xi_*) = b^2 / a - c$. However, the Cauchy-Schwartz inequality then requires $q(\xi) \leq q(\xi_*) \leq 0$, and $\vec{H}_f(\vec{p})$ is {\em negative} semi-definite. $(D_f\nabla_f\hat{\pi}_f)(\vec{p})$ is negative definite then only if $\bsym{\Lambda}_f(\vec{p})(\vec{I} - \bsym{\Omega}_f(\vec{p}))$ is ``more'' negative-definite than $\bsym{\Lambda}_f(\vec{p})\vec{H}_f(\vec{p})\bsym{\Lambda}_f(\vec{p})$ is, in the sense that
\begin{equation*}
	\vec{x}^\top\bsym{\Lambda}_f(\vec{p})(\vec{I} - \bsym{\Omega}_f(\vec{p}))\vec{x}
		< \vec{x}^\top\bsym{\Lambda}_f(\vec{p})\vec{H}_f(\vec{p})\bsym{\Lambda}_f(\vec{p})\vec{x}.
\end{equation*}

\section{Finite Purchasing Power}
\label{SEC:FiniteIncomes}

This section considers models in which there exists some limit on the population's purchasing power. That is, there exists $\varsigma \in [0,\infty)$ such that if $p_j \geq \varsigma$, no individual can purchase product $j$. However, important empirical examples of {\em Mixed} Logit models have finite purchasing power; see, e.g., \cite{Berry95, Petrin02}. Thus it is important to consider this case in the analysis of existence and uniqueness of equilibrium prices. Section \ref{SEC:LogitPriceEquilibrium} above proves that if $\varsigma = \infty$ (no purchasing power limit) and there is an outside good, it is not possible for any product's price to be $\infty$ in equilibrium. However, when $\varsigma < \infty$, it is possible for some$-$but not all$-$prices to be equal to $\varsigma$; in other words, firms may ``price some products out of the market''. Theoretical and computational treatments must be specially adapted to this case to account for this qualitatively different behavior. 

The analysis in this section proves the following theorem: 
\begin{theorem}
	Suppose $\vartheta > -\infty$, costs satisfy Assumption \ref{ASS:NonConstUnitCostsAssum} with $(D^2C)(\vec{y},P)$ finite as $P \downarrow 0$ for any $\vec{y} \in \set{Y}$, and Assumption \ref{LogitUtilityAssumption2} holds with $\varsigma < \infty$ and a $w$ that eventually decreases sufficiently quickly,  has finite sub-quadratic second derivatives, and $\lim_{p \uparrow \varsigma} (D^2w)(\vec{y},p)$ exists for any $\vec{y} \in \set{Y}$. Then there exists at least one equilibrium $\vec{p} \in [0,\varsigma]^J$, and any equilibrium satisfies $\vec{p} > \vec{c}(\vec{P}(\vec{p}))$. 
\end{theorem}

\subsection{Assumptions}

	Assumption \ref{LogitUtilityAssumption} must be revised to account for finite $\varsigma$:
	\begin{assumption}
		\label{LogitUtilityAssumption2}
		There exists $\varsigma \in (c_*,\infty]$, where $c_* = \max_j \{ \sup_{P \in [0,1]} c_j(P) \} < \infty$, and functions $w : \set{Y} \times [0,\varsigma) \to (-\infty,\infty)$ and $v : \set{Y} \to (-\infty,\infty)$ such that the utility $u : \set{Y} \times [0,\infty) \to \R$ can be written $u(\vec{y},p) = w(\vec{y},p) + v(\vec{y})$ for all $p < \varsigma$ and $u(\vec{y},p) = -\infty$ for all $p \geq \varsigma$. Moreover, assume that, for all $\vec{y} \in \set{Y}$, $w(\vec{y},\cdot) : [0,\varsigma) \to (-\infty,\infty)$ is (a) strictly decreasing, and (b) continuously differentiable, and (c) $\lim_{p \uparrow \varsigma} w(\vec{y},p) = -\infty$.  
	\end{assumption}
	
	The basic properties so useful in the case of infinite purchasing power must also be generalized:
	\begin{definition}
		\label{DEF:EDSQ2}
		If $\varsigma < \infty$, $w(\vec{y},\cdot)$ {\bfseries eventually decreases sufficiently quickly} if there exists $\delta(\vec{y}) > 0$ and $\alpha(\vec{y}) > 1$ such that $(Dw)(\vec{y},p) \leq - \alpha(\vec{y}) / (\varsigma - p)$ for all $p \in [\varsigma-\delta(\vec{y}),\varsigma)$. 
	\end{definition}
	
	\begin{definition}
		\label{DEF:SQSD2}
		$w(\vec{y},\cdot)$ has {\bfseries sub-quadratic second derivatives} if $\omega(\vec{y},p) < 1$ for all $p \in (0,\varsigma)$. 
	\end{definition}
	
	Gallego et. al \cite{Gallego06} consider equilibrium pricing under the attraction demand model, equivalent to the Logit model with nonlinear utilities, and allow $\varsigma < \infty$. Specifically, Gallego et. al formulate their model in terms of the ``attraction function'' $a_j(p) = e^{u_j(p)}$ and make what amount to the following assumptions on $u_j$: $u_j : [0,\varsigma) \to \R$ is continuously differentiable, strictly decreasing, and $p(D^2u_j)(p) \geq (Du_j)(p)$. This last assumption is violated by the ``BLP''-type utility $u_j(p) = \alpha \log( \varsigma - p ) + v_j$ \cite{Berry95, Morrow08, Morrow10a, Morrow10b}, as can be easily checked. Assumption \ref{LogitUtilityAssumption2} and Defs. \ref{DEF:EDSQ2} and \ref{DEF:SQSD2} above are weak enough to allow an analysis of this important case. 
	
	The following basic observations, stated without proof, are needed:
	\begin{lemma}
		Under Assumption \ref{LogitUtilityAssumption2} with $\vartheta > -\infty$ and $\varsigma < \infty$, 
		\begin{itemize}
			\item $P_j^L$ can be continuously extended to $[0,\varsigma]^J$, with $P_j^L(\vec{p}) \downarrow 0$ as $p_j \uparrow \varsigma$. Specifically, 
			\begin{equation*}
				P_j^L(\vec{p}) = \frac{e^{u_j(p_j)}}{e^{\vartheta} + \sum_{ k : p_k < \varsigma } e^{u_k(p_k)}}
			\end{equation*}
			\item $\hat{\pi}_f(\vec{p})$ can be continuously extended to $[0,\varsigma]^J$, with $\hat{\pi}_f(\vec{p}) \downarrow 0$ as $\vec{p}_f \uparrow \varsigma\vec{1}$ with
			\begin{equation*}
				\hat{\pi}_f(\vec{p}) 
					= \sum_{ j \in \set{J}_f, p_j < \varsigma }
						P_j^L(\vec{p})(p_j-c_j(P_j^L(\vec{p})))
%			\end{equation*}
%			Moreover, 
%			\begin{equation*}
				\quad\text{and}\quad
				0 < \sup_{\vec{p}_f \in [0,\varsigma]^{J_f}} \hat{\pi}_f(\vec{p}) < \infty
			\end{equation*}
			\item If $w$ eventually decreases sufficiently quickly, $\lambda_j(\vec{p}) = (Dw_j)(p_j)P_j^L(\vec{p})$ can be continuously extended to $[0,\varsigma]^J$ with $\lambda_j(\vec{p}) \uparrow 0$ as $p_j \uparrow \varsigma$. 
		\end{itemize}
	\end{lemma}
	If $\vartheta = -\infty$, $P_j^L$ {\em cannot} be continuously extended to $[0,\varsigma]^J$, for reasons analogous to those discussed in Section \ref{SEC:LogitModels}. 
	
The following consequences of Assumption \ref{LogitUtilityAssumption2} and Defs. \ref{DEF:EDSQ2} and \ref{DEF:SQSD2} are also used below. 
\begin{lemma}
	Suppose Assumption \ref{LogitUtilityAssumption2} holds with $\varsigma < \infty$. (i) If $\lim_{p\uparrow\varsigma} (Dw)(\vec{y},p)$ exists, then $(Dw)(\vec{y},p) = - \infty$ as $p \uparrow \varsigma$, and thus $(Dw)(\vec{y},p)\inv \uparrow 0$ as $p \uparrow \varsigma$. (ii) If $w$ eventually decreases sufficiently quickly, then $(Dw)(\vec{y},p) = - \infty$ as $p \uparrow \varsigma$. (iii) If $w(\vec{y},\cdot)$ is twice continuously differentiable and $\lim_{p\uparrow\varsigma} (D^2w)(\vec{y},p)$ exists, then it is ``eventually concave'' in the sense that there exists some $\epsilon \in (0,\varsigma)$ such that $(D^2w)(\vec{y},\cdot) < 0$ on $(\varsigma-\epsilon,\varsigma)$. (iv) If $w$ is twice continuously differentiable and $\lim_{p\uparrow\varsigma} (D^2w)(\vec{y},p)$ exists, then $\limsup_{p\uparrow\varsigma} \omega(\vec{y},p) \leq 0$. (v) If $w$ is twice continuously differentiable and eventually decreases sufficiently quickly, then $\liminf_{p\uparrow\varsigma} \omega(\vec{y},p) > - 1$. 
\end{lemma}
\proof
	(ii) is trivial. (i) and (iii) are both consequences of the following technical result: If $f : (0,1) \to (-\infty,0)$ is continuously differentiable, $\lim_{x \uparrow 1} f(x) = - \infty$, and $\lim_{x \uparrow 1} (Df)(x)$ exists, then $\lim_{x \uparrow 1} (Df)(x) = - \infty$. For proof, note that the fundamental theorem of calculus requires that
	\begin{equation*}
		\lim_{\delta \downarrow 0} \int_x^{1-\delta} (Df)(y)dy = -\infty. 
	\end{equation*}
%	Without knowing that $\lim_{x \uparrow 1} (Df)(x)$ exists, all that can be drawn from this result is that $\limsup_{x \uparrow 1} (Df)(x) = - \infty$ and 
%	\begin{equation*}
%		\int_{R(x,M)} (Df)(y)dy = -\infty
%		\quad\text{where}\quad
%		R(x,M) = \{ y \in [x,1) : (Df)(y) > - M \}
%	\end{equation*}
%	for any $M > 0$. For we can write 
%	\begin{equation*}
%		\int_x^{1-\delta} (Df)(y)dy 
%			= \int_{S(x,\delta,M)} (Df)(y)dy + \int_{R(x,\delta,M)} (Df)(y)dy
%	\end{equation*}
%	where $S(x,\delta,M) = \{ y \in [x,1-\delta] : (Df)(y) \geq - M \}$ and $R(x,\delta,M) = \{ y \in [x,1-\delta] : (Df)(y) < - M \}$. Because $\int_{S(x,\delta,M)} (Df)(y)dy$ is a decreasing function of $\delta$ and is bounded below by $- M \int_{S(x,M)} dy$, where $S(x,M) = \{ y \in [x,1) : (Df)(y) \geq - M \}$, $\lim_{\delta\downarrow 0} \int_{S(x,\delta,M)} (Df)(y)dy$ exists and is finite. Thus $\int_{R(x,M)} (Df)(y)dy = -\infty$. 
	If $\lim_{x \uparrow 1} (Df)(x)$ exists and were finite, then $\limsup_{x \uparrow 1} (Df)(x) > -\infty$ and this integral would be finite. Thus if $\lim_{x \uparrow 1} (Df)(x)$ exists then $\lim_{x \uparrow 1} (Df)(x) = -\infty$. The assumption that the limit exists is required because $(Df)$ could be highly oscillatory (in a manner similar to $\sin(x\inv)$) and still generate $\lim_{x \uparrow 1} f(x) = - \infty$. 
	
%	Alternatively, the mean value theorem states that for any $x,y \in (0,1)$, $x<y$, there exists some $z \in (x,y)$ such that $f(y) = f(x) + (Df)(z)(y-x)$. Because $f(y) \downarrow -\infty$ as $y \uparrow 1$, for any sequence $\{ y_n \}_{n=1}^\infty$, $y_n \uparrow 1$, and any $M > 0$, there exists some index $N(M) \in \N$ such that $f(y_n) \leq - M$ for {\em all} $n \geq N(M)$. There is a corresponding sequence $\{ z_n \}_{n = 1}^\infty$, $z_n \in [x,y_n] \subset [x,1)$, such that 
%	\begin{equation*}
%		(Df)(z_n) = \frac{ f(y_n) - f(x) }{ y_n - x } \leq - \frac{ M + f(x) }{ y_n - x } \leq - \frac{ M + f(x) }{ 1 - x }
%	\end{equation*}
%	for all $n \geq N(M)$. Because $(Df)(z)$ is finite for all $z \in (0,1)$, $\limsup_n z_n = 1$. 
	
	(iv): By (iii), $(D^2w)(\vec{y},p)$ is negative for all $p$ sufficiently close to $\varsigma$. Thus $\omega(\vec{y},p) < 0$ for all $p$ sufficiently close to $\varsigma$. 
	
	(v): Note that $- \omega(\vec{y},p) = D[(Dw)(\vec{y},p)\inv]$. The mean value theorem then states that for all $\delta \in (0,\varsigma)$, there exists some $\epsilon \in (0,\delta]$ such that
	\begin{equation*}
		\frac{1}{(Dw)(\vec{y},\varsigma-\delta)} 
			= - \left( \frac{1}{(Dw)(\vec{y},\varsigma)} - \frac{1}{(Dw)(\vec{y},\varsigma-\delta)} \right)
			= - \Big( - \omega(\vec{y},\varsigma-\epsilon)\delta \Big)
			= \omega(\vec{y},\varsigma-\epsilon)\delta. 
	\end{equation*}
	Because $w$ eventually decreases sufficiently quickly, there exists $\gamma > 0$ and $\alpha > 1$ such that
	\begin{equation*}
		\omega(\vec{y},\varsigma-\epsilon)\delta
			= \frac{1}{(Dw)(\vec{y},\varsigma-\delta)} \geq - \left( \frac{\delta}{\alpha} \right). 
	\end{equation*}
	Thus $\omega(\vec{y},\varsigma-\epsilon) \geq - \alpha\inv > -1$, and thus $\liminf_{p\uparrow\varsigma} \omega(\vec{y},p) > -1$. Note that we have not assumed $\lim_{p\uparrow\varsigma} \omega(\vec{y},p)$ exists in proving that $\liminf_{p\uparrow\varsigma} \omega(\vec{y},p) > -1$. 
\endproof

\subsection{A Variational Approach}

The natural approach to characterizing profit-maximizing prices when $\varsigma < \infty$ would be to assume firms solve the bound-constrained optimization problem
\begin{equation*}
	\begin{aligned}
		\mxm \quad \hat{\pi}_f(\vec{p})
		\quad \wrt \quad \vec{p}_f \in [0,\varsigma]^{J_f}. 
	\end{aligned}
\end{equation*}
The KKT conditions are the Variational Inequality (VI)
\begin{equation}
	\label{EQN:ProfVI}
	(\nabla_f\hat{\pi}_f)(\vec{p})^\top(\vec{p}_f - \vec{q}_f) \geq \vec{0}
		\quad\text{for all}\quad 
		\vec{q}_f \in [0,\varsigma]^{J_f}
\end{equation}
This approach is, unfortunately, not useful because of the following result:
\begin{lemma}
	Suppose $\vartheta > -\infty$, costs satisfy Assumption \ref{ASS:NonConstUnitCostsAssum}, Assumption \ref{LogitUtilityAssumption2} holds with $\varsigma < \infty$, and $w$ eventually decreases sufficiently quickly. Then $\hat{\pi}_f(\vec{p})$ can be continuously differentiably extended to $[0,\infty)^J$, where $(D_k\hat{\pi}_f)(\vec{p}) = 0$ whenever $p_k \geq \varsigma$, $k \in \set{J}_f$. 
\end{lemma}
\proof
	Suppose $\vec{q} \in (\vec{0},\varsigma\vec{1})$. Then
	\begin{equation*}
		(D_k\hat{\pi}_f)(\vec{q})
			=\lambda_k(\vec{q}) 
				\left( 
					q_k - c_k(P_k(\vec{q})) 
						- \sum_{j \in \set{J}_f} P_j^L(\vec{q})(q_j-c_j(P_j(\vec{q})) ) - \frac{1}{\abs{(Dw_k)(q_k)}} 
				\right).
	\end{equation*}
	Each quantity can be extended, continuously, to $[0,\varsigma]^J$, and thus so can $(D_k\hat{\pi}_f)$. Note in particular that the term in parentheses tends to 
	\begin{equation*}
		\varsigma - c_k(0) 
			- \sum_{j \in \set{J}_f, p_j < \varsigma } P_j^L(\vec{p})(p_j-c_j(P_j^L(\vec{p})) )
	\end{equation*}
	which is finite, and $\lambda_k(\vec{p}) \uparrow 0$ as $p_k \uparrow \varsigma$. Thus $(D_k\hat{\pi}_f)(\vec{p}) = 0$ when $p_k = \varsigma$. 
\endproof

As a result, the standard VI (\ref{EQN:ProfVI}) does not provide any information about profit-optimal prices or equilibria that have some prices equal to $\varsigma$. For an extreme illustration of this fact, note that $\vec{p}_f = \varsigma\vec{1}$ (trivially) solves (\ref{EQN:ProfVI}). However, $\varsigma\vec{1}$ cannot possible be profit-maximizing, because $\hat{\pi}_f(\varsigma,\vec{1},\vec{p}_{-f}) = 0$ while $\hat{\pi}_f(\vec{p}_f,\vec{p}_{-f}) > 0$ for any $\vec{p}_f$ satisfying $\vec{c}_f(\vec{P}_f(\vec{p})) < \vec{p}_f < \varsigma\vec{1}$. 

The $\bsym{\zeta}$ fixed-point form introduced above provides a convenient solution to this problem. 
\begin{theorem}
	\label{THM:ZetaVI}
	Suppose $\vartheta > -\infty$, Assumptions \ref{ASS:NonConstUnitCostsAssum} and \ref{LogitUtilityAssumption2} hold, $w$ eventually decreases sufficiently quickly and has sub-quadratic second derivatives. For any $\vec{p} \in [0,\varsigma]^J$, let $\set{J}_f^\circ = \{ j \in \set{J}_f : p_j < \varsigma \}$, $\set{J}_f^* = \{ j \in \set{J}_f : p_j = \varsigma \}$ and if $j \in \set{J}_f^*$, define 
	\begin{equation*}
		\zeta_j(\vec{p})
			= \lim_{p_j\uparrow\varsigma} \zeta_j(\vec{p})
			= \sum_{j \in \set{J}_f^\circ} P_j^L(\vec{p})(p_j-c_j(P_j^L(\vec{p})) ). 
	\end{equation*}
	
	(i) If $\vec{p}_f \in [0,\varsigma]^{J_f}$ locally maximizes $\hat{\pi}_f(\cdot,\vec{p}_{-f})$ then $p_j = c_j(P_j(\vec{p})) + \zeta_j(\vec{p})$ for all $j \in \set{J}_f^\circ$ and $\varsigma - c_j(0) - \zeta_j(\vec{p}) \leq 0$ for all $j \in \set{J}_f^*$. (ii) If, in addition, $\lim_{p \uparrow \varsigma} (D^2w)(\vec{y},\cdot)$ exists and $(D^2C)(\vec{y},P)$ is finite as $P \downarrow 0$ for all $\vec{y} \in \set{Y}$, then $p_j = c_j(P_j(\vec{p})) + \zeta_j(\vec{p})$ for all $j \in \set{J}_f^\circ$ and $\varsigma - c_j(0) - \zeta_j(\vec{p}) \leq 0$ for all $j \in \set{J}_f^*$ is  sufficient $\vec{p}_f \in [0,\varsigma]^{J_f}$ to maximize $\hat{\pi}_f(\cdot,\vec{p}_{-f})$. 
	
\end{theorem}
\proof
	Note that $\hat{\pi}_f(\cdot,\vec{p}_{-f})$ is continuously differentiable on $(0,\infty)^{J_f}$, and thus we can apply the vector mean value theorem to $\hat{\pi}_f(\cdot,\vec{p}_{-f})$ on $(0,\varsigma]^{J_f}$. 
	
	We also note that there exist neighborhoods $\set{U}_f^*$ and $\set{U}_f^\circ$, of $\vec{p}_f^*$ and $\vec{p}_f^\circ$ respectively, and a map $\vec{p}_f^\circ : ( \set{U}_f^* \cap [0,\varsigma]^{J_f^*} ) \to \set{U}_f^\circ$ such that $(\vec{p}_f^\circ(\vec{q}_f^*),\vec{q}_f^*) \in \set{U}_f^\circ \times ( \set{U}_f^* \cap [0,\varsigma]^{J_f^*} )$ and $\vec{p}_f^\circ(\vec{q}_f^*)$ is the {\em unique} solution to 
	\begin{equation*}
		(\nabla_f^\circ\hat{\pi}_f)\big(\vec{q}_f^\circ,\vec{q}_f^*,\vec{p}_{-f})
			= \vec{0}
	\end{equation*}
	as a problem in $\vec{q}_f^\circ$ only. This actually follows from the Implicit Function Theorem, applied to the continuously differentiable extension of $\bsym{\varphi}$ to all of $[0,\varsigma]^J$ given below in Lemma \ref{LEM:ZetaExtension}. Because of the sufficiency of stationarity, when $w$ has sub-quadratic second derivatives, $\vec{p}_f^\circ(\vec{q}_f^*)$ is, in fact, the unique local maximizer of $\hat{\pi}_f(\cdot,\vec{q}_f^*,\vec{p}_{-f})$ on $\set{U}_f^\circ$. Thus
	\begin{equation*}
		\hat{\pi}_f\big(\vec{q}_f^\circ,\vec{q}_f^*,\vec{p}_{-f})
			< \hat{\pi}_f\big(\vec{p}_f^\circ(\vec{q}_f^*),\vec{q}_f^*,\vec{p}_{-f})
	\end{equation*}
	for all $\vec{q}_f^* \in \set{U}_f^* \cap [0,\varsigma]^{J_f^*}$ and $\vec{q}_f^\circ \in \set{U}_f^\circ$. Furthermore, $(D_f^*\vec{p}_f^\circ)(\vec{q}_f^*) \to \vec{0}$ as $\vec{q}_f^* \uparrow \varsigma\vec{1}$. For $(D_f^*\vec{p}_f^\circ)(\vec{q}_f^*)$ solves
	\begin{equation*}
		(D_f^\circ\bsym{\varphi}_f^\circ)(\vec{q}_f^*,\vec{p}_f^\circ(\vec{q}_f^*),\vec{p}_{-f})
				(D_f^*\vec{p}_f^\circ)(\vec{q}_f^*)
			= (D_f^*\bsym{\varphi}_f^\circ)(\vec{q}_f^*,\vec{p}_f^\circ(\vec{q}_f^*),\vec{p}_{-f})
	\end{equation*}
	while $(D_f^*\bsym{\varphi}_f^\circ)(\vec{p}) = \lim_{\vec{q}_f^* \uparrow \varsigma\vec{1}} (D_f^*\bsym{\varphi}_f^\circ)(\vec{q}_f^*,\vec{p}_f^\circ(\vec{q}_f^*),\vec{p}_{-f}) \to \vec{0}$, and $(D_f^\circ\bsym{\varphi}_f^\circ)(\vec{p})$ is nonsingular. 

	(i): The necessity of $\varphi_j(\vec{p}) = p_j - c_j(P_j(\vec{p})) - \zeta_j(\vec{p}) = 0$ for $j \in \set{J}_f^\circ$ is obvious. Suppose then that there is {\em some} $j \in \set{J}_f^*$ such that $\varsigma - c_j(0) - \zeta_j(\vec{p}) > 0$. We can choose the neighborhood $\set{U}_f^*$ above so that $\varphi_j(\vec{q}_f,\vec{p}_{-f}) > 0$ for all $\vec{q}_f = (\vec{q}_f^\circ,\vec{q}_f^*) \in \set{U}_f^\circ \times ( \set{U}_f^* \cap [0,\varsigma]^{J_f^*} )$. Letting $\vec{q}_f^* = \varsigma\vec{1} - (\varsigma - q_j)\vec{e}_j$ for some $q_j < \varsigma$ (that is, changing only the $j\ith$ products' price), the vector mean value theorem states that there exists some $\vec{r}_f^* = \varsigma\vec{1} - ( \varsigma - \tau )\vec{e}_j$, $\tau \in (q_j,\varsigma)$, such that
	\begin{align*}
		\hat{\pi}_f(\vec{p}_f^\circ(\vec{q}_f^*),\vec{q}_f^*,\vec{p}_{-f})
			= \hat{\pi}_f(\vec{p}) 
				+ (D_j\hat{\pi}_f)(\vec{r}_f^*,\vec{p}_f^\circ(\vec{r}_f^*),\vec{p}_{-f}) (\varsigma - q_j)
			> \hat{\pi}_f(\vec{p}). 
	\end{align*}
	Thus $\vec{p}_f$ is not locally profit-maximizing for $\hat{\pi}_f(\cdot,\vec{p}_{-f})$. By contraposition, (i) holds. 

	(ii): Define $\hat{\pi}_f^*(\vec{q}_f^*) = \hat{\pi}_f(\vec{q}_f^*,\vec{p}_f^\circ(\vec{q}_f^*),\vec{p}_{-f})$. Also let $\nabla_f^*\hat{\pi}_f$ and $\nabla_f^\circ\hat{\pi}_f$ denote the derivatives of firm $f$'s profits with respect to the prices of products in $\set{J}_f^*$ and $\set{J}_f^\circ$, respectively. Note that
	\begin{align*}
		(\nabla^*\hat{\pi}_f^*)(\vec{q}_f^*)
			&= (\nabla^*\hat{\pi}_f)(\vec{q}_f^*,\vec{p}_f^\circ(\vec{q}_f^*),\vec{p}_{-f})
				+ (D_f^*\vec{p}_f^\circ)(\vec{q}_f^*)^\top(\nabla^\circ\hat{\pi}_f)(\vec{q}_f^*,\vec{p}_f^\circ(\vec{q}_f^*),\vec{p}_{-f}) \\
			&= (\nabla^*\hat{\pi}_f)(\vec{q}_f^*,\vec{p}_f^\circ(\vec{q}_f^*),\vec{p}_{-f})
	\end{align*}
	because $(\nabla^\circ\hat{\pi}_f)(\vec{q}_f^*,\vec{p}_f^\circ(\vec{q}_f^*),\vec{p}_{-f}) = \vec{0}$, by definition. Let $\vec{q}_f^* = \varsigma\vec{1} - \bsym{\delta}$ for some $\bsym{\delta} \geq 0$, $\bsym{\delta} \neq \vec{0}$. Then the vector mean value theorem states that there exists $\vec{r}_f^* = \varsigma\vec{1} - \tau\bsym{\delta}$, $\tau \in (0,1)$, such that 
	\begin{equation*}
		\hat{\pi}_f(\vec{q}_f^*,\vec{q}_f^\circ,\vec{p}_{-f})
			\leq \hat{\pi}_f^*(\vec{q}_f^*)
			= \hat{\pi}_f^*(\vec{p}_f^*) - (\nabla_f^*\hat{\pi}_f^*)(\vec{r}_f^*)^\top\bsym{\delta}
			= \hat{\pi}_f(\vec{p}) - (\nabla_f^*\hat{\pi}_f)(\vec{r}_f^*,\vec{p}_f^\circ(\vec{r}_f^*),\vec{p}_{-f})^\top\bsym{\delta}
	\end{equation*}
	
	Note also that 
	\begin{align*}
		\bsym{\varphi}_f^*(\vec{q}_f^*,\vec{p}_f^\circ(\vec{q}_f^*),\vec{p}_{-f})
			&= \bsym{\varphi}_f^*(\vec{p})
				- \Big( (D_f^*\bsym{\varphi}_f^*)(\vec{p}) + (D_f^\circ\bsym{\varphi}_f^*)(\vec{p})(D_f^*\vec{p}_f^\circ)(\varsigma\vec{1}) \Big) \bsym{\delta} + \order(\norm{ \bsym{\delta} }^2) \\
			&= \bsym{\varphi}_f^*(\vec{p})
				- (D_f^*\bsym{\varphi}_f^*)(\vec{p}) \bsym{\delta}
				+ \order(\norm{ \bsym{\delta} }^2) \\
			&= \bsym{\varphi}_f^*(\vec{p})
				- ( \vec{I} - \bsym{\Omega}_f^*(\varsigma\vec{1}) ) \bsym{\delta}
				+ \order(\norm{ \bsym{\delta} }^2)
	\end{align*}
	Because $\bsym{\varphi}_f^*(\vec{p}) \leq \vec{0}$ and $1 - \omega_j(\varsigma) > 0$ for all $j \in \set{J}_f^*$, $\bsym{\varphi}_f^*(\vec{q}_f^*,\vec{p}_f^\circ(\vec{q}_f^*),\vec{p}_{-f}) < \vec{0}$ for all $\bsym{\delta} \neq \vec{0}$ sufficiently small. 
	
	For such $\bsym{\delta} \geq 0$, $\bsym{\delta} \neq \vec{0}$, also satisfying $\vec{q}_f^* = \varsigma\vec{1} - \bsym{\delta} \in \set{U}_f^* \cap [0,\varsigma]^{J_f^*}$, 
	\begin{equation*}
		(\nabla_f^*\hat{\pi}_f)(\vec{r}_f^*,\vec{p}_f^\circ(\vec{r}_f^*),\vec{p}_{-f})
			= \bsym{\Lambda}_f^*(\vec{r}_f^*,\vec{p}_f^\circ(\vec{r}_f^*),\vec{p}_{-f})
				\bsym{\varphi}_f^*(\vec{r}_f^*,\vec{p}_f^\circ(\vec{r}_f^*),\vec{p}_{-f})
			\geq \vec{0}
	\end{equation*}
	with at least one positive component. Thus $(\nabla_f^*\hat{\pi}_f)(\vec{r}_f^*,\vec{p}_f^\circ(\vec{r}_f^*),\vec{p}_{-f})^\top\bsym{\delta} > 0$, and 
	\begin{align*}
		\hat{\pi}_f(\vec{q}_f^*,\vec{q}_f^\circ,\vec{p}_{-f})
			\leq \hat{\pi}_f(\vec{p}) - (\nabla_f^*\hat{\pi}_f)(\vec{r}_f^*,\vec{p}_f^\circ(\vec{r}_f^*),\vec{p}_{-f})^\top\bsym{\delta} < \hat{\pi}_f(\vec{p}). 
	\end{align*}
	$\vec{p}_{-f}$ is thus a local maximizer of $\hat{\pi}_f(\cdot,\vec{p}_{-f})$. 
	\endproof

\begin{corollary}
	\label{COR:ZetaVI}
	Suppose $\vartheta > -\infty$, Assumption \ref{ASS:NonConstUnitCostsAssum} holds with $D^2C$ finite as $P \downarrow 0$, and \ref{LogitUtilityAssumption2} holds with $w$ that eventually decreases sufficiently quickly, has sub-quadratic second derivatives, and $\lim_{p \uparrow \varsigma} (D^2w)(\vec{y},p)$ exists. 
	(i) $\vec{p}_{-f}$ locally maximizes $\hat{\pi}_f(\cdot,\vec{p}_{-f})$, for any $\vec{p}_{-f} \in [0,\varsigma]^{J_{-f}}$, if, and only if, $\vec{p}_{-f}$ solves the VI
	\begin{equation}
		\label{EQN:ZetaVIf}
		\bsym{\varphi}_f(\vec{p})^\top(\vec{p}_f - \vec{q}_f) \leq \vec{0}
			\quad\text{for all}\quad 
			\vec{q}_f \in [0,\varsigma]^{J_f}. 
	\end{equation}
	(ii) $\vec{p}$ is a local equilibrium if, and only if, $\vec{p}$ solves the VI
	\begin{equation}
		\label{EQN:ZetaVI}
		\bsym{\varphi}(\vec{p})^\top(\vec{p} - \vec{q}) \leq \vec{0}
			\quad\text{for all}\quad 
			\vec{q} \in [0,\varsigma]^J. 
	\end{equation}
\end{corollary}

Theorem \ref{THM:ZetaVI} also suggests the following general demonstration that it is possible for prices to be equal to $\varsigma$ in equilibrium if unit costs to be constant and differ within firms. Let $c_j = \varsigma - \kappa$ be constant unit costs for some $\kappa > 0$ and observe that $\varsigma - c_j - \hat{\pi}_f(\vec{p}) < 0$ if, and only if, $\kappa < \hat{\pi}_f(\vec{p})$. Because $\hat{\pi}_f(\vec{p})$ is independent of $c_j$ when $p_j = \varsigma$, $\kappa$ can be made small enough so that it is less than any lower bound on the optimal profits for firm $f$ {\em excluding} product $j$ from their set of offerings. Setting $c_j = \varsigma - \kappa$ with such a value of $\kappa$ then ensures that $p_j = \varsigma$ is locally profit-optimal for the original problem including product $j$. 

\subsection{Existence of Equilibrium}

To prove the existence of equilibrium, it remains to show that profit-maximizing prices are unique. While the modified VI (\ref{EQN:ZetaVI}) can be used to characterize profit-maximizing prices, smooth nonlinear systems are often easier to analyze. Particularly, establishing the uniqueness of profit-maximizing prices with (\ref{EQN:ZetaVIf}) would traditionally require strict monotonicity of $\bsym{\varphi}_f$ \cite{Harker90}, a property that may be difficult to verify. These obstacles can be overcome by {\em continuously} extending the $\bsym{\zeta}$ map, and thus $\bsym{\varphi}$, to all of $[0,\infty)^J$ in such a way that solutions of the nonlinear system with the extended $\bsym{\varphi}$ are solutions to the VIs (\ref{EQN:ZetaVIf}) and (\ref{EQN:ZetaVI}). This enables an existence and uniqueness proofs using the same process applied above. Another approach, enabled by the analysis below, is to apply a VI uniqueness theorem due to Simsek et. al \cite[Proposition 5.1]{Simsek07} also based on the Poincare-Hopf Theorem. 

\begin{lemma}
	\label{LEM:ZetaExtension}
	Suppose $\vartheta > -\infty$, Assumption \ref{ASS:NonConstUnitCostsAssum} holds with $D^2C$ finite as $P \downarrow 0$, and \ref{LogitUtilityAssumption2} holds with $w$ that eventually decreases sufficiently quickly, has sub-quadratic second derivatives, and $\lim_{p \uparrow \varsigma} (D^2w)(\vec{y},p)$ exists. Define the map $\vec{z} : [0,\infty)^J \to \R^J$ componentwise by
	\begin{equation*}
		z_k(\vec{p})
			= \left\{ \begin{aligned} 
				&\sum_{j \in \set{J}_{f(j)}^\circ} P_j^L(\vec{p}) (p_j - c_j(P_j^L(\vec{p}))) + \frac{1}{\abs{(Dw_j)(p_j)}} 
					&&\quad\text{if } p_k < \varsigma \\
				&\omega_k(\varsigma)(p_k - \varsigma) + \sum_{j \in \set{J}_{f(k)}^\circ} P_j^L(\vec{p}) (p_j - c_j(P_j^L(\vec{p})))
					&&\quad\text{if } p_k \geq \varsigma
			\end{aligned} \right.
	\end{equation*}
	and the map $\bsym{\Phi} : [0,\infty)^J \to \R^J$ by $\bsym{\Phi}(\vec{p}) = \vec{p} - \vec{c}(\vec{P}(\vec{p})) - \vec{z}(\vec{p})$. 
	
	(i) $\vec{z}$ (or $\bsym{\Phi}$) is a continuously differentiable extension of $\bsym{\zeta}$ (or $\bsym{\varphi}$) from $[0,\varsigma]^J$ to $[0,\infty)^J$. (ii) For all $j$, $\Phi_j(\vec{p}) < 0$ when $\vec{p} \geq \vec{c}(\vec{P}(\vec{p}))$ and $p_j = c_j(P_j(\vec{p}))$ and there exists $\bar{p}_j$ such that $\Phi_j(\vec{p}) > 0$ for all $p_j > \bar{p}_j$, regardless of $\vec{p}_{-j}$. (iii) $\bsym{\Phi}_f(\vec{p}_f,\vec{p}_{-f}) = \vec{0}$ if, and only if, $\mathrm{proj}_{[0,\varsigma]}(\vec{p}_f)$ solves the VI (\ref{EQN:ZetaVIf}), where ``$\mathrm{proj}_{[0,\varsigma]}$'' denotes the projection onto $[0,\varsigma]^{J_f}$. (iv) $\bsym{\Phi}(\vec{p}) = \vec{0}$ if, and only if, $\mathrm{proj}_{[0,\varsigma]}(\vec{p})$ solves the VI (\ref{EQN:ZetaVI}), where ``$\mathrm{proj}_{[0,\varsigma]}$'' denotes the projection onto $[0,\varsigma]^J$. 
\end{lemma}
\proof
	(i): The claim concerning $\vec{z}$ and $\bsym{\zeta}$ follow by taking derivatives for any prices in $(0,\varsigma)^J$ and taking limits. Specifically, $(D_l \zeta_k)(\vec{p}) = \omega_k(p_k) \delta_{k,l} + (D_l\bar{\pi}_f)(\vec{p})$. Now, $(D_l\bar{\pi}_f)(\vec{p}) \to 0$ as $p_l \uparrow \varsigma$, from which we can deduce the following: 
	\begin{itemize}
		\item $(D_l\zeta_k)(\vec{p}) = (D_lz_k)(\vec{p})$ when $k,l \in \set{J}^\circ$, 
		\item $(D_l\zeta_k)(\vec{p}) \to \omega_k(\varsigma)\delta_{k,l} = (D_lz_k)(\vec{p})$ as $p_k,p_l \uparrow \varsigma$, 
		\item $(D_l\zeta_k)(\vec{p}) \to 0 = (D_lz_k)(\vec{p})$ when $p_k < \varsigma$ but $p_l \uparrow \varsigma$, and 
		\item $(D_l\zeta_k)(\vec{p}) \to (D_l\bar{\pi}_{f(k)})(\vec{p}) = (D_lz_k)(\vec{p})$ when $p_k \uparrow \varsigma$ but $p_l < \varsigma$. 
	\end{itemize}
	
	The claim concerning $\bsym{\Phi}$ and $\bsym{\varphi}$ is an obvious consequence, noting that $D_l [ c_k(P_k(\vec{p})) ] \downarrow 0$ as $p_k$ or $p_l \uparrow \varsigma$, and thus $\vec{c}(\vec{P}(\vec{p}))$ is continuously differentiable on $[0,\infty)^J$. 
	
	(ii): The first part of this claim follows from the corresponding result for $\varphi_j$. To prove the second part, note that, by definition, 
	\begin{equation*}
		\Phi_k(\vec{p})
			= ( 1 - \omega_k(\varsigma) ) ( p_k - \varsigma )
					+ \varsigma - c_k(0) -  \sum_{j \in \set{J}_{f(k)}^\circ} P_j^L(\vec{p}) (p_j - c_j(P_j^L(\vec{p})))
	\end{equation*}
	for all $p_j \geq \varsigma$. Because $\sum_{j \in \set{J}_{f(k)}^\circ} P_j^L(\vec{p}) (p_j - c_j(P_j^L(\vec{p})))$ is bounded over $[0,\infty)^J$ and $w_k$ has finite sub-quadratic second derivatives, $\bar{p}_k \geq \varsigma$ can be chosen large enough so that $\Phi_k(\vec{p}) > 0$ for all $p_k \geq \bar{p}_k$, regardless of $\vec{p}_{-k}$. 
	
	(iii) and (iv): We prove (iv), the proof for $(iii)$ being nearly identical. Let $\bsym{\Phi}(\vec{p}) = \vec{0}$. Then $\varphi_j(\vec{p}) = 0$ for all $j \in \set{J}^\circ$ and
	\begin{equation*}
		p_k - c_k(0) - \omega_k(\varsigma)(p_k - \varsigma) - \sum_{j \in \set{J}_{f(k)}^\circ} P_j^L(\vec{p}) (p_j - c_j(P_j^L(\vec{p}))) = 0. 
	\end{equation*}
	Therefore 
	\begin{equation*}
		\varsigma - c_k(0) - \sum_{j \in \set{J}_{f(k)}^\circ} P_j^L(\vec{p}) (p_j - c_j(P_j^L(\vec{p}))) 
			= - (1-\omega_k)(p_k-\varsigma)
			\leq 0. 
	\end{equation*}
	This implies that $\vec{p}$ solves the VI (\ref{EQN:ZetaVI}). Conversely, suppose $\vec{q} \in [0,\varsigma]^J$ solves the VI (\ref{EQN:ZetaVI}). Define $p_k = q_k$ for all $k \in \set{J}^\circ$ and $p_k = \varsigma -\varphi_k(\vec{q})/(1 - \omega_k(\varsigma))$ for all $k \in \set{J}^*$. Note that $\Phi_k(\vec{p}) = \varphi_k(\vec{q}) = 0$, $p_k \geq \varsigma$ for all $k \in \set{J}^*$ (because $\varphi_k(\vec{q}) \leq 0$ for all such $k$), and, for all $k \in \set{J}^*$, 
	\begin{align*}
		\Phi_k(\vec{p})
			&= p_k - c_k(0) - \omega_k(\varsigma)(p_k - \varsigma) - \sum_{j \in \set{J}_{f(k)}^\circ} P_j^L(\vec{q}) (p_j - c_j(P_j^L(\vec{q}))) \\
			&= (1-\omega_k(\varsigma))(p_k-\varsigma) + \varphi_k(\vec{q})
%			\\
%			&= - \varphi_k(\vec{q}) + \varsigma - c_k(0) - \sum_{j \in \set{J}_{f(k)}^\circ} P_j^L(\vec{q}) (p_j - c_j(P_j^L(\vec{q}))) 
			= 0. 
	\end{align*}
	Thus $\vec{q} = \mathrm{proj}_{[0,\varsigma]}(\vec{p})$ and $\bsym{\Phi}(\vec{p}) = \vec{0}$. 
\endproof

The obvious corollary is as follows: 
\begin{corollary}
	Assume the hypotheses of Lemma \ref{LEM:ZetaExtension}, and let $\vec{p} \in [0,\infty)^J$. (i) $\mathrm{proj}_{[0,\varsigma]}(\vec{p}_f)$ locally maximizes $\hat{\pi}_f(\cdot,\vec{p}_{-f})$ if, and only if, $\bsym{\Phi}_f(\vec{p}) = \vec{0}$. (ii) $\mathrm{proj}_{[0,\varsigma]}(\vec{p})$ is a local equilibrium if, and only if, $\bsym{\Phi}(\vec{p}) = \vec{0}$. 
\end{corollary}

An adaptation of the techniques in Sections \ref{SEC:LogitPriceEquilibrium} and \ref{SEC:QuantityCosts} again establishes the uniqueness of profit-maximizing prices. 
\begin{lemma}
	Assume the hypotheses of Lemma \ref{LEM:ZetaExtension}. For any $\vec{p}_{-f}$, the nonlinear system $\bsym{\Phi}_f(\vec{p}_f,\vec{p}_{-f}) = \vec{0}$ has a unique solution $\vec{p}_f \in [0,\infty)^{J_f}$ satisfying $c_j(P_j(\vec{p})) < p_j$ for all $j \in \set{J}_f$ with $p_j < \varsigma$ for at least one $j \in \set{J}_f$. 
\end{lemma}
\proof
	For any $\vec{p}_{-f}$, $\bsym{\Phi}_f(\cdot,\vec{p}_{-f})$ has a zero in the interior of $\{ \vec{p}_f : \vec{p}_f \geq \vec{c}_f(\vec{P}_f(\vec{p})) \}$; the proof is exactly analogous to the proof of existence in the case $\varsigma = \infty$: We first show that the homotopy $\bsym{\rho}_f$ between $[0,\infty)^{J_f}$ and $\{ \vec{p}_f : \vec{p}_f \geq \vec{c}_f(\vec{P}_f(\vec{p})) \}$ still exists. The inverse map $\vec{p}_f \mapsto \bsym{\epsilon}_f = \vec{p}_f - \vec{c}_f(\vec{P}_f(\vec{p}))$ is again well-defined and continuous, trivially. $\bsym{\rho}_f(\bsym{\epsilon}_f)$ is ostensibly defined by $\vec{p}_f = \vec{c}_f(\vec{P}_f^L(\vec{p})) + \bsym{\epsilon}_f$. For $\epsilon_k = \varsigma - c_k(0) + \delta$, $\delta \geq 0$, $\rho_k(\bsym{\epsilon}) = \varsigma + \delta$ solves this fixed-point equation regardless of $\epsilon_j$, $j \neq k$: 
	\begin{equation*}
		p_k = \varsigma + \delta
			= c_k(0) + (\varsigma - c_k(0) + \delta)
			= c_k(P_k(\vec{p})) + \epsilon_k
	\end{equation*}
	Note that $(D_k^\epsilon\rho_k)(\bsym{\epsilon}_f) = \delta_{k,l}$ for $\epsilon_k > \varsigma$. 
	
	Supposing $\vec{p}_f < \varsigma\vec{1}$, 
	\begin{equation*}
		D_f \Big[ \vec{c}_f(\vec{P}_f^L(\vec{p})) + \bsym{\epsilon}_f \Big]
			= (D^2\vec{C}_f)(\vec{P}_f^L(\vec{p}))\bsym{\Lambda}_f(\vec{p})
				- (D^2\vec{C}_f)(\vec{P}_f^L(\vec{p}))\vec{P}_f^L(\vec{p})\bsym{\lambda}_f(\vec{p})^\top. 
	\end{equation*}
	As $p_k \uparrow \varsigma$, the $k\ith$ row {\em and} $k\ith$ column of this matrix vanish because $\lim_{P \downarrow 0}(D^2C_j)(P) < \infty$. In particular, the proof that the spectrum of the fixed-point map does not contain 1 given before holds on $[0,\infty)^{J_f}$. Constructing an upper bound on the magnitude of the fixed point for any $\bsym{\epsilon}$ then proves the fixed point is unique, again by Kellogg's uniqueness theorem. 
	
	The vanishing of the derivatives also proves that $(D_l\rho_k)(\bsym{\epsilon}_f) \to \delta_{k,l}$ as either $p_k$ or $p_l$ $\uparrow \varsigma$, $k,l\in\set{J}_f$, and thus $\bsym{\rho}_f$ is continuously differentiable on $[0,\infty)^{J_f}$. $\bsym{\rho}_f$ must then be continuous on $[0,\infty)^{J_f}$. 
	
	As before, consider the vector field $\bsym{\Psi}_f = \bsym{\Phi}_f \circ \bsym{\rho}_f: [0,\infty)^{J_f} \to \R^{J_f}$. This vector field points outward on the boundary of $[\vec{0},\bar{\vec{p}}_f]$, where the existence of $\bar{\vec{p}}_f$ was established in Lemma \ref{LEM:ZetaExtension}. By the Poincare-Hopf Theorem, $\bsym{\Psi}_f$ has a zero $\bsym{\epsilon}_f \in (\vec{0},\bar{\vec{p}}_f)$, which is uniquely related to a zero $\vec{p}_f = \bsym{\rho}_f(\bsym{\epsilon}_f)$ satisfying $c_j(P_j(\vec{p})) < p_j$ for all $j \in \set{J}_f$. Note that this zero cannot have $p_j \geq \varsigma$ for all $j \in \set{J}_f$: For if that were true, then
	\begin{equation*}
		\varsigma 
			\leq \left( \frac{1}{1-\omega_j(\varsigma)} \right) c_j(0) 
				- \left( \frac{\omega_j(\varsigma)}{1-\omega_j(\varsigma)} \right) \varsigma 
			= - \left( \frac{1}{1-\omega_j(\varsigma)} \right) ( \varsigma - c_j(0) ) + \varsigma
			< \varsigma. 
	\end{equation*}
	By contradiction, there must exist some $j \in \set{J}_f$ such that $p_j < \varsigma$. 
	
	Uniqueness of profit-maximizing prices follows from a very similar approach to that used previously. Here we show that any zero of $\bsym{\Phi}_f$ has index one, and the rest of the proof proceeds exactly the same way. Note that 
	\begin{itemize}
		\item $(D_l\Phi_k)(\vec{p}) = (D_l\varphi_k)(\vec{p})$ when $k,l \in \set{J}_f^\circ$, 
		\item $(D_l\Phi_k)(\vec{p}) = (1-\omega_k(\varsigma)) \delta_{k,l}$ when $k,l \in \set{J}_f^*$
		\item $(D_l\Phi_k)(\vec{p}) = 0$ when $k \in \set{J}_f^\circ$ but $l \in \set{J}_f^*$
	\end{itemize}
	These relations imply that there exists a symmetric permutation $\vec{T}_f$ to make $(D_f\bsym{\Phi}_f)(\vec{p})$ block-triangular: 
	\begin{equation*}
		(D_f\bsym{\Phi}_f)(\vec{p})
			= \vec{T}_f 
				\begin{bmatrix} 
					(D_f^\circ\bsym{\varphi}_f^\circ)(\vec{p}) & \vec{0} \\
					\vec{A} & \vec{I} - \bsym{\Omega}_f^*(\varsigma)
				\end{bmatrix}
				\vec{T}_f
	\end{equation*}
	where $\vec{A}$ is some matrix. Thus
	\begin{equation*}
		\det (D_f\bsym{\Phi}_f)(\vec{p})
			= ( \det \vec{T}_f )^2
				\det \Big( (D_f^\circ\bsym{\varphi}_f^\circ)(\vec{p}) \Big)
				\det \Big( \vec{I} - \bsym{\Omega}_f^*(\varsigma) \Big)
	\end{equation*}
	and, because $1 - \omega_k(\varsigma) > 0$, 
	\begin{equation*}
		\mathrm{sign} \det (D_f\bsym{\Phi}_f)(\vec{p})
			= \mathrm{sign} \det \Big( (D_f^\circ\bsym{\varphi}_f^\circ)(\vec{p}) \Big). 
	\end{equation*}
	But $\bsym{\varphi}_f^\circ$ is identical to the $\bsym{\varphi}_f$ map if we exclude the products $\set{J}_f^*$ from $\set{J}_f$, and $\vec{p}_f^\circ$ are profit-maximizing prices strictly in the interior of $[0,\varsigma]^{J_f^\circ}$ where $J_f^\circ = | \set{J}_f^\circ |$. Then
	\begin{equation*}
		(-1)^{J_f^\circ}
			= \mathrm{ sign } \det (D_f^\circ\hat{\pi}_f^\circ)(\vec{p})
			= \mathrm{ sign } \det \bsym{\Lambda}_f^\circ(\vec{p})
				\mathrm{ sign } \det \Big( (D_f^\circ\bsym{\varphi}_f^\circ)(\vec{p}) \Big)
			= (-1)^{J_f^\circ} \mathrm{ sign } \det \Big( (D_f^\circ\bsym{\varphi}_f^\circ)(\vec{p}) \Big)
	\end{equation*}
	implies
	\begin{equation*}
		\mathrm{sign} \det (D_f\bsym{\Phi}_f)(\vec{p})
			= \mathrm{sign} \det \Big( (D_f^\circ\bsym{\varphi}_f^\circ)(\vec{p}) \Big)
			= 1. 
	\end{equation*}
\endproof

This result can also be proved using the uniqueness theorem for VI's given by Simsek et. al \cite[Proposition 5.1]{Simsek07}.

%%%%%%%%%%%%%%%%%%%%%%%%%%%%%%%%%%%%%%%%%%%%%%%%%%%%%%
%%%%%%%%%%%%%%%%%%%%%%%%%%%%%%%%%%%%%%%%%%%%%%%%%%%%%%
%%%%%%%%%%%%%%%%%%%%%%%%%%%%%%%%%%%%%%%%%%%%%%%%%%%%%%
\section{Properties of Equilibrium Prices}
\label{SEC:LogitStructure}

	This section establishes properties that the finite prices and markups of any equilibrium must satisfy based on properties of $\bsym{\zeta}$. The most general result is Corollary \ref{LogitEQMarkupDiffs}, which states that the difference between profit optimal markups for two products offered by the same firm with prices less than $\varsigma$ depends only on the prices and characteristics of those two products when unit costs are constant. This property is very similar to the embodiment of the ``Independence of Irrelevant Alternatives'' (IIA) property in Logit models: the ratio of choice probabilities depends only on the characteristics and prices of those two products \cite{Train03}. In Corollaries \ref{ConcaveLogitEQMarkupDiffs} through \ref{ConcaveLogitEQPricesValue} this result is applied to concave-in-price utility functions under hypotheses on the unit cost and value functions to illuminate some counterintuitive properties of equilibrium prices under Logit. This is the only section of the article that focuses somewhat on concave-in-price utility functions.

	\subsection{Properties of Profit-Optimal Prices}
	\label{SUBSEC:IntraFirmStructure}

	 For this subsection, we focus on a single firm $f \in \N(F)$ and derive our results as properties of locally profit-optimal prices. Naturally, these properties will be manifest in locally equilibrium prices as well. This section is also the only portion of this article in which we focus heavily on concave in price utilities, which will satisfy our existence conditions. 
	 
	The basic observation is as follows.
	\begin{corollary}
		\label{LogitEQMarkupDiffs}
		Suppose Assumptions \ref{ASS:NonConstUnitCostsAssum} and \ref{LogitUtilityAssumption2} hold. Let $\vec{p}_f$ be profit maximizing. For any $j,k \in \set{J}_f^\circ$, 
		\begin{equation}
			\label{EQN:MarkupDiffEqn}
			\Big( p_j - c_j(P_j(\vec{p})) \Big) - \Big( p_k - c_k(P_k(\vec{p})) \Big) 
				= - \left( \frac{1}{(Dw_j)(p_j)} - \frac{1}{(Dw_k)(p_k)} \right). 
		\end{equation}
	\end{corollary}
	That is, the difference between profit optimal markups for any two products offered by a single firm depends only on the corresponding utility derivatives. If unit costs are constant, this implies that the difference between profit optimal markups for any two products offered by a single firm depends only on the characteristics and prices of those products. 
	\proof Eqn. (\ref{EQN:MarkupDiffEqn}) follows immediately from the fixed-point equation $p_j = c_j(P_j(\vec{p})) + \bar{\pi}_f(\vec{p}) + \abs{(Dw_j)(p_j)}\inv$ for all $j \in \set{J}_f^\circ$. \endproof
	
	One application is motivated by the frequent application of constant coefficient linear in price utility functions. 
	\begin{corollary}
		Suppose Assumption \ref{ASS:NonConstUnitCostsAssum} holds. If $w(\vec{y},p) \equiv - \alpha p$ for some $\alpha > 0$, $p_j - c_j(P_j(\vec{p})) = p_k - c_k(P_k(\vec{p}))$ for all $j,k \in \set{J}_f$ when $\vec{p}_f$ is profit-maximizing. 
	\end{corollary}
	In other words, profit-optimal markups are constant regardless of product costs or the value of product characteristics. Constant intra-firm markups have appeared as an assumption \cite{Rossi06,Doraszelski06}, but not often proven to be an equilibrium outcome.
	
	The following example motivates the more general propositions on profit-optimal markups given below. Consider the quadratic in price utility $w(\vec{y},p) \equiv w(p) = - \alpha p^2$ and constant unit costs. Then 
	\begin{equation*}
		(p_j - c_j) - (p_k - c_k) = \left( \frac{1}{2\alpha} \right) \left( \frac{1}{p_j} - \frac{1}{p_k} \right), 
	\end{equation*}
	demonstrating that {\em locally profit optimal markups decrease with the corresponding prices} (i.e., $p_j - c_j > p_k - c_k$ if and only if $p_j < p_k$). Rearranging and setting $\lambda = 1 / ( 2\alpha )$, we obtain 
	\begin{equation*}
		\left( p_j - \frac{\lambda}{p_j} \right)
				- \left( p_k -  \frac{\lambda}{p_k} \right)
			= c_j - c_k. 
	\end{equation*}
	The function $\eta_\lambda(p) = p - \lambda/p$ is strictly increasing in $p$ for non-negative $\lambda$, and thus $c_j > c_k$ implies $p_j > p_k$. Thus, {\em locally profit optimal prices increase with costs while the corresponding markups decrease with costs}. Additionally, we note that if $c_j = c_k$ then $p_j = p_k$, even if $\vec{y}_j \neq \vec{y}_k$; that is, {\em profit-optimal prices reflect only product costs, not value}. 
	
	We first generalize the counterintuitive property that differences in characteristics that do not impact costs or (local) willingness to pay do not impact prices, even if they impact product value. 
	\begin{corollary}
		\label{ValueInvariantPrices}
		Suppose unit costs are constant, \ref{LogitUtilityAssumption2} holds and $w$ has sub-quadratic second derivatives. Let $\vec{p}_f$ be profit-maximizing, and suppose that $c_j = c_k$ and $(Dw_j)(p) = (Dw_k)(p)$ for all $p \in (0,\varsigma)$ for some $j,k \in \set{J}_f^\circ$, even if $\vec{y}_j \neq \vec{y}_k$. Then $p_j = p_k$. 
	\end{corollary}
	In other words, for any separable utility with sub-quadratic second derivatives, profit-optimal prices are determined by costs, not value. One would expect that real firms would not follow this rule, charging higher markups for the more valued product. 
	\proof
		Corollary \ref{LogitEQMarkupDiffs} implies
		\begin{equation*}
			p_j - \frac{1}{\abs{(Dw_j)(p_j)}} = \theta(p_j) = \theta(p_k) = p_k - \frac{1}{\abs{(Dw_k)(p_j)}}. 
		\end{equation*}
		Because the map $\theta(p) = p - 1/\abs{(Dw_j)(p)} = p - 1/\abs{(Dw_k)(p)}$ is strictly increasing when $w$ has sub-quadratic second derivatives, $p_j = p_k$. 
	\endproof
	
	This proposition, as stated, must be restricted to constant unit costs. A weaker result applies for non-constant unit costs: 
	\begin{corollary}
		\label{ValueInvariantPrices}
		Assume Assumption \ref{ASS:NonConstUnitCostsAssum} holds, unit costs are strictly convex, \ref{LogitUtilityAssumption2} holds, $w$ has sub-quadratic second derivatives. Let $\vec{p}_f$ be profit-optimal, and suppose that $c_j(P) = c_k(P) = c(P)$ for all $P \in [0,1]$ and $(Dw_j)(p) = (Dw_k)(p)$ for all $p \in [0,\varsigma)$ for some $j,k \in \set{J}_f^\circ$, even if $\vec{y}_j \neq \vec{y}_k$. Then $p_j > p_k$ if, and only if, $P_j(\vec{p}) > P_k(\vec{p})$, $p_j = p_k$ if, and only if, $P_j(\vec{p}) = P_k(\vec{p})$, and $p_j < p_k$ if, and only if, $P_j(\vec{p}) < P_k(\vec{p})$. 
	\end{corollary}
	This result may also be seen as slightly counterintuitive, as higher-priced products are intuitively associated with lower choice probabilities. 
	\proof
		Corollary \ref{LogitEQMarkupDiffs} implies
		\begin{equation*}
			\theta(p_j) - \theta(p_k)
				= c(P_j(\vec{p})) - c(P_k(\vec{p}))
		\end{equation*}
		Because total costs are strictly convex, unit costs are strictly increasing in $P$. Thus, 
		\begin{align*}
			\begin{matrix}
			p_j > p_k 
				& \iff & \theta(p_j) > \theta(p_k) 
				& \iff & c(P_j(\vec{p})) > c(P_k(\vec{p})) 
				& \iff & P_j(\vec{p}) > P_k(\vec{p}) \\
			p_j = p_k 
				& \iff & \theta(p_j) = \theta(p_k) 
				& \iff & c(P_j(\vec{p})) = c(P_k(\vec{p}))
				& \iff & P_j(\vec{p}) = P_k(\vec{p}) \\
			p_j < p_k
				& \iff & \theta(p_j) < \theta(p_k) 
				& \iff & c(P_j(\vec{p})) < c(P_k(\vec{p}))
				& \iff & P_j(\vec{p}) < P_k(\vec{p})
			\end{matrix}.
		\end{align*}
	\endproof
	
	 Corollary \ref{LogitEQMarkupDiffs} also implies the second counterintuitive property of locally profit optimal markups $-$ that they decrease with costs $-$ under Logit with any utility function that is both strictly concave in price and separable in price and characteristics. 
	\begin{corollary}
		\label{ConcaveLogitEQMarkupDiffs}
		Suppose that $w$ is separable in price and characteristics and strictly concave in price. Then firm $f$'s higher unit cost products (at optimality) have lower locally profit optimal markups. That is, if $j,k \in \set{J}_f^\circ$ and $c_j(P_j(\vec{p})) > c_k(P_k(\vec{p}))$, then $p_j - c_j(P_j(\vec{p})) < p_k - c_k(P_k(\vec{p}))$. 
	\end{corollary}
	\proof We prove that $p_j - c_j(P_j(\vec{p})) \geq p_k - c_k(P_k(\vec{p}))$ implies $c_j(P_j(\vec{p})) \leq c_k(P_k(\vec{p}))$. By Corollary \ref{LogitEQMarkupDiffs}, $p_j - c_j(P_k(\vec{p})) \geq p_k - c_k(P_k(\vec{p}))$ implies $(Dw)(p_j)\inv \leq (Dw)(p_k)\inv$ or, equivalently, $(Dw)(p_j) \geq (Dw)(p_k)$. By strict concavity, this implies that $p_j \leq p_k$. But then $p_j - c_j(P_j(\vec{p})) \geq p_k - c_k(P_k(\vec{p}))$ implies that
	\begin{equation*}
		c_j(P_j(\vec{p})) - c_k(P_k(\vec{p})) \leq p_j - p_k \leq 0
	\end{equation*}
	\endproof
	
	When unit costs are constant, these propositions can be easily connected to value. Intuition holds that both locally profit optimal markups and costs should increase with value, if not costs. The following assumption makes this connection explicit. 
	\begin{assumption}[Value Costs Hypothesis]
		More valued products cost more per unit to offer; that is, $v(\vec{y}) > v(\vec{y}^\prime)$ implies that $c(\vec{y}) > c(\vec{y}^\prime)$ for all $\vec{y},\vec{y}^\prime \in \set{Y}$. 
	\end{assumption}
	Mussa \& Rosen \cite{Mussa78} include this as a basic feature of cost functions. Bresnahan \cite{Bresnahan87} has also remarked that this is a natural condition. When considering equilibrium prices, this assumption need only be applied within firms. That is, there may be firm-specific cost functions each independently satisfying the value costs hypothesis, while the value costs hypothesis is violated across firms. This states that two distinct firms can produce a value-equivalent product at distinct unit costs without violating the results that apply this hypothesis. With this definition, we provide the following restatement of Corollary \ref{ConcaveLogitEQMarkupDiffs}. 
	\begin{corollary}
		\label{ConcaveLogitEQMarkupDiffsValue}
		Suppose unit costs are constant, Assumption \ref{LogitUtilityAssumption2} holds, $w$ is separable in price and characteristics, strictly concave in price, and that the value costs hypothesis holds. Then firm $f$'s higher value products have lower locally profit optimal markups. That is, if $j,k \in \set{J}_f$ and $v_j > v_k$, then $p_j - c_j < p_k - c_k$. 
	\end{corollary}
	\proof $v_j > v_k$ implies $c_j > c_k$, and the result follows from Corollary \ref{ConcaveLogitEQMarkupDiffs}. 
	\endproof
	
	Markups can increase with value when $w$ is convex in price. Consider $w(\vec{y},p) \equiv w(p) = - \alpha \log p$ and constant unit costs, for which
	\begin{equation*}
		( p_j - c_j ) - ( p_k - c_k ) = \left( \frac{1}{\alpha} \right) \left( p_j - p_k \right). 
	\end{equation*}
	Thus, locally profit optimal markups {\em increase} with the corresponding prices. This implies
	\begin{align*}
%		( p_j - c_j ) - ( p_k - c_k ) 
%			&= \left( \frac{\sigma}{\alpha} \right) \left( p_j - p_k \right) \\
%		p_j - p_k - \left( \frac{\sigma}{\alpha} \right) \left( p_j - p_k \right)
%			&= ( c_j - c_k ) \\
%		\left( 1 - \frac{\sigma}{\alpha} \right) ( p_j - p_k )
%			&= ( c_j - c_k ) \\
		p_j - p_k &= \left( \frac{1}{\alpha - 1} \right) ( c_j - c_k ). 
	\end{align*}
	Hence if $\alpha > 1$, locally profit optimal prices increase with costs, and locally profit optimal markups {\em increase} with costs. 
%	This too is general. 
%	\begin{cor}
%		\label{ConvexLogitEQMarkupDiffsValue}
%		Suppose that $w$ is separable in price and characteristics and strictly convex in price; suppose also that the value costs hypothesis holds. Then firm $f$'s higher value products have higher locally profit optimal markups. That is, if $j,k \in \set{J}_f$ and $v_j > v_k$, then $p_j - c_j > p_k - c_k$. 
%	\end{cor}
%	\begin{pf}
%		Suppose, on the contrary, that $p_j - c_j \leq p_k - c_k$. Then Corollary \ref{} implies that $(Dw)(p_j)\inv \geq (Dw)(p_k)\inv$. But because $w$ is strictly convex, this implies that $p_j \leq p_k$. Hence $c_k - c_j \leq p_k - p_j$
%	\end{pf}
	While this is a more intuitive outcome, it comes from a less intuitive utility specification. 
	
	Another assumption, the ``unique value hypothesis,'' further connects value with profit-optimal prices. As defined by Nagle \cite{Nagle87}, the unique value hypothesis postulates that as a product's combination of characteristics becomes more valued, individuals are less sensitive to price changes. This is transcribed to our framework as follows. 
	\begin{assumption}[Unique Value Hypothesis]
		For any $\vec{y},\vec{y}^\prime \in \set{Y}$, $v(\vec{y}) > v(\vec{y}^\prime)$ implies 
		\begin{equation*}
			\abs{(Dw)(\vec{y},p)} \leq \abs{(Dw)(\vec{y}^\prime,p)}
			\quad\text{ for all }\quad 
			p \in (0,\varsigma).
		\end{equation*} 
	\end{assumption}
	
	This definition suggests an example of a non-separable but convex in price utility for which markups increase with value. Consider $w(\vec{y},p) = - \alpha(\vec{y})p$, where $\alpha : \set{Y} \to (0,\infty)$, and assume unit costs are constant. Then
	\begin{equation*}
		(p_j - c(\vec{y}_j)) - (p_k - c(\vec{y}_k)) = \left( \frac{1}{\alpha(\vec{y}_j)} - \frac{1}{\alpha(\vec{y}_k)} \right), 
	\end{equation*}
	and $(p_j - c(\vec{y}_j)) \geq (p_k - c(\vec{y}_k))$ if and only if $\alpha(\vec{y}_j) \leq \alpha(\vec{y}_k)$. The unique value hypothesis mandates that $v(\vec{y}_j) > v(\vec{y}_k)$ implies $\alpha(\vec{y}_j) \leq \alpha(\vec{y}_k)$, and hence markups do not decrease with value if this hypothesis holds. Note that this is consistent with our previous result for constant coefficient linear in price utility where $\alpha(\vec{y}) \equiv \alpha \in (0,\infty)$. Whenever $\alpha(\vec{y}_j) < \alpha(\vec{y}_k)$, that is whenever the unique value hypothesis holds in a non-trivial way, markups can strictly increase with value. 

%	 A related and important question is whether higher value products have higher locally profit optimal prices. By Corollary \ref{ValueInvariantPrices}, this cannot hold without additional hypothesis. We show that this indeed holds, for concave in price utility functions, using the value costs and unique value hypotheses. This is based on the following extension to Lemma \ref{PhiIncreasing}. 
%	\begin{lemma}
%		\label{PsiIncreasing}
%		Let $\psi(\vec{y},p) = p + ( Dw)(\vec{y},p)\inv - c(\vec{y})$. (i) $\psi(\vec{y},\cdot)$ is a strictly increasing function of $p$ if, and only if, $w$ has sub-quadratic second derivatives. (ii) If both the value costs and unique value hypotheses hold, then $v(\vec{y}) > v(\vec{y}^\prime)$ implies $\psi(\vec{y},p) < \psi(\vec{y}^\prime,p)$. 
%	\end{lemma}
	
	 A related and important question is whether higher value products have higher locally profit optimal prices. By Corollary \ref{ValueInvariantPrices}, this cannot hold without an additional hypothesis. 
	\begin{corollary}
		\label{UniqueValueInequality}
		Suppose unit costs are constant, Assumption \ref{LogitUtilityAssumption2} holds, $w$ has sub-quadratic second derivatives, satisfies the unique value hypothesis, and the value costs hypothesis holds. Then firm $f$'s higher value products have higher locally profit optimal prices. That is, for any $j,k \in \set{J}_f^\circ$, $v_j > v_k$ implies that $p_j > p_k$ when $\vec{p}_f$ are profit-maximizing.
	\end{corollary}
	\proof
		The unique value hypothesis implies that when $v(\vec{y}) > v(\vec{y}^\prime)$, $\theta(\vec{y},p) \leq \theta(\vec{y}^\prime,p)$ for all $p \in (0,\varsigma)$, where $\theta(\vec{y},p) = p - \abs{(Dw)(\vec{y},p)}\inv$. Specifically, if $v(\vec{y}_j) > v(\vec{y}_k)$ then $\theta_j(p_k) \leq \theta_k(p_k)$. Suppose that $v(\vec{y}_j) > v(\vec{y}_k)$ while $p_j \leq p_k$. Because $\theta_j(p)$ is a strictly increasing function of $p$, we have 
		\begin{equation*}
			\theta_j(p_j) \leq \theta_j(p_k) \leq \theta_k(p_k). 
		\end{equation*}
		Thus Eqn. (\ref{EQN:MarkupDiffEqn}) implies that $c_j - c_k = \theta_j(p_j) - \theta_k(p_k) \leq 0$, in contradiction to the value costs hypothesis. 
	\endproof
	
	Because any separable utility trivially satisfies the unique value hypotheses, the following is a direct consequence of Corollary \ref{UniqueValueInequality}. 
	\begin{corollary}
		\label{ConcaveLogitEQPricesValue}
		Suppose unit costs are constant, Assumption \ref{LogitUtilityAssumption2} holds, $w$ is separable in price and characteristics, has sub-quadratic second derivatives, and that the value costs hypothesis holds. Then firm $f$'s higher value products have higher locally profit optimal prices. That is, $v_j > v_k$ implies that $p_j > p_k$ for any $j,k \in \set{J}_f^\circ$. 
	\end{corollary}
	
	%%%%%%%%%%%%%%%%%%%%%%%%%%%%%%%%%%%%%%%%%%%%%%%%%%%%%%
	%%%%%%%%%%%%%%%%%%%%%%%%%%%%%%%%%%%%%%%%%%%%%%%%%%%%%%
	%%%%%%%%%%%%%%%%%%%%%%%%%%%%%%%%%%%%%%%%%%%%%%%%%%%%%%
	
	\subsection{An Inter-Firm Property of Equilibrium Prices}
	\label{SUBSEC:InterFirmStructure}
	
	Eqn. (\ref{EQN:MarkupDiffEqn}) is a special case of the following: 
	\begin{corollary}
		\label{MPFMarkupDiffs}
		Suppose Assumptions \ref{ASS:NonConstUnitCostsAssum} and \ref{LogitUtilityAssumption2} hold, and let $\vec{p} \in (0,\infty)^J$ be equilibrium prices. For any $f,g \in \N(F)$, $j \in \set{J}_f^\circ$, and $k \in \set{J}_g^\circ$, 
		\begin{equation}
			\label{EQN:MPFMarkupDiffsEqn}
			( p_j - c_j(P_j(\vec{p})) ) - ( p_k - c_k(P_k(\vec{p})) )
				= \left( \bar{\pi}_f(\vec{p}) - \frac{1}{(Dw_j)(p_j)} \right)
					- \left( \bar{\pi}_g(\vec{p}) - \frac{1}{(Dw_k)(p_k)} \right). 
		\end{equation}
	\end{corollary}
	
	This equation expresses the existence of a {\em portfolio effect} present in equilibrium pricing with multi-product firms, constant unit costs, and even the simplest Logit model. 
	\begin{corollary}
		\label{PortfolioEffect}
		Assume unit costs are constant, Assumption \ref{LogitUtilityAssumption2} holds, $w$ have sub-quadratic second derivatives, and $\vec{p} \in (0,\infty)^J$ are equilibrium prices. Suppose that $\vec{y}_j = \vec{y}_k$ and $c_f(\vec{y}_j) = c_g(\vec{y}_k)$ for some $j \in \set{J}_f^\circ$ and $k \in \set{J}_g^\circ$. Then $p_j > p_k$ if, and only if, $\hat{\pi}_f(\vec{p}) > \hat{\pi}_g(\vec{p})$, $p_j < p_k$ if, and only if, $\hat{\pi}_f(\vec{p}) < \hat{\pi}_g(\vec{p})$, and $p_j = p_k$ if, and only if, $\hat{\pi}_f(\vec{p}) = \hat{\pi}_g(\vec{p})$. 
	\end{corollary}
	That is, equilibrium prices for the same product offered at the same cost but by different firms are influenced by the profitability of other products in these firms' portfolios. Stated another way, if the other products offered by a particular firm did {\em not} matter in determining equilibrium prices, then we would expect $\vec{y}_j = \vec{y}_k$ and $c_f(\vec{y}_j) = c_g(\vec{y}_k)$ for some $j \in \set{J}_f$ and $k \in \set{J}_g$ to imply that $p_j = p_k$. 
	\proof The proof follows by observing that Eqn. (\ref{EQN:MPFMarkupDiffsEqn}) can be written $\theta_j(p_j) - \theta_k(p_k) = \hat{\pi}_f(\vec{p}) - \hat{\pi}_g(\vec{p})$, and $\theta_j = \theta_k$. The result follows. 
	\endproof

\section{Conclusions}

	This article has proved the existence of equilibrium prices for Bertrand competition with multi-product firms under the Logit model without restrictive assumptions on the firms or their products. Instead of studying a particular utility function, general conditions on the utility function are identified under which existence holds. The proofs circumvents fundamental obstacles to the extension of existing equilibrium proofs for single-product firms by applying the Poincare-Hopf theorem. One of the fixed-point equations explicitly demonstrates that Logit price equilibrium problems are ``single-parameter problems'' when unit costs are constant, even when firms offer many products. By invoking the conventional assumption that utility is concave in price and separable in price and characteristics along with the reasonable assumption that more valued products always cost more to make per unit, a counterintuitive result is obtained: the more the population values a product's characteristics, the lower its profit-optimal markup. 
	
	There are at least two important areas for future research. One is establishing the uniqueness of equilibrium prices. Kellogg's uniqueness condition for Brouwer-Schauder fixed-point theorem \cite{Kellogg76}, used in Section \ref{SEC:QuantityCosts}, can be applied to show that equilibrium prices under {\em linear-in-price} utility Logit models are unique, a result already known for both single-product \cite{Milgrom90,Caplin91} and multi-product \cite{Sandor01, Konovalov10} firms. Generalizing this analysis to nonlinear utility functions and non-constant costs may be a promising direction. As suggested in the introduction, another important area is the extension of this analysis to non-Logit RUMs, especially those with heterogeneity. Formally the $\bsym{\eta}$ and $\bsym{\zeta}$ characterizations presented in this article extend to both any GEV and Mixed Logit models; see \cite{Morrow08, Morrow10a, Morrow10b} for the extension to Mixed-Logit models, subsequent analysis, and application in large-scale computations of equilibrium prices. Establishing the existence of simultaneously stationary prices using these characterizations is straightforward, but not enough to ensure the existence of equilibrium \cite{Morrow10b}. 

%%%%%%%%%%%%%%%%%%%%%%%%%%%%%%%%%%%%%%%%%%%%%%%%%%%%%
%%%%%%%%%%%%%%%%%%%%%%%%%%%%%%%%%%%%%%%%%%%%%%%%%%%%%
%%%%%%%%%%%%%%%%%%%%%%%%%%%%%%%%%%%%%%%%%%%%%%%%%%%%%

\appendix

%%%%%%%%%%%%%%%%%%%%%%%%%%%%%%%%%%%%%%%%%%%%%%
%%%%%%%%%%%%%%%%%%%%%%%%%%%%%%%%%%%%%%%%%%%%%%
%%%%%%%%%%%%%%%%%%%%%%%%%%%%%%%%%%%%%%%%%%%%%%

\section{Mathematical Notation} 
\label{APP:MathNotation}
	
	{\em Sets.} $\N$ denotes the natural numbers $\{1,2,\dotsc\}$, and $\N(N)$ denotes the natural numbers up to $N$, that is, $\N(N) = \{1,\dotsc,N\}$. $\R$ denotes the set of real numbers $(-\infty,\infty)$, $[0,\infty)$ denotes the non-negative real numbers, and $[0,\infty]$ denotes the extended non-negative half-line. We denote the $(J-1)$-dimensional simplex $\{ (x_1,\dotsc,x_N) \in [0,1]^N : \sum_{n=1}^N x_n = 1 \}$ by $\mathbb{S}(N)$, and the $J$-dimensional ``pyramid'' $\{ (x_1,\dotsc,x_N) \in [0,1]^N : \sum_{n=1}^N x_n \leq 1 \}$ by $\triangle(J)$. Hyper-rectangles in $\R^N$, i.e. sets of the form $[a_1,b_1] \times \dotsb \times [a_N,b_N]$ for some $a_n,b_n \in \R$ with $a_n < b_n$ for all $n \in \N(N)$, are denoted by $[\vec{a},\vec{b}]$ where $\vec{a} = (a_1,\dotsc,a_N)$ and $\vec{b} = (b_1,\dotsc,b_N)$. For other sets, we typically use calligraphic upper case letters such as ``$\set{A}$''. For any set $\set{A}$, $\abs{\set{A}}$ denotes its cardinality. For any $\set{B} \subset \set{A}$, $\set{A} \setminus \set{B}$ denotes the set $\{ b \in \set{A} : b \notin \set{B} \}$. For any set $\set{A}$, $\mathfrak{F}(\set{A})$ denotes the collection of finite subsets of $\set{A}$. 
	
	{\em Symbols.} Bold, un-italicized symbols (e.g., ``$\vec{x}$'') denote vectors and matrices; typically we reserve lower case letters to refer to vectors and use upper case letters to refer to matrices; the vector of choice probabilities ``$\vec{P}$'' is an exception made to conform with existing notation of these quantities. Throughout we use $\vec{1}$ to denote a vector of ones of the appropriate size for the context in which it appears. $\vec{I}$ always denotes the identity matrix of a size appropriate for the context. For any $\vec{x} \in \R^N$, $\diag(\vec{x})$ denotes the $N \times N$ diagonal matrix whose diagonal is $\vec{x}$. Any vector inequalities between vectors are to be taken componentwise: for example, $\vec{x} < \vec{y}$ means $x_n < y_n$ for all $n$. Random variables are denoted with capital letters ``$X$'', with random vectors being denoted with bold capital letters (e.g., ``$\vec{Q}$''). While this overlaps with our notation for matrices, it should not cause any confusion. $\Prob$ denotes a probability and $\Expect$ denotes an expectation. ``$\log$'' always denotes the natural (base $e$) logarithm. ``\esssup'' denotes the essential supremum of a measurable function, where the measure on measurable subsets of the domain should always be clear. 

	{\em Differentiation.} Our conventions for denoting differentiation follow \cite{Munkres91}. We use the symbol ``$D$'' to denote differentiation using subscripts to invoke additional specificity. Letting $\vec{f} : \R^M \to \R^N$, $(D_m f_n)(\vec{x})$ denotes the derivative of the $n\ith$ component function with respect to the $m\ith$ variable and $(D\vec{f})(\vec{x})$ is the $N \times M$ derivative matrix of $\vec{f}$ at $\vec{x}$ with components $( (D\vec{f})(\vec{x}) )_{n,m} = (D_m f_n)(\vec{x})$. Thus for $f:\R^M \to \R$, $(Df)(\vec{x})$ is a row vector. If $f : \R^M \to \R$, we define the gradient $(\nabla f)(\vec{x}) \in \R^M$ as the transposed derivative: $(\nabla f) (\vec{x}) = (Df)(\vec{x})^\top$. 
	
	{\em Other Definitions.} Let $\set{X}$ be any topological space and let $f : \set{X} \to \R$. We say $\vec{x}^* \in \set{X}$ is a {\em local maximizer (over $\set{X}$)} of $f$ if there exists a neighborhood of $\vec{x}^*$, say $\set{U}$, such that $f(\vec{x}^*) \geq f(\vec{x})$ for all $\vec{x} \in \set{U}$. We say $\vec{x}^* \in \set{X}$ is a {\em maximizer (over $\set{X}$)} of $f$ if $f(\vec{x}^*) \geq f(\vec{x})$ for all $\vec{x} \in \set{X}$.
	
%%%%%%%%%%%%%%%%%%%%%%%%%%%%%%%%%%%%%%%%%%%%%%%%%%%%%%
%%%%%%%%%%%%%%%%%%%%%%%%%%%%%%%%%%%%%%%%%%%%%%%%%%%%%%
%%%%%%%%%%%%%%%%%%%%%%%%%%%%%%%%%%%%%%%%%%%%%%%%%%%%%%
	
\section{Examples for the Logit Model}
\label{APP:LogitExamples}

	%%%%%%%%%%%%%%%%%%%%%%%%%%%%%%%%%%%%%%%%%%%%%%%%%%%%%%
	%%%%%%%%%%%%%%%%%%%%%%%%%%%%%%%%%%%%%%%%%%%%%%%%%%%%%%
	%%%%%%%%%%%%%%%%%%%%%%%%%%%%%%%%%%%%%%%%%%%%%%%%%%%%%%

	We first provide some examples of indirect utilities to illustrate properties (a-c). A linear in price utility, given by $w(\vec{y},p) = - \alpha(\vec{y})p$ for some $\alpha : \set{Y} \to (0,\infty)$, satisfies (a-c). More generally, any ``Cobb-Douglas'' in price utility, given by $w(\vec{y},p) = - \alpha(\vec{y})p^{\beta(\vec{y})}$ with $\alpha,\beta : \set{Y} \to (0,\infty)$, satisfies (a-c). A ``Cobb-Douglas'' specification for ``remaining income,'' $w(\vec{y},p) = \alpha(\vec{y})(\varsigma - p)^{\beta(\vec{y})}$ is a bit more complicated, being a function finite for all finite prices and satisfying (a-c) only for $\beta : \set{Y} \to (2\N + 1)$, where $2\N + 1$ denotes the set of odd positive integers: if $\beta(\vec{y}) : \set{Y} \to (-\infty,0)$ then $w$ is not finite for all finite $p$; clearly $w$ violates (a) if $\beta(\vec{y}) = 0$; if $\beta(\vec{y}) > 0$ is not an integer, then $w$ is complex for $p > \varsigma$; finally, if $\beta(\vec{y}) \in \N$ is not an {\em odd} positive integer then $w$ violates (a). The common ``log-transformed'' Cobb-Douglas in ``remaining income'' utility $w(\vec{y},p) = \alpha(\vec{y}) \log(\varsigma - p)$ for $p < \varsigma < \infty$, $\alpha : \set{Y} \to (0,\infty)$,\footnote{This log transformation usually occurs (see \cite{Berry95}, \cite{Rossi06}) based on the observation that choices are invariant over increasing utility transformations, so that $u^\prime(\vec{y},p) = e^{w(\vec{y},p)} e^{v(\vec{y})}$ yields the same {\em random} choices as the specification introduced in the text, with the caveat that the {\em additive} errors introduced in the text are taken as {\em multiplicative} errors (with a related distribution) in the former specification. In a Cobb-Douglas specification for the former, $u^\prime(\vec{y},p) \propto (\varsigma - p)^{\alpha(\vec{y})} = e^{ \alpha(\vec{y}) \log (\varsigma - p)}$, illustrating that the logarithm of this utility has the log-transformed specification for the price component.} is not finite for all finite prices. Allenby \& Rossi's negative log of price utility, given by $w(\vec{y},p) = - \alpha(\vec{y}) \log p$ for $\alpha : \set{Y} \to (0,\infty)$ satisfies (a-c) \cite{Allenby91}. Finally, the utility $w(p) = - \alpha ( \log p - \varepsilon \sin \log p )$, where $\alpha > 1$ and $\varepsilon \in (0,1)$, satisfies (a-c).
	
	We now demonstrate which of these utility functions is eventually log bounded and/or eventually decreases sufficiently quickly. Any linear in price or Cobb-Douglas in price utility is both eventually log bounded and eventually decreases sufficiently quickly. For if $\beta(\vec{y}) \geq 1$, $(Dw)(\vec{y},p) = - \alpha(\vec{y}) \beta(\vec{y}) p^{ \beta(\vec{y}) - 1 } \downarrow -\infty$ as $p \to \infty$. If $\beta(\vec{y}) < 1$, then although $(Dw)(\vec{y},p) = - \alpha(\vec{y}) \beta(\vec{y}) p^{ - (1 - \beta(\vec{y})) } \uparrow 0$ as $p \to \infty$, 
	\begin{equation*}
		(Dw)(\vec{y},p) - \frac{r}{p} 
			= - \alpha(\vec{y}) \beta(\vec{y})  \frac{1}{p^{1-\beta(\vec{y}) }} + \frac{r}{p}
			= \left( \frac{1}{p} \right) \left[ r - \alpha(\vec{y}) \beta(\vec{y})  p^{\beta(\vec{y})} \right]
			\leq 0
	\end{equation*}
	if $p \geq \sqrt[\beta(\vec{y})]{\alpha(\vec{y}) \beta(\vec{y}) /r}$ and hence $w(\vec{y},p)$ eventually decreases sufficiently quickly for any $r$. The class of negative log of price utility functions contain the most obvious examples of utilities that are neither eventually log bounded nor eventually decrease sufficiently quickly; particularly $w(\vec{y},p) \leq - \alpha(\vec{y}) \log p$ with $\alpha(\vec{y}) \leq 1$. If $\alpha(\vec{y}) < 1$ there are no finite profit maximizing prices under this utility. 
	
	In the text we defined utilities with sub-quadratic second derivatives. Any linear in price utility has sub-quadratic second derivatives, since $(D^2w)(\vec{y},p) \equiv 0$. More generally, under any Cobb-Douglas in price utility
	\begin{equation*}
		\frac{ (D^2w)(\vec{y},p) }{ (Dw)(\vec{y},p)^2 } 
			= - \left( \frac{1}{\alpha(\vec{y})} \right) \left( \frac{\beta(\vec{y}) - 1}{\beta(\vec{y})} \right) \left( \frac{1}{p^{\beta(\vec{y})}} \right) \\
			= \left( \frac{\beta(\vec{y}) - 1}{\beta(\vec{y})} \right) \left( \frac{1}{w(\vec{y},p)} \right), 
	\end{equation*}
	and hence $w$ has sub-quadratic second derivatives if $\beta(\vec{y}) \geq 1$. If $\beta(\vec{y}) < 1$, then $w$ has sub-quadratic second derivatives only at $(\vec{y},p)$ such that $\abs{w(\vec{y},p)} > (1-\beta(\vec{y}))/\beta(\vec{y})$, i.e. $p > \sqrt[\beta(\vec{y})]{(1-\beta(\vec{y}))/(\alpha(\vec{y})\beta(\vec{y}))}$. Finally, if $w(\vec{y},p) = - \alpha(\vec{y}) \log p$ then $(D^2w)(\vec{y},p)/(Dw)(\vec{y},p)^2 \equiv 1/\alpha(\vec{y})$ and hence $w$ has sub-quadratic second derivatives if $\alpha(\vec{y}) > 1$. Hence far from requiring concavity, some convex utility functions have sub-quadratic second derivatives.

	%%%%%%%%%%%%%%%%%%%%%%%%%%%%%%%%%%%%%%%%%%%%%%%%%%%%%%
	%%%%%%%%%%%%%%%%%%%%%%%%%%%%%%%%%%%%%%%%%%%%%%%%%%%%%%
	%%%%%%%%%%%%%%%%%%%%%%%%%%%%%%%%%%%%%%%%%%%%%%%%%%%%%%
	
	Let $\alpha(\vec{y}) \equiv \alpha > 0$. For the linear-in-price utility, $\zeta_j(\vec{p}) = \hat{\pi}_f(\vec{p}) + 1/\alpha$ with the fixed-point equation being $p_j = c_j + \hat{\pi}_f(\vec{p}) + 1/\alpha$. For any Cobb-Douglas in price utility, $\zeta_j(\vec{p}) = \hat{\pi}_f(\vec{p}) + (1/(\alpha\beta)) p_j^{1 - \beta}$ with the fixed-point equation being $p_j = c_j + \hat{\pi}_f(\vec{p}) + (1/(\alpha\beta)) p_j^{1 - \beta}$. For negative log of price, $\zeta_j(\vec{p}) = \hat{\pi}_f(\vec{p}) + (1/\alpha)p_j$ with the fixed-point equation being $p_j = c_j + \hat{\pi}_f(\vec{p}) + (1/\alpha)p_j$. 
	
	Our proof that the negative log of price utility has no finite profit maximizing prices can be strengthened using the relationship between $\bsym{\zeta}$ and the profit gradients. We already know that $w(\vec{y},p)/1 = - (\alpha(\vec{y})/1) \log p$ does not eventually decrease sufficiently quickly when $\alpha(\vec{y}) \leq 1$. We have also observed that $\zeta_j(\vec{p}) = \hat{\pi}_f(\vec{p}) + (1/\alpha_j)p_j$, which implies that
	\begin{equation*}
		\zeta_j(\vec{p}) - ( p_j - c_j )
			= ( \hat{\pi}_f(\vec{p}) + c_j ) + \left( \frac{1}{\alpha_j}  - 1 \right) p_j
			= ( \hat{\pi}_f(\vec{p}) + c_j ) + \left( \frac{1 - \alpha_j}{\alpha_j} \right) p_j. 
	\end{equation*}
	Thus, if $\alpha_j = \alpha(\vec{y}_j) \leq 1$, the $j\ith$ price derivative of profit is always positive. While we have already shown that only infinite prices maximize profits under this utility when $\alpha_j < 1$, this shows the same holds for $\alpha_j = 1$ as well even though the corresponding maximal profits are finite. 
	
%	The relationship between the fixed-point map $\bsym{\zeta}$ and the profit derivatives can be used to show that utilities that eventually decrease sufficiently quickly are necessary for satisfaction of a local criterion for finite profit-maximizing prices. 
%	\begin{proposition}
%		Let $w$ satisfy (a-c), and suppose that for some $j \in \set{J}_f$ and all $\vec{p}_{-j}$, there exists some $\bar{p}(\vec{p}_{-j})$ such that $(D_j\hat{\pi}_f)(p_j,\vec{p}_{-j}) < 0$ if $p_j > \bar{p}(\vec{p}_{-j})$. Then $w_j/1$ eventually decreases sufficiently quickly in the sense that there exists $r_j > 1$ and $\bar{p}_j > 0$ such that $(Dw_j)(p_j)/1 \leq - r /\bar{p}_j$ for $p_j > \bar{p}_j$.
%	\end{proposition}
%	
%	If $(D_j\hat{\pi}_f)(p_j,\vec{p}_{-j}) < 0$, then 
%	\begin{equation*}
%		(Dw_j)(p_j)\inv > \hat{\pi}_f(p_j,\vec{p}_{-j}) + c_j + p_j. 
%	\end{equation*}
%	Moreover, if $w_j$ does not eventually decrease sufficiently quickly, then for all $r > 1$ there exists some $p^\prime(r) > \bar{p}(\vec{p}_{-j})$ such that $(Dw_j)(p^\prime(r)) > - r/p^\prime(r)$; that is, $- p^\prime(r)/r > (Dw_j)(p_j)\inv$. Hence
%	\begin{equation*}
%		\left( \frac{r-1}{r} \right) p^\prime(r)
%			> \hat{\pi}_f(p^\prime(r),\vec{p}_{-j}) + c_j > c_j > 0. 
%	\end{equation*}
%	Clearly, if $p^\prime(r)$ is bounded as $r \downarrow 1$ then we can choose some $r$ so that these inequalities provide a contradiction. In principle, however, it is possible to construct a function $w$ such that $p^\prime(r) \to \infty$ like $(r-1)\inv$ as $r \downarrow 1$, and thus $(r-1)p^\prime(r)/r$ remains positive as $r \downarrow 1$. 
	
	We now present an example of a utility function for which has finite profit-maximizing prices but for which a ``local'' criterion restricting profit maximization at infinity fails. This local criterion is simply that profits decrease for all sufficiently large prices. Let $w(p) = - \alpha( \log p - \varepsilon \sin \log p )$ with $\alpha > 1$ and $\varepsilon \in [1 - \alpha\inv , 1)$. Then $p_j - c_j - \zeta_j(\vec{p}) \geq 0$ if and only if
	\begin{equation}
		\label{Ineq}
		p_j \left( 1 - \frac{1}{\alpha(1 - \varepsilon \cos \log p_j)} \right) \geq c_j + \hat{\pi}_f(\vec{p}). 
	\end{equation}
	But based on our choice of $\varepsilon$, there exist arbitrarily large $p_j$ such that the left hand side above is non-positive: For all $\bar{p}$ there exists some $p_j > \bar{p}$ such that $\alpha( 1 - \varepsilon \cos \log p_j ) = \alpha(1-\varepsilon) \leq 1$, which implies the claim. Since $c_j + \hat{\pi}_f(\vec{p})$ is positive (or rather is for all $\vec{p}$ that matter), the inequality (\ref{Ineq}) is violated and there exist arbitrarily large $p_j$ such that profits {\em increase}, locally, with $p_j$, despite the fact that profits must vanish as $\vec{p}_f \to \bsym{\infty}$ since this utility is eventually log bounded. That is, the {\em local} criterion for finite profit maximizing prices is violated.
	%%%%%%%%%%%%%%%%%%%%%%%%%%%%%%%%%%%%%%%%%%%%%%
%%%%%%%%%%%%%%%%%%%%%%%%%%%%%%%%%%%%%%%%%%%%%%
%%%%%%%%%%%%%%%%%%%%%%%%%%%%%%%%%%%%%%%%%%%%%%

\section{Inapplicability of Supermodularity}
	\label{SUBSEC:Supermodular}
	
	This appendix states a generalization of Sandor's \cite{Sandor01} result that profits are neither supermodular nor log-supermodular {\em arbitrarily close to equilibrium prices} under Logit with linear in price utility \cite[Chapter 4]{Sandor01}. Such a result rules out the applicability of the approach developed by Milgrom \& Roberts \cite{Milgrom90} to proving existence of equilibrium prices in the multi-product firm setting by implying that there cannot exist a compact set with non-empty interior containing any equilibrium on which Logit profits are supermodular or log-supermodular. 
	\begin{lemma}
		\label{NotSupermodular}
		Let $\vartheta > -\infty$, unit costs be constant, and Assumption \ref{LogitUtilityAssumption} hold with a $w$ with sub-quadratic second derivatives. Suppose $\vec{p}_f^* \in (0,\infty)^{J_f}$ maximizes $\hat{\pi}_f(\cdot,\vec{p}_{-f})$. Then for any $\varepsilon > 0$, there exists a $\vec{p}_f$ such that $\lvert\lvert \vec{p}_f - \vec{p}_f^* \rvert\rvert < \varepsilon$, and $(D_lD_k\hat{\pi}_f)(\vec{p}) < 0$, and $(D_lD_k \log \hat{\pi}_f)(\vec{p}) < 0$ for all $k,l \in \set{J}_f$, $k \neq l$, where $\vec{p} = (\vec{p}_f,\vec{p}_{-f})$.  
	\end{lemma}
	Naturally, because supermodularity has been used to prove the existence of equilibrium prices under Logit for single-product firms, the proof relies on the fact that firms produce more than one product. 
	\proof
		It can be shown that when $k,l \in \set{J}_f$ and $k \neq l$, the second derivatives of profits are given by
		\begin{align*}
			&(D_lD_k\hat{\pi}_f)(\vec{q}) \\
			& \quad\quad 
				= \abs{(Dw_k)(q_k)}P_k^L(\vec{q}) \big( \hat{\pi}_f(\vec{q}) - (q_k-c_k) - (Dw_k)(q_k)\inv \big) P_l^L(\vec{q}) \abs{(Dw_l)(q_l)} \\
			&\quad\quad\quad\quad
				+ \abs{(Dw_k)(q_k)}P_k^L(\vec{q}) \big( \hat{\pi}_f(\vec{q}) - (q_l-c_l) - (Dw_l)(q_l)\inv \big) P_l^L(\vec{q}) \abs{(Dw_l)(q_l)}
		\end{align*}
		for {\em any} $\vec{q}$. The goal is to choose $\vec{q}$, $\norm{\vec{q} - \vec{p}} < \varepsilon$, so that $\hat{\pi}_f(\vec{q}) - (q_k-c_k) - (Dw_k)(q_k)\inv < 0$ and $\hat{\pi}_f(\vec{q}) - (q_l-c_l) - (Dw_l)(q_l)\inv < 0$ for any $k,l \in \set{J}_f$, $k \neq l$. 
		
		By the $\bsym{\zeta}$ fixed-point characterization,
		\begin{align*}
			&\hat{\pi}_f(\vec{p}_f,\vec{p}_{-f}) - ( p_k - c_k ) - (Dw_k)(p_k)\inv \\
			&\quad\quad\quad\quad
				< \hat{\pi}_f(\vec{p}_f^*,\vec{p}_{-f}) - ( p_k - c_k ) - (Dw_k)(p_k)\inv \\
			&\quad\quad\quad\quad
				= \hat{\pi}_f(\vec{p}_f^*,\vec{p}_{-f}) - ( p_k^* - c_k ) - (p_k - p_k^*) - (Dw_k)(p_k)\inv \\
			&\quad\quad\quad\quad
				= (Dw_k)(p_k^*)\inv - (Dw_k)(p_k)\inv - (p_k - p_k^*). 
		\end{align*}
		Thus, $\hat{\pi}_f(\vec{p}_f,\vec{p}_{-f}) - ( p_k - c_k ) - (Dw_k)(p_k)\inv < 0$ if $\theta_k(p_k) \leq \theta_k(p_k^*)$. Because $w$ has sub-quadratic second derivatives, $\theta_k$ is strictly increasing and any $p_k < p_k^* - \varepsilon$ will do. The same logic goes for $l \in \set{J}_f$, and the claim follows. 
		
		For the second claim, note that
		\begin{align*}
			(D_l D_k \log \hat{\pi}_f)(\vec{p})
				&= \frac{ (D_lD_k\hat{\pi}_f)(\vec{p}) \hat{\pi}_f(\vec{p}) - (D_k\hat{\pi}_f)(\vec{p})(D_l\hat{\pi}_f)(\vec{p}) }{\hat{\pi}_f(\vec{p})^2}. 
%				&= \hat{\pi}_f(\vec{q})^{-2} \left( ( D_k \hat{\pi}_f )(\vec{q}) P_l(\vec{q}) (Dw_l)(q_l) \hat{\pi}_f(\vec{q}) 
%										+ (Dw_k)(q_k)P_k(\vec{q})( D_l \hat{\pi}_f )(\vec{q}) \hat{\pi}_f(\vec{q}) 
%										- (D_k\hat{\pi}_f)(D_l\hat{\pi}_f) \right) \\
		\end{align*}
		We have already established that the first term in the numerator is negative at $\vec{p}$ as defined above. Furthermore, 
		\begin{equation*}
			(D_k\hat{\pi}_f)(\vec{p})
				= \abs{ (Dw_k)(p_k) } P_k^L(\vec{p})( \hat{\pi}_f(\vec{p}) - (p_k - c_k) - (Dw_k)(p_k)\inv ) < 0
		\end{equation*}
		by the same argument and hence $(D_k\hat{\pi}_f)(\vec{p})(D_l\hat{\pi}_f)(\vec{p}) > 0$, making the second term in the numerator also negative. This completes the proof. 
	\endproof

%%%%%%%%%%%%%%%%%%%%%%%%%%%%%%%%%%%%%%%%%%%%%%%%%%%%%
%%%%%%%%%%%%%%%%%%%%%%%%%%%%%%%%%%%%%%%%%%%%%%%%%%%%%
%%%%%%%%%%%%%%%%%%%%%%%%%%%%%%%%%%%%%%%%%%%%%%%%%%%%%

\bibliographystyle{amsplain}
\bibliography{logit}

\end{document}